%% file: paper.tex
\documentclass[twoside,12pt]{article}
\renewcommand{\theequation}{\thesection.\arabic{equation}}
\newlength{\extraspace}
\newlength{\extraspaces}
\newcommand{\be}{\begin{equation}
\addtolength{\abovedisplayskip}{\extraspaces}
\addtolength{\belowdisplayskip}{\extraspaces}
\addtolength{\abovedisplayshortskip}{\extraspace}
\addtolength{\belowdisplayshortskip}{\extraspace}}
\newcommand{\ee}{\end{equation}}
\newcommand{\ba}{\begin{eqnarray}
\addtolength{\abovedisplayskip}{\extraspaces}
\addtolength{\belowdisplayskip}{\extraspaces}
\addtolength{\abovedisplayshortskip}{\extraspace}
\addtolength{\belowdisplayshortskip}{\extraspace}}
\newcommand{\ea}{\end{eqnarray}}
\newcommand{\bas}{\begin{eqnarray*}
\addtolength{\abovedisplayskip}{\extraspaces}
\addtolength{\belowdisplayskip}{\extraspaces}
\addtolength{\abovedisplayshortskip}{\extraspace}
\addtolength{\belowdisplayshortskip}{\extraspace}}
\newcommand{\eas}{\end{eqnarray*}}
\newcounter{subequation}[equation]
\makeatletter
\expandafter\let\expandafter\reset@font\csname reset@font\endcsname
\def\subeqnarray{\arraycolsep1pt
    \def\@eqnnum\stepcounter##1{\stepcounter{subequation}
        {\reset@font\rm(\theequation\alph{subequation})}}\eqnarray}

\makeatother
\newenvironment{theorem}[1]
{\vspace{3mm}\noindent {\bf #1 :} }{\vspace{2mm}}
\newcommand{\bt}[1]{\begin{theorem}{#1}}
\newcommand{\et}{\end{theorem}}
\newcommand{\newsection}[1]{
\vspace{12mm}
\pagebreak[3]
\addtocounter{section}{1}
\setcounter{equation}{0}
\setcounter{subsection}{0}
\addcontentsline{toc}{section}{\protect\numberline{\arabic{section}}{#1}}
\begin{flushleft}
{\large\bf \thesection. #1}
\end{flushleft}
\nopagebreak
\medskip
\nopagebreak}
 
\newcommand{\newsubsection}[1]{
\vspace{1cm}
\pagebreak[3]
 
\addtocounter{subsection}{1}
\addcontentsline{toc}{subsection}{\protect
\numberline{\arabic{section}.\arabic{subsection}}{#1}}
\noindent{ \bf \thesubsection. #1}
\nopagebreak
\vspace{2mm}
\nopagebreak}
 
\ifx\undefined\pdfoutput
  \usepackage{graphicx}
\else
  \usepackage[pdftex]{graphicx}
\fi
\newcommand{\N}{\mbox{I\hspace{-.4ex}N}}

\newcommand{\Z}{\mbox{{\sf Z}\hspace{-1ex}{\sf Z}}}

\newcommand{\eins}{\mbox{\rm 1\hspace{-.6ex}I}}

\renewcommand{\Box}{\rule{3mm}{3mm}}
\newcommand{\cA}{{\cal A}}

\newcommand{\cC}{{\cal C}}
\newcommand{\cF}{{\cal F}}

\newcommand{\cL}{{\cal L}}

\newcommand{\cS}{{\cal S}}
\begin{document}
%
%
\begin{titlepage}
%
%
\begin{flushright}
MZ-TH/01-25
\end{flushright}
\vspace*{1.5cm}
\begin{center}
{\Large {\bf Quantum Field Theory On A Discrete Space}}\\[4mm]
{\Large {\bf And}}\\[4mm]
{\Large {\bf Noncommutative Geometry}}\\[2.5cm]
Rainer H\"au\ss ling\\[4mm]
Institut f\"ur Physik, WA THEP\\[4mm]
Johannes Gutenberg-Universit\"at Mainz\\[4mm]
D-55099 Mainz, Germany\\[3cm]
{\bf Abstract}\\[6mm]
\end{center}
We analyse in detail the quantization of a simple noncommutative model
of spontaneous symmetry breaking in zero dimensions taking into account
the noncommutative setting seriously. The connection to the counting
argument of Feynman diagrams of the corresponding theory in four dimensions
is worked out explicitly. Special emphasis is put on the motivation as well
as the presentation of some well-known basic notions of quantum field theory
which in the zero-dimensional theory can be studied without being spoiled
by technical complications due to the absence of divergencies.
\end{titlepage}
%
%
\newpage
\thispagestyle{empty}
\input{leerseite}
\newpage
\pagenumbering{roman}
\tableofcontents
\input{chap0}
\input{chap1}

\input{chap2}
\input{chap3}

\input{chap4}

\input{chap5}

\appendix
\renewcommand{\newsection}[1]{
\vspace{12mm}
\pagebreak[3]
\addtocounter{section}{1}
\setcounter{equation}{0}
\setcounter{subsection}{0}
\addcontentsline{toc}{section}
{\protect\numberline{\Alph{section}}{#1}}
\begin{flushleft}
{\large\bf \thesection. #1}
\end{flushleft}
\nopagebreak
\medskip
\nopagebreak}
\input{appa}

\input{appb}

\input{appc}

\input{appd}

\input{appe}
\input{appf}
\input{appg}
\input{apph}
\input{appi}

\input{appj}
\newpage
{

}
%
%
\input{leerseite}
\input{dank}
\end{document}

%% file: leerseite.tex
\newpage
\thispagestyle{empty}
\vspace*{1cm}

%% file: chap0.tex
\newpage
\newsection{Introduction}
\pagenumbering{arabic}

Without any doubt, one of the most impressive successes of high energy 
particle physics in the recently bygone century consists of the description 
of fundamental interactions in terms of local quantum gauge theories 
culminating in the formulation of the standard model of electroweak and 
strong interactions. However, in spite of its almost perfect
agreement with experiment, from a theoretical point of view the standard
model contains some serious drawbacks like, for instance, the large number
of free parameters and the unsettled origin of the Higgs mechanism and
the Yukawa couplings. Beside many other theoretical approaches that
we do not want to follow up, a promising ray of hope showed up with the
development of Noncommutative Geometry about 15 years ago. Already at its
beginnings several attempts were started concerning the application of ideas
borrowed from this new mathematical branch to elementary particle physics
in order to overcome eventually the unpleasant features of the ordinary
description. As a result, a whole battery of such noncommutative gauge
models emerged each of which, on the one hand, more or less resolves the
disadvantages indicated above. On the other hand, however, this noncommutative
model building is mainly restricted to the derivation of the classical theory,
and second quantization of the models is afterwards realized in a habitual
manner, thus disregarding completely the noncommutative origin of the
formulation. Obviously, such a treatment is quite unsatisfactory.\\
Before further pursuing the argument just addressed we would like to make
a step backwards in order to state explicitly the purpose of the present
thesis which, in fact, is a threefold one:\\
Since long ago it is well-known that the counting of the number of Feynman
diagrams in a given theory can be achieved in an easy manner by studying
the zero-dimensional version of the four-dimensional theory in question.
This connection holds true as long as the model under investigation does
not include spontaneous symmetry breaking or, more precisely, as long as
there are no massless modes involved in the game. The inquiry of the question
whether or not the counting argument keeps its validity for theories with
spontaneous breakdown of symmetry, too, constitutes the first purpose of
this thesis.\\
The second aim is a pedagogical one: The zero-dimensional models treated
below feature the fact that all relevant quantities to be calculated are
finite from the outset in contrast to the general behaviour of those
quantities in four dimensions. Thus, due to the complete absence of
divergencies the zero-dimensional models provide a well-suited ``field''
theoretical stage for motivating and studying some basic notions of
ordinary quantum field theory such as renormalization, gauge fixing,
ghost contributions, Slavnov-Taylor identity and so on in an elementary
framework that is not perturbed by complications of a more technical
nature.\\
Finally, let us return to the introductory remarks given above. Because
tackling the problem of finding an adequate quantization of physically
significant noncommutative gauge models in full generality is a really
hard task it is certainly more promising to start the investigation of
this problem at first by looking at simple but nontrivial examples of
such noncommutative models. The zero-dimensional theory of spontaneous
symmetry breaking studied later on represents such a model, and,
hence, as our third goal we try to analyze how the original noncommutative
structure enters the quantization procedure.\\[.3cm]
In detail, this thesis is organized as follows:\\
In chapter~2 we briefly review $\varphi^4$-theory in zero dimensions both
for refreshing the basic reasoning underlying the counting argument of
Feynman diagrams and for illustrating the notion and the necessity of
renormalization in a very simple ``field'' theoretical framework even though
there are no divergencies at all in the present case. The methods of 
investigation developed in the first part of this chapter are to some extent
orthogonal to the ones in the existing literature and will serve as a guide
line for the more involved considerations to follow.\\
The question whether or not the counting argument of Feynman diagrams holds
true also for theories including spontaneous symmetry breaking (SSB) is posed
and answered in the affirmative in chapter~3 which starts off with the naive
definition of the zero-dimensional version of a simple model with SSB
consisting of one complex scalar ``field'' $\Phi$ with a self-interaction
of the Mexican hat type but not comprising ordinary gauge fields. The problems
in connection with the presence of a massless Goldstone mode are discussed
extensively entailing the introduction of a gauge fixing term 
and a Faddeev-Popov
term which even in zero dimensions are necessary for a proper
definition of the theory. Chapter~3 again ends with a study of 
renormalization.\\
Chapter~4 at first deals with the equivalent formulation of the model just
mentioned within a special noncommutative language which emerged as a
mathematically less involved alternative to the model building due to Connes
and Lott from a collaboration of physicists from Marseille and Mainz. In the
Marseille-Mainz approach the basic noncommutative object is a generalized
gauge potential which in general, i.e. in more complicated physical theories,
unifies both ordinary gauge fields and Higgs fields 
into one single matrix-valued quantity.
An adequate quantization of such a noncommutative model obviously should take
into account the original noncommutative structure seriously. For these
purposes, in chapter~3 a well-adapted matrix calculus is developed which
simplifies the subsequent determination of the higher orders of the 
zero-dimensional theory substantially. All this is explained in chapter~3
at length which, on the other hand, is intended to present only a first
insight into the functionality of the matrix calculus. Accordingly, in 
chapter~4 the gauge fixing and the ghost contributions are completely
neglected. These contributions, however, are essential and indispensable
for a consistent and proper treatment of the whole theory.\\
Chapter~5 closes this gap once more. In a first part the problems with a really
unifying incorporation of ghost contributions - this unification taken
in a strictly noncommutative sense - are discussed in detail. The negative
outcome of this discussion notwithstanding an ad-hoc inclusion of these
contributions is nevertheless possible, and chapter~5 concludes with the
description of how higher orders of the by now well-defined theory can be
calculated within the noncommutative setting.\\
Finally, chapter~6 summarizes the results and raises some unsettled questions
which will need further investigation.\\
In order not to disturb the line of argument in the main text unnecessarily
technical details of the calculations are moved to several appendices as far
as possible.

%% file: chap1.tex
\newpage
\newsection{Zero-dimensional $\varphi^4$-theory}

We want to begin our investigations with a short review of a 
quite simple example, namely $\varphi^4$-theory in zero spacetime
dimensions. As it is well-known since long ago (see e.g. 
\cite{CLP}, \cite{ZJ}, \cite{IZ}, and references therein), studying
zero-dimensional ``field'' theories amounts to nothing else but
counting the number of Feynman diagrams (weighted with possible
symmetry factors) of the corresponding field theory in four
spacetime dimensions. This result is mainly due to the fact that
in zero dimensions each Feynman propagator appearing in the
Feynman diagrams just contributes a trivial factor of $1$ (up to
normalization) to the
evaluation of the diagram, see also the end of 
section~1 of this chapter for some details.\\
The aforementioned statement holds true as long as there is no
spontaneous symmetry breaking and as long as the zero-dimensional
theory is not renormalized. Of course, renormalization of the
zero-dimensional theory is only in need if this theory is to have
a real {\it physical} meaning, like, e.g., in the case of discrete
(parts of) models within noncommutative geometry. In the present
context of $\varphi^4$-theory, the renormalization of the
theory in zero dimensions, which we will study in section~3, is purely for
illustrative pedagogical reasons, showing, however, in an extremely
simple field theoretical framework, what renormalization in general
is needed for in a (here hypothetically) physical theory, even if
there are {\it no divergencies} at all.\\
In section~1 we explicitly calculate the $n$-point functions of
zero-dimensional $\varphi^4$-theory and compare the results with
the counting of Feynman diagrams of the theory in four dimensions.
Of course, this kind of considerations has already been done in the
literature at length. The methods, however, we are going to employ
are to some extent different from the ones of previous works and will
also give us a hint for the considerations to follow. 
Especially, we will obtain a closed analytical expression for
the $n$-point functions valid to all orders in perturbation theory.
Furthermore, we will
have the opportunity to fix some notation.\\
Section~2 is devoted to the study of connected and 1 PI 
(one-particle-irreducible) Green's functions thereby establishing
in the present simple example of $\varphi^4$-theory some
techniques which will prove very useful later on when we turn our
attention to the more complicated model including spontaneous
symmetry breaking.   

\noindent
\newsubsection{$n$-point functions of zero-dimensional
$\varphi^4$-theory}

In four-dimensional Euclidean space(time) the classical action of
the $\varphi^4$-theory is given by
\be \label{action4}
S [\varphi ] = \int \; d^4 x \; \left\{
\frac{1}{2} (\partial_\mu \varphi (x))(\partial^\mu \varphi (x)) \; + \;
\frac{1}{2} m^2 \varphi^2 (x) \; + \;
\frac{\lambda}{4!} \varphi^4 (x) \right\} \; \; \; .
\ee
The corresponding action $S (y)$ in zero dimensions is derived from
(\ref{action4}) by suppressing the argument $x$ of $\varphi (x)$ and,
in addition, by omitting all derivatives of the field $\varphi$ as well
as the overall integration, i.e.:
\be \label{action0}
S (y) = \frac{1}{2} m^2 y^2 \; + \; \frac{\lambda}{4!} y^4
\ee
In (\ref{action0}) $\lambda$ is a parameter tuning the strength of the
(self)interaction and $m^2$ a further parameter of dimension
$(mass)^2$, which is kept for later purposes.\\
As usual, the generating functional $Z (j)$ of general Green's functions
is obtained by coupling the theory to an external source $j$ and summing
over all possible ``paths'':
\be \label{zzz}
Z (j) = \; {\cal N} \; \int_{- \infty}^{+ \infty} \; d y \;
\mbox{exp} \left\{ \frac{1}{\hbar} \left( -S(y) + jy
\right) \right\}
\ee
Because we are in zero dimensions the integration in (\ref{zzz}) is just
an ordinary integration. Please note that we will keep factors of
$\hbar$ explicitly, in order to allow for a loop expansion eventually.
On the other hand, we put $c \equiv 1$ throughout the whole thesis.
The normalization constant $\cal N$ (well-defined in zero dimensions)
is fixed by the requirement
\be \label{zzznorm}
Z (j = 0) = 1 \; \; \; .
\ee
The $n$-point functions $G_n (\lambda, m^2)$, i.e. the Green's functions
with $n$ external legs, result from (\ref{zzz}) by multiple differentiation
with respect to $j$ and final evaluation at the point $j = 0$:
\ba \label{npoint}
G_n (\lambda, m^2) & = & \left. \left( \frac{\hbar \partial}{\partial j} 
                         \right)^n \; Z (j) \; \right|_{j = 0} \\[.5ex]
& = & {\cal N} \; \int_{- \infty}^{+ \infty} \; d y \; y^n \;
      \mbox{exp} \left\{ - \frac{1}{\hbar} S (y) \right\} \nonumber
\ea
Hence, at the moment we are interested in an exact analytical
solution of an integral of the form
\be \label{intn}
I_n (\lambda, m^2) = \int_{- \infty}^{+ \infty} \; d y \; y^n \;
\mbox{exp} \left\{ \frac{1}{\hbar} \left(
- \frac{1}{2} m^2 y^2 - \frac{\lambda}{4!} y^4 \right) \right\} \; \; \; ,
\ee
the calculation of which will be briefly sketched in the following.\\
In a first step we substitute $y$ according to
$x = \left( \frac{\lambda}{\hbar 4!} \right)^\frac{1}{4} y$.
This leads to
\ba \label{inthn}
I_n (\lambda, m^2) & = & \left( \frac{\lambda}{\hbar 4!} 
                     \right)^{- \frac{n + 1}{4}} \;
\hat{I}_n (\lambda, m^2) \\
\mbox{with } \; \;
\hat{I}_n (\lambda, m^2) & = & 
\int_{- \infty}^{+ \infty} \; d x \; x^n \;
\mbox{exp} \left\{ - \frac{1}{2} m^2 \sqrt{\frac{4!}{\hbar \lambda}} \; x^2
- x^4 \right\} \; \; \; . \nonumber
\ea
Using the explicit integral representation for $\hat{I}_n$ and integration
by parts once one easily verifies that $\hat{I}_n$ obeys an ordinary
linear differential equation of second order,
\be \label{dgl1}
\frac{8}{3} \; \hbar \; \lambda^3 \; \frac{d^2 \hat{I}_n}{d \lambda^2} \; + \;
4 \; \lambda \; (m^4 + \hbar \lambda) \; \frac{d \hat{I}_n}{d \lambda} \; - \;
m^4 \; (n + 1) \; \hat{I}_n \; = \; 0 \; \; \; ,
\ee
which by means of the further substitution
\be \label{lamlamh}
\hat{\lambda} = \frac{3 m^4}{2 \hbar \lambda}
\ee
transforms into a differential equation of Kummer's type (see appendix~A):
\be \label{dgl2}
\hat{\lambda} \; \frac{d^2 \hat{I}_n}{d \hat{\lambda}^2} \; + \;
(\frac{1}{2} - \hat{\lambda}) \; \frac{d \hat{I}_n}{d \hat{\lambda}} \; - \;
\frac{n + 1}{4} \; \hat{I}_n \; = \; 0
\ee
Hence, making use of (\ref{solkum}) as well as (\ref{lamlamh}) and
(\ref{inthn}) we find for $I_n (\lambda, m^2)$ the expression
\be \label{loesintn}
I_n (\lambda, m^2) = c_1 \; \lambda^{- \frac{n +1}{4}} \;
{}_1 F_1 \left( \frac{n + 1}{4}; \frac{1}{2}; 
\frac{3 m^4}{2 \hbar \lambda} \right) \;
+ \; c_2 \; \lambda^{- \frac{n + 3}{4}} \; m^2 \;
{}_1 F_1 \left( \frac{n + 3}{4}; 
\frac{3}{2}; \frac{3 m^4}{2 \hbar \lambda} \right)
\; \; \; ,
\ee
where $c_1$ and $c_2$ are two integration constants, which can be fixed
by means of (\ref{intnent}) taking into account the asymptotic
behaviour of ${}_1 F_1$ (\ref{f11asy}). The final answer for
$I_n (\lambda, m^2)$ is therefore:
\be \label{loesintnf}
I_n (\lambda, m^2) = \sqrt{2} \; \pi \; \frac{n!}{n!!} 
\left( \frac{3 \hbar}{2 \lambda} \right)^\frac{n + 1}{4} 
\left\{ \displaystyle\frac{{}_1 F_1 (\frac{n + 1}{4}; \frac{1}{2};
\frac{3 m^4}{2 \hbar \lambda} )}{\Gamma (\frac{n + 3}{4} )} \; - \;
2 \left( \frac{3 m^4}{2 \hbar \lambda} \right)^\frac{1}{2} \;
\frac{{}_1 F_1 (\frac{n + 3}{4}; \frac{3}{2}; \frac{3 m^4}{2 \hbar 
\lambda})}{\Gamma (\frac{n + 1}{4} )} \right\}
\ee
This result can be rewritten in a more compact form due to a relation
between the confluent hypergeometric function ${}_1 F_1$ and
Weber's parabolic cylinder function $U$, see (\ref{f11u}):
\be \label{loesintnfa}
I_n (\lambda, m^2) = \sqrt{2 \pi} \; \frac{n!}{n!!} \;
\left( \frac{3 \hbar}{\lambda} \right)^\frac{n + 1}{4}
\mbox{exp} \left\{ \frac{3 m^4}{4 \hbar \lambda} \right\} \; 
U \! \left( \frac{n}{2},
\sqrt{\frac{3 m^4}{\hbar \lambda}} \right)
\ee
Thus taking into account the correct normalization factor $\cal N$,
see (\ref{zzznorm}), the $n$-point functions $G_n (\lambda, m^2 )$,
that we are looking for, are finally given by the following
completely analytical closed expression:
\ba \label{npointf}
G_n (\lambda, m^2 ) & = & \frac{I_n (\lambda, m^2 )}{I_0 (\lambda, m^2 )} \\
\label{npointff}
& = & \frac{n!}{n!!} \left( \frac{3 \hbar}{\lambda} \right)^{\! \frac{n}{4}}
\frac{U \! \left( \frac{n}{2}, \sqrt{\frac{3 m^4}{\hbar \lambda}} 
\right)}{U \! \left( 0, \sqrt{\frac{3 m^4}{\hbar \lambda}} \right)}
\; \; \; (n \mbox{ even})
\ea
For $n$ odd, we find $G_n (\lambda, m^2 ) \equiv 0$ because of
$I_n (\lambda, m^2 ) \equiv 0$ for $n$ odd, as can be seen directly
from (\ref{intn}). Of course, this last statement is in agreement
with the fact that there are no Green's functions in $\varphi^4$-theory
for an odd number of external legs.\\[0.3cm]
Let us now turn for a while to the connection between the
considerations above for zero-dimensional $\varphi^4$-theory and
the counting of Feynman diagrams of the corresponding theory in four
dimensions. To this end we first need the free Feynman propagator $\Delta_F$
of the zero-dimensional theory. In fact, $\Delta_F$ can be calculated
in various ways, all of which lead to the same result. For example, we
could use (\ref{npointff}) in the case $n = 2$ and $\lambda = 0$
bearing in mind the asymptotic expansion of ${}_1 F_1$ or
$U$, respectively. Another possibility is to read off $\Delta_F$
from (\ref{npointp}) setting $n = 2$ and $\lambda = 0$ in
(\ref{npointp}). 
Alternatively, we can determine $Z_{free} (j)$ (i.e. $Z (j)$ for
$\lambda = 0$),
\be \label{zzzfrei}
Z_{free} (j) \; = \; \frac{\int_{- \infty}^{+ \infty} d y \;
\mbox{exp} \{ - \frac{1}{2} \frac{m^2}{\hbar} y^2 \; + \;
\frac{j}{\hbar} y \}}{\int_{- \infty}^{+ \infty} dy \;
\mbox{exp} \{ - \frac{1}{2} \frac{m^2}{\hbar} y^2 \}}
\; = \; \mbox{exp} \left( \frac{1}{2 m^2 \hbar} j^2 \right)
\; \; \; ,
\ee
and make use of the general definition:
\be \label{feynpropd}
\Delta_F = \left. \left( \frac{\hbar \partial}{\partial j} \right)^2
Z_{free} (j) \; \right|_{j = 0}
\ee
In either case we find
\be \label{feynprop}
\Delta_F = \frac{\hbar}{m^2} \; \; \; .
\ee 
Hence, setting $\hbar \equiv 1 \equiv m^2$ each Feynman propagator
$\Delta_F$ contributes a trivial factor of $1$ to the evaluation
of the Feynman diagrams. The analytical expression $J_\gamma$ of a 
Feynman diagram $\gamma$ being the product of propagators and
some factors stemming from the vertices (with {\it no} integrations
over internal loops because we are in zero dimensions), in zero
dimensions every Feynman
diagram $J_\gamma$ is therefore given by
\begin{displaymath}
J_\gamma = \lambda^n \; \; \; ,
\end{displaymath}
where $n$ is the number of vertices of $\gamma$. The only extra factors
that have to be taken into account when looking at a specific physical
process are some combinatorial factors, at least in the present case
of $\varphi^4$-theory, where there is no additional symmetry. These
combinatorial factors originate from Wick's theorem when expanding the
physical process in question into a sum of Feynman diagrams up to
a given order in the coupling constant.\\
In order to illustrate this point let us have a look at the following
example: According to (\ref{npointp}) (or, equivalently, (\ref{npointff})),
the $4$-point function $G_4 (\lambda)$ is analytically given by (for
$\hbar \equiv 1 \equiv m^2$):
\be \label{fourp}
G_4 (\lambda ) = 3 \; - \; 4 \lambda \; + \;
\frac{33}{4} \lambda^2 \; + \; {\cal O} (\lambda^3 )
\ee
On the other hand, as can be taken from almost every text book on
quantum field theory (see e.g. \cite{ZJ}, \cite{PR}), 
the diagrams contributing to $G_4 (\lambda )$ read:\\
\par
\begin{figure}[ht]
\begin{center}
\includegraphics[width=150mm,angle=0]{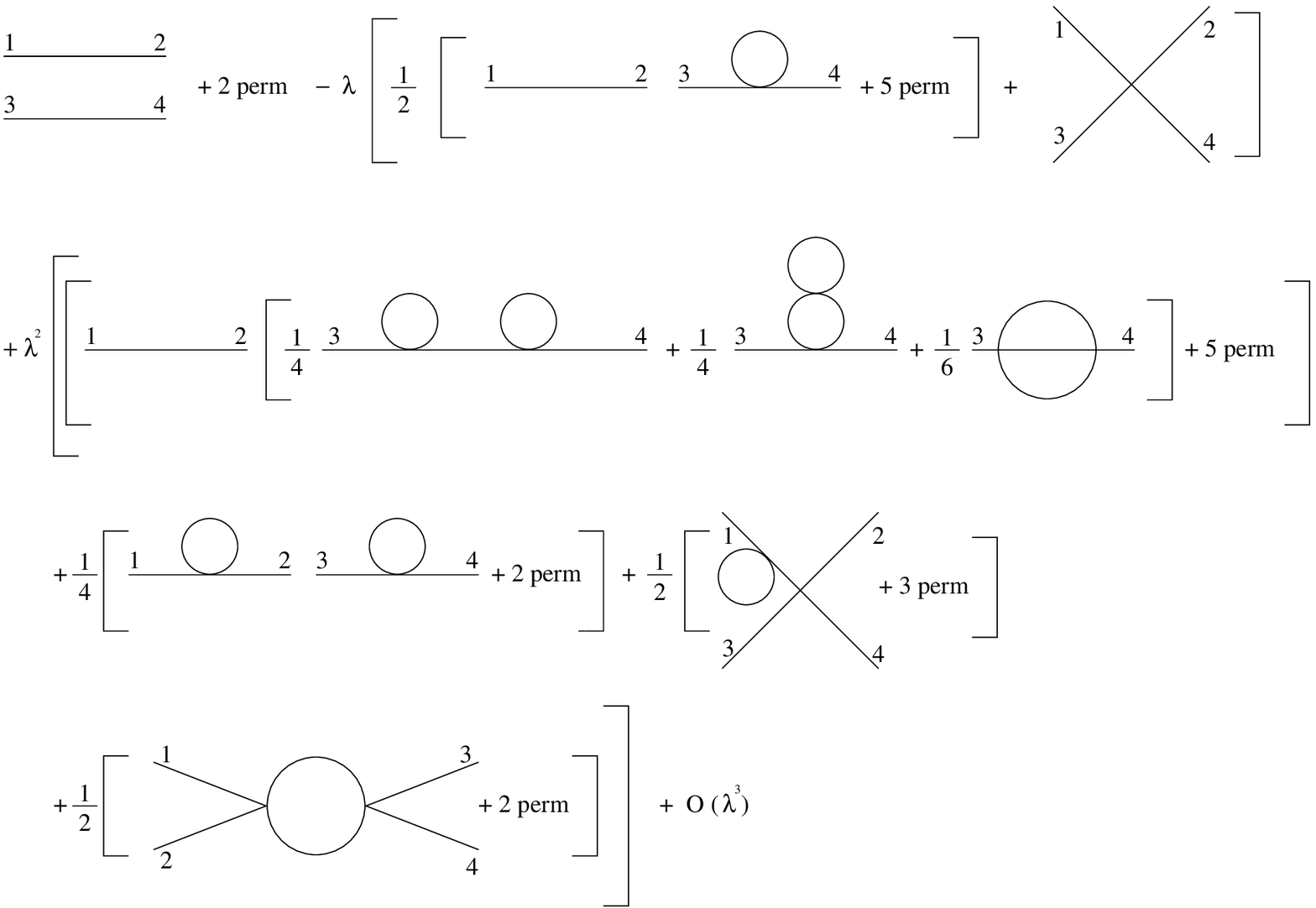}
\end{center}
\caption{\small Feynman diagrams of $\varphi^4$-theory 
contributing to the $4$-point
function up to and including order $\lambda^2$. \label{subline}}
\end{figure}
\noindent
In figure 1 the powers of the coupling constant $\lambda$ are
made explicit. The combinatorial factors in front of the diagrams
stem from Wick's theorem. With the above interpretation in mind
we find complete agreement.

\newsubsection{The various generating functionals}

In the previous section we directly calculated the $n$-point functions
of zero-dimensional $\varphi^4$-theory, see (\ref{npointff}). There is a
question that arises naturally in this context: Is it possible to 
determine all Green's functions at once in an explicit sense, i.e. not
in the form of a bunch of formulae labeled by the number $n$ of external
legs? Or, stated in a different way: Is it possible to
find a closed and/or analytical expression for the generating functional
$Z (j)$ of general Green's functions (\ref{zzz}) {\it itself}?\\
From the considerations up to now the following conclusion can be drawn:
In order to answer the above question the most promising and perhaps best
strategy to choose is to look at a differential equation that $Z (j)$ is to
fulfill and to try to solve this differential equation. Of course, the
desired equation for $Z (j)$ will be an equation of Dyson-Schwinger type
obtained by elaborating the right hand side of the trivial equation
\be \label{dysstra}
0 \; = \;
{\cal N} \; \int_{- \infty}^{+ \infty} d y \;
\frac{\partial}{\partial y} \; \mbox{exp} \left\{
\frac{1}{\hbar} \left( - S (y) + j y \right) \right\} \; \; \; .
\ee
Carrying out this recipe using the identity
\be \label{ident}
\int_{- \infty}^{+ \infty} d y \; y^n \; \mbox{exp}
\left\{ \frac{1}{\hbar} \left( - S(y) + j y \right) \right\} =
\left( \frac{\hbar \partial}{\partial j} \right)^n
\int_{- \infty}^{+ \infty} d y \; \mbox{exp}
\left\{ \frac{1}{\hbar} \left( - S(y) + j y \right) \right\}
\ee
several times leads to an ordinary linear differential
equation of third order for $Z (j)$:
\be \label{diffz}
\hbar^3 \; \frac{\lambda}{6} \; \frac{\partial^3 Z}{\partial j^3} \; + \;
\hbar \; m^2 \; \frac{\partial Z}{\partial j} \; - \;
j \; Z \; = \; 0
\ee
Formula (\ref{ident}), in the present context, is mathematically rigorous 
and not formal at all.\\
As a matter of fact, consulting the mathematical literature on
differential equations (see e.g. \cite{KA}), it turns out that
there is no known analytical solution in the form of a {\it closed}
expression to the differential equation (\ref{diffz}). Hence, the
answer to the question posed at the beginning of the
section is partially {\it no}, but nevertheless
it is, of course, possible to find a solution in the form of an 
infinite power series. Indeed, $Z(j)$ is given as a power series
in the source $j$ by the following formula
\ba \label{loeszj}
Z (j) & = & \displaystyle\frac{\sum_{n = 0}^\infty 
\displaystyle\frac{1}{(2n)!!} \left( \frac{3}{\lambda} \right)^\frac{n}{2} \;
\hbar^{- \frac{3 n}{2}} \;
U(n, \sqrt{\frac{3 m^4}{\hbar \lambda}} ) \;
j^{2n}}{U(0, \sqrt{\frac{3 m^4}{\hbar \lambda}} ) }\\[.5ex]
& = & \sum_{n = 0}^\infty \frac{G_{2n} (\lambda, m^2)}{\hbar^{2n}} \;
\frac{j^{2n}}{(2n)!} \; \; \; ,\nonumber
\ea
which, in fact, is nothing else but a collection of the results
obtained previously.\\
Alternatively, $Z(j)$ can be expressed as the quotient of two power
series in the coupling constant $\lambda$ keeping this way the
$j$-dependence of $Z$ in closed form:
\ba
Z(j) & = & \mbox{exp} \left\{ \frac{j^2}{2 \hbar m^2} \right\}
\displaystyle\frac{\sum_{n = 0}^\infty 
\displaystyle\frac{(-1)^n}{(4!)^n 4^n} 
\left( \displaystyle\frac{\hbar}{m^4} \right)^n
H_{4n} (i \; \displaystyle\frac{j}{\sqrt{2 \hbar m^2}} ) \; 
\displaystyle\frac{\lambda^n}{n!}}{\sum_{p = 0}^\infty
\displaystyle\frac{(-1)^p}{(4!)^p} 
\left( \displaystyle\frac{\hbar}{m^4} \right)^p 
\displaystyle\frac{(4p)!}{(4p)!!} \;
\displaystyle\frac{\lambda^p}{p!}} \nonumber \\[.5ex]
\label{loeszl}
& = & \displaystyle\frac{\sum_{n = 0}^\infty 
\displaystyle\frac{(-1)^n}{(4!)^n} 
\left( \displaystyle\frac{\hbar}{m^4} \right)^n 
\displaystyle\frac{(4n)!}{(4n)!!} \; {}_1 F_1 
(2n + \frac{1}{2}; \frac{1}{2};
\frac{j^2}{2 \hbar m^2} ) \;
\displaystyle\frac{\lambda^n}{n!}}{\sum_{p = 0}^\infty 
\displaystyle\frac{(-1)^p}{(4!)^p} 
\left( \displaystyle\frac{\hbar}{m^4} \right)^p 
\displaystyle\frac{(4p)!}{(4p)!!} \; 
\displaystyle\frac{\lambda^p}{p!}}
\ea
On the r.h.s. of the first line of (\ref{loeszl})
$H_m$ denotes as usual the Hermite
polynomial of order $m$. The above result is most easily derived
from the integral representation of $Z(j)$ (\ref{zzz}) by expanding
the $\lambda$-dependent term of the integrand into a power series
in $\lambda$ and by making use of the identity (\ref{ident}) as well
as (\ref{gauss0}) and Rodrigues' formula for the Hermite polynomials.
The second equality in (\ref{loeszl}) exploits the connection
between the confluent hypergeometric functions ${}_1 F_1$ and the
Hermite polynomials $H_m$, see (\ref{fherm}).\\[0.3cm]
Especially with regard to renormalization which will be the topic
of the next section, but also in view of counting of special
types of Feynman diagrams, in the following we are mainly interested
in 1~PI diagrams and the respective generating functional 
$\Gamma$ of 1~PI diagrams.
Thus, in a first step we perform the transition from the generating
functional \nolinebreak $Z(j)$ of general Green's functions to the generating
functional $W(j)$ of connected Green's functions via
\be \label{zw}
Z(j) = e^{\frac{1}{\hbar} W(j)} \; \; \; .
\ee
Accordingly the differential equation (\ref{diffz}) for $Z(j)$ transforms
into a nonlinear differential equation of second order for
$\frac{\partial W}{\partial j}$:
\be \label{diffw}
\hbar^2 \; \frac{\lambda}{6} \; \frac{\partial^3 W}{\partial j^3} \; + \;
\hbar \; \frac{\lambda}{2} \; \frac{\partial W}{\partial j} \;
\frac{\partial^2 W}{\partial j^2} \; + \;
\frac{\lambda}{6} \left( \frac{\partial W}{\partial j} \right)^3 \; + \;
m^2 \; \frac{\partial W}{\partial j} \; = \; j
\ee
In a second and final step we pass from $W(j)$ to the generating
functional $\Gamma (\varphi)$ of 1 PI Green's functions; the relation
between $W(j)$ and $\Gamma (\varphi )$ is, as usual, given by a 
Legendre transformation from $j$ to $\varphi$,
\ba \label{legendre}
& \Gamma(\varphi ) \; + \; W(j(\varphi )) \; - \;
j(\varphi ) \varphi \; = \; 0 & \\[.5ex]
\label{legendrey}
& \mbox{with } \; \; \; \varphi = \varphi (j) := 
\displaystyle\frac{\partial W}{\partial j} & \\[.5ex]
\label{legendrej}
& \mbox{and } \; \; \; j = j(\varphi ) = 
\displaystyle\frac{\partial \Gamma}{\partial \varphi} 
\; \; , \mbox{ respectively} \; . &
\ea
Hence, calculating the Legendre transformations of the second and third
derivative of $W$ with respect to $j$ by differentiating appropriately
the identity $\varphi (j(\varphi )) = \varphi$ using the
chain rule as well as (\ref{legendrey}), (\ref{legendrej}), we end up
with the following highly nonlinear differential equation for
$\Gamma (\varphi )$ (or in order to be precise for 
$\frac{\partial \Gamma}{\partial \varphi}
\equiv \Gamma'$):
\be \label{diffg}
- \; \hbar^2 \; \frac{\lambda}{6} \; \Gamma'''   \; + \;
\hbar \; \frac{\lambda}{2} \; \varphi \; (\Gamma'' )^2 \; + \;
\left( \frac{\lambda}{6} \varphi^3 + m^2 \varphi \right) \;
(\Gamma'' )^3 \; = \; \Gamma' \; (\Gamma'' )^3
\ee
Again, it seems to be impossible to find a {\it closed} analytical
expression for $\Gamma (\varphi )$ obeying (\ref{diffg}). However,
(\ref{diffg}) can be solved by recurrence yielding the loop expansion
of $\Gamma (\varphi )$ or, equivalently, the expansion of 
$\Gamma (\varphi )$ in powers of $\hbar$ as follows:\\
Evaluating (\ref{diffg}) consistently in order $\hbar^0$ immediately leads
to (because the first two terms on the l.h.s. of (\ref{diffg}) drop out)
\begin{displaymath}
\frac{\lambda}{6} \varphi^3 + m^2 \varphi \; = \; {\Gamma^{(0)}}' \; \; \; ,
\end{displaymath}
where the superscript ${}^{(0)}$ of $\Gamma^{(0)}$ indicates that 
$\Gamma$ is to be taken in zeroth order in $\hbar$. Simple integration
then yields
\begin{displaymath}
\Gamma^{(0)} = \frac{\lambda}{4!} \varphi^4 +
\frac{1}{2} m^2 \varphi^2 + C^{(0)} \; \; \; ,
\end{displaymath}
and the integration constant $C^{(0)}$ is fixed by the requirement
\be \label{gfix}
\Gamma (\varphi = 0 ) = 0 \; \; \; \mbox{ to all orders } \; \; ,
\ee
i.e. $C^{(0)} \equiv 0$. Hence we obtain the expected result (see
(\ref{action0}))
\be \label{g0}
\Gamma^{(0)} = \frac{1}{2} m^2 \varphi^2 + \frac{\lambda}{4!} \varphi^4 
\equiv S(\varphi ) \; \; \; .
\ee
In the next step of recurrence we consistently analyze (\ref{diffg}) 
in order $\hbar^1$:
\bas
\frac{\lambda}{2} \varphi ({\Gamma^{(0)}}'')^2 +
(\frac{\lambda}{6} \varphi^3 + m^2 \varphi) [(\Gamma'')^3]^{(1)} & = &
[\Gamma' \; (\Gamma'')^3]^{(1)} \\
\Leftrightarrow \; \; \frac{\lambda}{2} \varphi ({\Gamma^{(0)}}'')^2 +
{\Gamma^{(0)}}' \; [(\Gamma'')^3]^{(1)} & = &
{\Gamma^{(0)}}' \; [(\Gamma'')^3]^{(1)} +
{\Gamma^{(1)}}' \; ({\Gamma^{(0)}}'')^3
\eas
Therefrom (and from (\ref{gfix}), (\ref{g0})) we unambiguously get 
the answer for $\Gamma^{(1)}$:
\be \label{g1}
\Gamma^{(1)} = \frac{1}{2} \; \mbox{ln } (m^2 + 
\frac{\lambda}{2} \varphi^2 )
- \frac{1}{2} \; \mbox{ln } m^2
\ee
It is clear how the recursive process proceeds. Here we list the
results for two and three loops:
\ba \label{g2}
\Gamma^{(2)} & = &
- \frac{\lambda}{6} \; \frac{\lambda \varphi^2 - 6 m^2}{(\lambda \varphi^2 +
2 m^2)^3} \; - \; \frac{\lambda}{8 m^4} \\[.5ex]
\label{g3}
\Gamma^{(3)} & = &
\frac{16 \lambda^2 m^2}{3} \; \frac{2 \lambda \varphi^2 - m^2}{(\lambda 
\varphi^2 +
2 m^2)^6} \; + \; \frac{\lambda^2}{12 m^8}
\ea
(We use the convention 
\be \label{gnorm}
\Gamma = \sum_{n = 0}^\infty \hbar^n \Gamma^{(n)}
\ee
for the definition of $\Gamma^{(n)}$.)\\
The recursive determination of $\Gamma$ obviously calls for 
automatization in form of a computer routine. Such a routine written
for {\it mathematica} is given in appendix C where higher orders
of $\Gamma$ can be found, too. Appendix C also contains an alternative
possibility of finding closed analytical expressions for the 1 PI
$n$-point functions $\Gamma_{n}$ valid to all orders of the loop
expansion.\\[.3cm]
At this point, it is instructive to consider once again the connection
of zero-dimensional $\varphi^4$-theory with the counting of (special
types of) Feynman diagrams of the theory in four dimensions. Let us,
for example, look at the 1 PI 4-point function $\Gamma_4^{(\leq 1)}$
up to (and including) 1-loop order: Then besides the trivial contribution
from the ordinary $\varphi^4$-vertex (last diagram in the first line
of figure 1) only the last diagram of figure 1 contributes. This diagram
appears three times multiplied with a combinatorial factor of $\frac{1}{2}$.
Hence, from the counting argument of diagrams we expect
$\Gamma_4^{(\leq 1)}$ to be:
\begin{displaymath}
\Gamma_4^{(\leq 1)} = - \; \lambda \; + \; \frac{3}{2} \; \hbar \; \lambda^2
\end{displaymath}
One easily checks that this result indeed coincides with\footnote{The extra
minus sign is due to the convention we have chosen in the definition of
the Legendre transformation, see (\ref{legendre}): The 1~PI contributions
contained in $W$ (besides the connected but one-particle-reducible
contributions) pick up a minus sign when regarded as contributions 
to $\Gamma$. \label{fn1}}
\begin{displaymath}
- \frac{\partial^4}{\partial \varphi^4} 
\left. (\Gamma^{(0)} + \Gamma^{(1)} ) \; \right|_{\varphi = 0} \; \; \; ,
\end{displaymath}
where $\Gamma^{(0)}$ and $\Gamma^{(1)}$ are taken from (\ref{g0}) and
(\ref{g1}), respectively.\\
Generally speaking, if one is interested in the number of diagrams
(multiplied with the correct combinatorial factors) contributing to
the 1 PI $n$-point function $\Gamma_n^{(\leq m)}$ up to order \nolinebreak
$m$ in the loop expansion, one first has to calculate $\Gamma$ up to order
$m$ by means of the routine in appendix C, then to differentiate
the resulting expression $n$ times with respect to $\varphi$, and
finally to evaluate the result of differentiation
at the point $\varphi = 0$.

\newsubsection{Renormalization}

Suppose that zero-dimensional $\varphi^4$-theory is to have a {\it physical}
meaning on its own. Pursuing the consequences of this ad-hoc assumption
in the following will provide us with an extremely simple and pedagogical
toy model exhibiting the unavoidable necessity of renormalization in
a perturbatively defined quantum field theory, even if there are no
divergencies at all as in the present example: All the expressions
for $\Gamma^{(n)}$, $n = 0, 1, 2, \dots$, derived in the previous
section (see (\ref{g0}) - (\ref{g3}) and appendix C) are 
well-defined and finite. Indeed, not only the necessity of renormalization
can be explained considering the present model but also the procedure
of renormalization itself can be studied in detail in this context.\\
The classical action $S(\varphi ) = \Gamma^{(0)} (\varphi )$, see
(\ref{action0}) or (\ref{g0}), as well as the higher orders $\Gamma^{(n)}$ of
$\Gamma$ contain the parameter $\lambda$ describing the strength of the
self-interaction and the parameter $m^2$ of dimension $(mass)^2$. If the
theory is to be physical we have to reexpress these parameters in terms of
a {\it physical} coupling constant $g_{phys}$ and a {\it physical} mass
$m_{phys}^2$ accessible to physical measurements. This is already true
at the lowest order of the loop expansion, i.e. at the purely classical
level. As usual in quantum field theories, a connection between the set 
of parameters and the set of really physical measureable quantities is
established by a set of normalization conditions
which in the present context we choose to be:
\ba \label{normcm}
\Gamma_2 & = & m_{phys}^2 \\
\label{normcg}
\Gamma_4 & = & g_{phys}
\ea
($\Gamma_2$ denotes the 1 PI 2-point function, $\Gamma_4$ is the 1 PI
4-point function.)\\
Taking into account that we have $\Gamma_2^{(0)} = m^2$ and
$\Gamma_4^{(0)} = \lambda$ which directly follows from appropriate
differentiations of $\Gamma^{(0)}$ (\ref{g0}) with respect to
$\varphi$ and subsequent evaluation at $\varphi = 0$, the normalization
conditions (\ref{normcm}), (\ref{normcg}) clearly supply us with the
desired unambiguous interrelation between the parameters $\lambda, m^2$ and
the physical observables $g_{phys}, m_{phys}^2$ at lowest order, namely:
\be \label{parphys0}
m^2 = m_{phys}^2 \; \; \mbox{ and } \; \; \lambda = g_{phys}
\ee
However, moving on to the next order, i.e. 1-loop order, we realise that
the 1 PI 2-point function and the 1 PI 4-point function pick up
radiative corrections:
\be \label{gtowfour1}
\Gamma_2^{(\leq 1)} = m^2 + \frac{\lambda}{2 m^2} \hbar \; \; 
\mbox{ and } \; \;
\Gamma_4^{(\leq 1)} = \lambda - \frac{3 \lambda^2}{2 m^4} \hbar
\ee
(For this consider appropriate differentiations of $\Gamma^{(0)} +
\Gamma^{(1)}$ with $\Gamma^{(1)}$ according to (\ref{g1}).)\\
In order to still fulfill the normalization conditions (\ref{normcm}),
(\ref{normcg}) which are the essential ingredient for turning the formerly
unphysical model into a physical theory, the only possibility is to
introduce counterterms to the classical action,
\be \label{ct1}
\Gamma_{c.t.}^{(1)} = \frac{1}{2} \; \delta {m^2}^{(1)} \hbar \; 
\varphi^2 \; + \; 
\frac{\delta \lambda^{(1)} \hbar}{4!} \; \varphi^4 \; \; \; ,
\ee
i.e. to consider $\Gamma_{eff}^{(\leq 1)} = S(\varphi ) +
\Gamma_{c.t.}^{(1)}$ instead of $S(\varphi )$ as the starting point for
perturbative calculations, and to determine ${\delta m^2}^{(1)}$ and
$\delta \lambda^{(1)}$ in such a way that now the normalization
conditions hold true up to 1-loop order. In other words:
The parameters $\lambda$ and $m^2$ have to be power series in $\hbar$
by themselves:
\be \label{parpow}
m^2 = {m^2}^{(0)} + \sum_{n = 1}^\infty \delta {m^2}^{(n)} \hbar^n \; \;
\mbox{ and } \; \;
\lambda = \lambda^{(0)} + \sum_{n = 1}^\infty \delta \lambda^{(n)} \hbar^n
\ee
A short calculation shows that with (\ref{ct1}) the normalization
conditions indeed can be fulfilled yielding:
\ba \label{parphysm1}
m^2 & = & m_{phys}^2 - \frac{1}{2}
\frac{g_{phys}}{m_{phys}^2} \hbar + {\cal O} (\hbar^2 ) \\
\label{parphysg1}
\lambda & = & g_{phys} + \frac{3}{2} 
\frac{g_{phys}^2}{m_{phys}^4} \hbar + {\cal O} (\hbar^2 )
\ea
Hence, the renormalized action up to (and including) 1-loop order is given by
\ba \label{gren1}
\Gamma_{ren}^{(\leq 1)} & = & \frac{1}{2} m_{phys}^2 \varphi^2 +
\frac{g_{phys}}{4!} \varphi^4 \\[.5ex]
& & + \frac{\hbar}{2} \left( \mbox{ln } \left[
\frac{m_{phys}^2 + \frac{1}{2} g_{phys} \varphi^2}{m_{phys}^2} \right]
- \frac{1}{2} \frac{g_{phys}}{m_{phys}^2} \varphi^2
+ \frac{1}{8} \frac{g_{phys}^2}{m_{phys}^4} \varphi^4 \right) 
\; \; \; . \nonumber
\ea
Two remarks are of some importance here. Normally, in quantum field theory
the Fourier transformations of the $n$-point functions contributing to
a physical process with $n$ external particles are functions of the
external momenta of these particles starting from 1 loop on. As a 
consequence, the left hand sides of
the normalization conditions in momentum space have to be
stated at a specific value (or at several specific values) of the momenta
in order to guarantee momentum independence of the right hand sides
which represent certain physical measurements that are independent of
momentum a priori. In the present case of a zero-dimensional theory,
however, there is no momentum dependence of (the Fourier transformations
of) the $n$-point functions at all. This observation is basically due to
the fact that a zero-dimensional theory lives on just one point, say $x_0$,
and, accordingly, Fourier space also contains just one momentum $p_0$
which without loss of generality can be chosen to be $p_0 = 0$. This is
the reason for the simple form of the normalization conditions
(\ref{normcm}) and (\ref{normcg}).\\
The second remark concerns the number of normalization conditions and is
intimately related to the first one: The reader could perhaps wonder
that only two normalization conditions have been given above whereas
in four-dimensional $\varphi^4$-theory a third additional normalization
condition on the residue of the 2-point function is needed in order to
fix the wave function renormalization $z$ appearing in front of the
kinetic term $(\partial_\mu \varphi ) (\partial^\mu \varphi)$. In zero
dimensions there is no kinetic term and, hence, no space for an additional
free parameter like a wave function renormalization which in turn -- due
to its absence -- also, of course, does not have to be fixed.\\[.5cm]
\par
\begin{figure}[ht]
\begin{center}
\setlength{\unitlength}{1cm}
\begin{minipage}[t]{15cm}
\begin{picture}(15,9)
\put(7.8,9.5){$\Gamma_{ren}$}
\put(14.3,0.9){$\varphi$}
\includegraphics[width=150mm,angle=0]{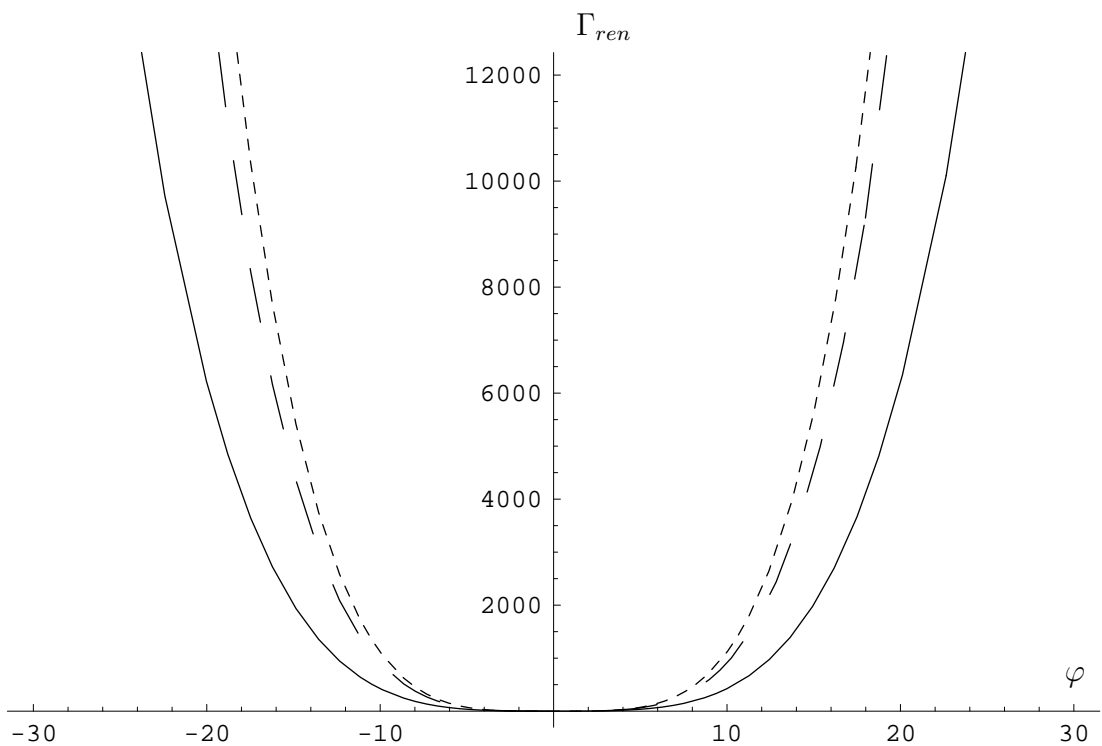}
\end{picture}
\end{minipage}
\end{center}
\caption{\small Plots of $\Gamma_{ren}^{(0)}$ (solid line),
                $\Gamma_{ren}^{(\leq 1)}$ (dashed line) and
$\Gamma_{ren}^{(\leq 2)}$ (dotted line) for $\hbar = 1$, $m_{phys} = 1$
and $g_{phys} = 0.9$. \label{graph1}}
\end{figure}
\noindent
Apart from these remarks, it is clear how the above considerations 
recursively generalize to higher orders in the loop expansion.
Especially, in 2-loop order $m^2$ (\ref{parphysm1}) and $\lambda$
(\ref{parphysg1}) are corrected by counterterms with coefficients
$\delta {m^2}^{(2)} \hbar^2$ and $\delta \lambda^{(2)} \hbar^2$,
respectively:
\be \label{parphys2}
\delta {m^2}^{(2)} = - \frac{7}{12} \frac{g_{phys}^2}{m_{phys}^6} \; \;
\mbox{ and } \; \;
\delta \lambda^{(2)} = \frac{3}{4} \frac{g_{phys}^3}{m_{phys}^8}
\ee
Accordingly, the renormalized action $\Gamma_{ren}^{(\leq 2)}$ up to 
2 loops reads
\ba \label{gren2}
\Gamma_{ren}^{(\leq 2)} & = & \Gamma_{ren}^{(\leq 1)} +
\frac{\hbar^2}{4} \left( - \frac{2 g_{phys}}{3} 
\frac{g_{phys} \varphi^2 - 6 m_{phys}^2}{(2 m_{phys}^2 + g_{phys}
\varphi^2 )^3} - \frac{g_{phys}}{2 m_{phys}^4} \right. \nonumber \\[.5ex]
& & \hspace*{.6cm} \left. - 
    \frac{7}{6} \frac{g_{phys}^2}{m_{phys}^6} \varphi^2 
+ \frac{1}{8} \frac{g_{phys}^3}{m_{phys}^8} \varphi^4 
+ 4 \frac{g_{phys}^2}{m_{phys}^4}
\frac{\varphi^2}{2 m_{phys}^2 + g_{phys} \varphi^2} \right) \; \; \; ,
\ea
where we already have taken into account that the 1-loop counterterms
(\ref{ct1}) in the 2-loop approximation develop radiative corrections
leading to the last term on the right hand side of (\ref{gren2}).\\
In figure~\ref{graph1} above, $\Gamma_{ren}^{(\leq n)}$ is represented for
$n = 0,1$ and $2$. In order to be able to visualize the small corrections
stemming from higher orders we, as an example, have chosen $g_{phys} = 0.9$ 
and $m_{phys} = 1$ as well as $\hbar = 1$.\\
The recursive nature of the determination of $\Gamma_{ren}$ again asks
for an implementation on the computer. Appendix D presents such a short
{\it mathematica} routine yielding in the case of zero-dimensional
$\varphi^4$-theory $m^2, \lambda$ and $\Gamma_{ren}$ up to an arbitrary
prescribed order of the loop expansion.

%% file: chap2.tex
\newpage
\newsection{A model with spontaneous symmetry breaking}

Among other things in the previous chapter we were concerned with the
counting of Feynman diagrams for a given theory by considering the
respective theory in zero dimensions. More precisely, we investigated
in this regard the simple example of $\varphi^4$-theory in quite some
detail by using techniques which are - to some extent - 
orthogonal to the ones existing in the literature. Of course,
this kind of considerations is generalizable to more complicated
models including gauge bosons and fermions like, for example,
QED or QCD. (See \cite{CLP} for the case of QED.) However, what happens
if the theory in question exhibits spontaneous symmetry breaking?
Does the connection between the counting of Feynman diagrams and the
theory in zero dimensions still hold, or has this connection to be
modified, or is it perhaps completely destroyed?\\
One tough problem that shows up immediately when tackling the above
question(s) pertains to the existence of massless modes due to the
residual symmetry surviving the spontaneous breakdown of symmetry: In zero
dimensions the Feynman propagators (in Fourier space) are something
like $\frac{1}{mass}$ or $\frac{1}{(mass)^2}$ because of the absence
of nonvanishing momenta. Hence, for massless fields, what are
the corresponding Feynman propagators? Are they undefined or infinite?
Or what else could they be? On the other hand, the
counting argument substantially relied on the fact that every Feynman
propagator appearing in the Feynman diagrams can be normalized to just
$1$. Is there a way out of this dilemma?\\
In order to study this problem and to answer the questions just
mentioned we again restrict ourselves to a simple model of spontaneous
symmetry breaking, namely the theory of one complex scalar field
$\Phi$ with a self-interaction of the Mexican hat type. This model will
be defined in the first section of this chapter where once more it is 
shown that a naive treatment of the zero-dimensional theory in higher
orders leads to inconsistencies related to the problem stated above.
Also two possibilities to circumvent these difficulties will be indicated
there. The second section then follows up the most promising and
el\-egant possibility in detail consisting in adding a gauge
fixing term of t'Hooft's type to the classical action accompanied by the
introduction of Faddeev-Popov ghost fields $c$ and 
\nolinebreak $\bar{c}$. Note that
this becomes necessary even though
there are no gauge fields present in the model under consideration. In other
words, inspired by the usual approach to quantum gauge field theories
BRS invariance will replace (hidden) local gauge invariance and serve
as {\it the} defining property also for higher orders in perturbation
theory. Having thus at our disposal a proper definition of the 
zero-dimensional theory to all orders we will continue to calculate
recursively the generating functional $\Gamma$ of 1 PI Green's functions in
line with the investigations of the previous chapter. Section~3.3 is
devoted to the question how far the counting argument can be transferred
to the case of spontaneous symmetry breaking. Several examples will be
given in this context for illustrative purposes. In the final section
we anew turn to the task of renormalizing the zero-dimensional model.\\
Besides the more pedagogical motivation for dealing with this kind of topic
there is another, even more important reason 
which we would like to mention qualitatively
already at this point:
In models based on Noncommutative Geometry ordinary gauge fields and Higgs
fields are combined into a single object, namely a generalized potential
or superpotential \nolinebreak $\cal A$. 
Starting from this extended gauge potential
$\cal A$ and calculating the noncommutative Yang-Mills action by also
generalizing the usual Yang-Mills procedure yields the by now well-known
result (among other things) that in such a noncommutative model the
mechanism of spontaneous symmetry breaking (SSB) is automatically included
and therefore a built-in feature of the model which, in addition, allows for
a geometrical interpretation of SSB. Especially, if in the noncommutative
approach the gauge fields are set to zero and if the Higgs fields are
restricted to live on just one point, we obtain a very simple 
zero-dimensional {\it noncommutative} model that (by appropriate
choice of the Higgs content) coincides with the model of the present
chapter at least at the classical level. Hence, all the considerations
in this chapter formulated in the common language of usual Higgs fields,
and thus avoiding the notion of a superpotential $\cal A$, will serve as
a reference point for the analogous considerations in the
subsequent chapter which deals with the quantum theory of that noncommutative
toy model with just Higgs ``fields'' appearing in 
\nolinebreak $\cal A$. Of course,
if we take the noncommutative point of view, things have to be
formulated as far as possible using the noncommutative language, i.e
the basic object $\cal A$. As it will turn out there is indeed a substantial
simplification in the noncommutative framework. But this will be a later
story.

\newsubsection{Definition of the model and the problems with the Goldstone
               particle}

In the following we will consider the theory of just one complex scalar
field $\Phi$ with self-interaction given by (in four dimensions):
\ba \label{acssb4}
S[\Phi, \overline{\Phi} ] & = & \int d^4 x \; \left\{
(\partial_\mu \Phi(x))(\partial^\mu \overline{\Phi}(x)) +
m^2 \Phi(x) \overline{\Phi}(x) + 
\frac{\lambda}{4} (\Phi(x) \overline{\Phi}(x) )^2 \right\} \\
& = & \int d^4 x \; \left\{ {\cal L}_{kin} [\Phi, \overline{\Phi} ] + 
V[\Phi, \overline{\Phi} ] \right\} \nonumber
\ea
In order to obtain the Mexican hat accounting for SSB we must have
\begin{displaymath}
m^2 \; < \; 0 \; \; \; ,
\end{displaymath}
which will be assumed from now on. The absolute minima of 
$V[\Phi, \overline{\Phi} ]$ are taken on a circle around the
origin $(\Phi, \overline{\Phi} ) = (0,0)$ with radius
\be \label{absmin}
|\Phi| = \sqrt{\frac{- 2 m^2}{\lambda}} =: \frac{v}{\sqrt{2}} \; \; \; .
\ee
The theory in zero dimensions again results from the corresponding theory
in four dimensions by omitting all derivatives of the fields as well as
the overall integration and by suppressing the argument of the fields, i.e.
in the present case the zero-dimensional action takes the form:
\be \label{acssb0pre}
S(z, \bar{z} )\;  = \; m^2 \; z \bar{z} \; + \;
\frac{\lambda}{4} \; (z \bar{z} )^2
\ee
In order to eventually proceed to the quantum theory we furthermore
perform a shift from the unstable local maximum at the origin
$(z, \bar{z} ) = (0,0)$ of $S$ to one of the minima of $S$ by means of
a translation:
\ba \label{shift}
z & = & \frac{1}{\sqrt{2}} (x + v + i y) \nonumber \\
\bar{z} & = & \frac{1}{\sqrt{2}} (x + v - i y)
\ea
Looking for (1 PI) Green's functions with (amputated) incoming and outgoing
physical ``fields'' $x$ and $y$ it is better to work with the
real-valued variables $x$ and $y$ right from the beginning instead
of using the unshifted ``fields'' $z$ and $\overline{z}$, see also later. 
Hence, expressing $S(z, \bar{z} )$ (\ref{acssb0pre}) in terms of
$x$ and $y$ defined via (\ref{shift})
yields after a short calculation (up to a constant 
$- \frac{m^4}{\lambda}$ which is the value of $S(z, \overline{z} )$
at the minima and therefore can be dropped without loss of 
generality\footnote{Taking along the constant $- \frac{m^4}{\lambda}$ leads
to an additional {\it constant} factor on the r.h.s. of (\ref{nssbz}) which
can be absorbed into a redefinition of the normalization constant $\cal N$.
This normalization constant is afterwards fixed by means of the requirement
(\ref{ssbnormf}).}) the
starting point for all considerations to follow:
\be \label{acssb0}
S(x,y) \; = \; \frac{\lambda}{16} \; (x^2 + y^2)^2 \; + \;
\frac{\lambda v}{4} \; x (x^2 + y^2) \; - \; m^2 \; x^2
\ee
$S(x,y)$ describes a massive scalar ``field'' $x$ with mass
$m_x^2 = - 2 m^2$, i.e. a Higgs field, interacting with itself and
another scalar, but massless ``field'' $y$, which is a 
Goldstone field.\\[.3cm]
The generating functional $Z(\rho, \tau )$ of general Green's functions
now reads
\be \label{nssbz}
Z(\rho, \tau ) = {\cal N} \; 
\int_{- \infty}^{+ \infty} dx \int_{- \infty}^{+ \infty} dy \;
\mbox{exp} \left\{ \frac{1}{\hbar} (- S(x,y) + \rho x + \tau y) 
\right\} \; \; \; .
\ee
$\rho$ and $\tau$ are two external sources to which the theory is coupled
in order to sum up all $n$-point functions into the single object $Z$.
The normalization constant (again well-defined in zero dimensions) is fixed
as usual by the requirement:
\be \label{ssbnormf}
\left. Z(\rho, \tau ) \right|_{\rho = 0 = \tau} \; = \; 1
\ee
Copying the procedure of section~2.2 and thus heading for the
differential equation(s) for $Z(\rho, \tau )$ we elaborate 
the right hand sides of
\bas
0 & = & 
{\cal N} \; \int_{- \infty}^{+ \infty} d y 
\int_{- \infty}^{+ \infty} d x \; \frac{\partial}{\partial x} \;
\mbox{exp} \left\{ \frac{1}{\hbar} \left(
- S(x,y) + \rho x + \tau y \right) \right\} \; \; , \\[.5ex]
0 & = &
{\cal N} \; \int_{- \infty}^{+ \infty} d x 
\int_{- \infty}^{+ \infty} d y \; \frac{\partial}{\partial y} \;
\mbox{exp} \left\{ \frac{1}{\hbar} \left(
- S(x,y) + \rho x + \tau y \right) \right\} \; \; .
\eas
This leads after an easy and short calculation to a system of two coupled
partial differential equations for $Z(\rho, \tau )$ which resembles
in zero dimensions the system of Dyson-Schwinger equations for 
the model under consideration:
\ba \label{nssbdiffz}
\hbar^3 \; \frac{\lambda}{4} \; \left(
\frac{\partial^3 Z}{\partial \rho^3} +
\frac{\partial^3 Z}{\partial \rho \partial \tau^2} \right) \; + \;
\hbar^2 \; \frac{\lambda v}{4} \; \left(
3 \frac{\partial^2 Z}{\partial \rho^2} +
\frac{\partial^2 Z}{\partial \tau^2} \right) \; - \;
\hbar \; 2 m^2 \; \frac{\partial Z}{\partial \rho} \; - \;
\rho Z & = & 0 \nonumber \\[.5ex]
\hbar^3 \; \frac{\lambda}{4} \; \left(
\frac{\partial^3 Z}{\partial \tau^3} +
\frac{\partial^3 Z}{\partial \rho^2 \partial \tau} \right) \; + \;
\hbar^2 \; \frac{\lambda v}{2} \; 
\frac{\partial^2 Z}{\partial \rho \partial \tau} \; - \;
\tau Z & = & 0
\ea
Being mainly interested in 1 PI diagrams we have
to perform the transition from the generating functional $Z(\rho, \tau )$
to the generating functional $\Gamma (x, y)$ of 1~PI Green's functions
according to\footnote{Please note that we are working with the shifted
physical fields $x$ and $y$ located at one of the minima of the Mexican
hat. As a consequence, the vacuum expectation values of the classical
fields $x$ and $y$ are zero at the classical level. This is the reason 
for the absence of a constant (constants) (vacuum expectation
value(s)) in (\ref{nssbleg}).}:
\ba \label{nssbzw}
Z(\rho, \tau ) & \longrightarrow &
W(\rho, \tau ) \; := \; \hbar \; \mbox{ln } Z(\rho, \tau ) \\[.5ex]
\label{nssbwg}
& \stackrel{Legendre}{\longrightarrow} &
\Gamma (x, y) \; := \; \rho (x,y) \; x \; + \; \tau (x,y) \; y \; - \;
W(\rho (x,y), \tau (x,y) ) \\[.5ex]
\label{nssbleg}
\mbox{with} & : & x \; := \; 
\frac{\partial W}{\partial \rho} \; \; ,
\; \; y \; := \; \frac{\partial W}{\partial \tau}
\ea
We skip the details of the calculation here and refer the interested
reader to appendix~E. Up to order $\hbar$ the following
two coupled differential equations for $\Gamma (x,y)$ result:
\ba \label{nssbdiffgh}
\frac{\partial \Gamma}{\partial x} & = &
\frac{\lambda}{4} \left( x^3 + x y^2 + 3 v x^2 + v y^2 \right) - 2 m^2 x
\nonumber \\
& & + \; \hbar \; \frac{\lambda}{4} \; \hat{h}^{- 1} (x,y) \left\{
(x + v) \left( 3 \frac{\partial^2 \Gamma}{\partial y^2} +
\frac{\partial^2 \Gamma}{\partial x^2} \right) -
2 y \frac{\partial^2 \Gamma}{\partial x \partial y} \right\}
+ {\cal O} (\hbar^2) \nonumber \\[.5ex]
\frac{\partial \Gamma}{\partial y} & = &
\frac{\lambda}{4} \left( y^3 + x^2 y + 2 v x y \right) \nonumber \\
& & + \; \hbar \; \frac{\lambda}{4} \; \hat{h}^{- 1} (x,y) \left\{
- 2 (x + v) \frac{\partial^2 \Gamma}{\partial x \partial y} +
y \left( \frac{\partial^2 \Gamma}{\partial y^2} +
3 \frac{\partial^2 \Gamma}{\partial x^2} \right) \right\}
+ {\cal O} (\hbar^2)
\ea
The function $\hat{h} (x,y)$ occuring on the right hand sides
is defined in (\ref{nssbh}):
\be \label{nssbhh}
\hat{h} (x,y) = \frac{\partial^2 \Gamma}{\partial x^2}
\frac{\partial^2 \Gamma}{\partial y^2} -
\left( \frac{\partial^2 \Gamma}{\partial x \partial y} \right)^2
\ee
Solving (\ref{nssbdiffgh}) recursively order by order in the loop expansion,
in zeroth order we find, as a matter of consistency, the expected answer
\be \label{nssbg0}
\Gamma^{(0)} (x,y) = S (x,y) \; \; \; ,
\ee
where $S(x,y)$ is given by (\ref{acssb0}). Proceeding to the next order
and, hence, evaluating (\ref{nssbdiffgh}) consistently in order $\hbar^1$,
a straightforward calculation (using (\ref{nssbg0})) shows that the
1-loop approximation of $\Gamma$ reads:
\ba \label{nssbg1}
\Gamma^{(1)} (x,y) & = & \frac{1}{2} \mbox{ ln }
\hat{h}^{(0)} (x,y) \; + \; C^{(1)} \\[.5ex]
\mbox{with } \; \hat{h}^{(0)} (x,y) & = &
\frac{1}{16} 
(3 \lambda x^2 + \lambda y^2 + 6 \lambda v x - 8 m^2)
( \lambda x^2 + 3 \lambda y^2 + 2 \lambda v x) -
\frac{\lambda^2}{4} (x + v)^2 y^2 \nonumber
\ea
In (\ref{nssbg1}) $C^{(1)}$ is a constant of integration that actually
should be fixed by the requirement\footnote{(\ref{nssbnormg}) is nothing
else but the translation of (\ref{ssbnormf}) to the functional $\Gamma$
taking into account that $x$ and $y$ are the physical fields.
In fact, (\ref{nssbnormg}) was implicitly used when deriving (\ref{nssbg0}).}
\be \label{nssbnormg} 
\left. \Gamma (x,y) \right|_{x = 0 = y} = 0 \; \; 
\mbox{ to all orders } \; \; .
\ee
However, as can be seen at once from (\ref{nssbg1}), $\Gamma^{(1)}$
is not defined in the limit $x \rightarrow 0$ and $y \rightarrow 0$.
Even worse, trying to calculate in 1-loop 
order the 1 PI $n$-point functions by
repeated differentiations of $\Gamma^{(1)}$ with respect to $x$ and $y$
and subsequent evaluation at $x = 0 = y$ we have to realise that none of those
$n$-point functions is defined, or, expressing the 
same statement differently, all possible
1 PI $n$-point functions are infinite starting from 1-loop on. This
unavoidable fact just reflects, of course, the problems mentioned
in the introduction of the present chapter concerning the masslessness
of the Goldstone boson $y$: In higher orders, all possible 1~PI
Green's functions necessarily involve diagrams with internal
$\langle yy\rangle$-propagators due to the nature of the couplings in $S(x,y)$.
But because the ``field'' $y$ is massless the Feynman propagator for
$y$ is just infinite in zero dimensions. As a consequence, even in zero
dimensions higher orders are spoiled by a remnant of what in four dimensions
would be called an infrared problem.\\[.3cm]
One possibility to circumvent this dilemma certainly would consist in an
ad-hoc addition of a mass term $m_y^2 y^2$ to the action $S(x,y)$. However,
proceeding this way the hidden symmetry of the model (see section~3.2) is
brutally destroyed and it will be difficult (and perhaps even impossible)
to properly control the higher orders and, especially, the dependence
of the theory on the auxiliary mass $m_y^2$ from 1 loop on.\\
Hence, we should look for an alternative possibility: The best strategy
would be to introduce a mass term for the Goldstone boson $y$ without
loosing track of the (hidden) symmetry. Borrowing ideas from ordinary
quantum field theory this can be acchieved by adding a gauge fixing term
of t'Hooft's type,
\be \label{ssbgf}
\Gamma_{g.f.} = \frac{1}{2} B^2 + \xi m B y \; \; \; ,
\ee
to $S(x,y)$ and enlarging (local) gauge invariance to BRS invariance.
In (\ref{ssbgf}) $\xi$ is a gauge parameter, and $B$ is an auxiliary
field of dimension 2. Eliminating $B$ by means of its equation of motion
produces the desired mass term for $y$:
\be \label{beqmot}
\frac{\partial \Gamma_{g.f.}}{\partial B} = 0 \; \Rightarrow \;
B = - \xi m y \; \Rightarrow \; 
\Gamma_{g.f.} = - \frac{1}{2} \xi^2 m^2 y^2
\ee
On the other hand, keeping the auxiliary field $B$ until the end of all
calculations and having at our disposal BRS invariance (as {\it the}
defining symmetry of the theory) everything will be well-defined.
Section~3.2 follows up this second promising possibility in detail.

\newsubsection{Proper definition of the theory and recursive
               determination of $\Gamma^{(n)}$}

It is evident that $S(z, \overline{z} )$ (\ref{acssb0pre}) is invariant
under the (gauge) transformations
\be \label{gaugetrz}
z \rightarrow z e^{i \alpha} \; \mbox{ and } \;
\overline{z} \rightarrow \overline{z} e^{- i \alpha} \; \mbox{ with } \;
\alpha \; \mbox{ arbitrary real } \; \; .
\ee
Carrying out the shift (\ref{shift}) from $z, \overline{z}$ to the
physical variables $x, y$ the above symmetry is hidden, but, of course,
nevertheless still present. When expressed for $x$ and $y$ the infinitesimal
version of the symmetry of $S(x,y)$ (\ref{acssb0}) takes the form
\be \label{gaugetr}
\delta_\alpha x = - \alpha y \; \mbox{ and } \;
\delta_\alpha y = \alpha (x + v) \; \; \; .
\ee
Obviously, adding the gauge fixing term $\Gamma_{g.f.}$ (\ref{ssbgf}) to
$S(x,y)$ breaks invariance under (\ref{gaugetr}) due to ($\delta_\alpha B
= 0$ by definition)
\be \label{gaugebr}
\delta_\alpha \Gamma_{g.f.} = \alpha \xi m B (x + v) \; \; \; .
\ee
In order to keep track of the original symmetry (\ref{gaugetr}) one has to
enlarge the space of ``fields'' $\{x,y\}$ by some unphysical ``fields''
$c$ and $\overline{c}$, the Faddeev-Popov ghosts, and, at the same time,
to go over from gauge invariance to BRS invariance:
\be \label{brs}
\begin{array}{lclcl} 
s x = - c y & , & s y = c (x + v) & , & \\
s c = 0 & , & s \overline{c} = B & , & s B = 0
\end{array} 
\ee
It has to be remarked that in the present context Faddeev-Popov ghosts
have to be introduced although there are no gauge fields in the model
at all. At first sight this seems unusual when compared to ordinary quantum
field theory. Taking, however, the noncommutative point of view (see later)
according to which the Higgs fields are part of a (generalized) gauge
potential and, hence, by themselves a sort of generalized gauge fields,
the above introduction of ghost ``fields'' becomes more plausible.\\
In a next step $S + \Gamma_{g.f.}$ has to be completed by a 
Faddeev-Popov ($\phi \pi$) part $\Gamma_{\phi \pi}$ chosen in such a way
that the variation of $\Gamma_{\phi \pi}$ under BRS exactly cancels
the variation of $\Gamma_{g.f.}$ under BRS (see (\ref{gaugebr}) with
$\delta_\alpha \rightarrow s$ and $\alpha \rightarrow c$). This leads
to
\be \label{phipip}
\Gamma_{\phi \pi} = - \xi m \overline{c} (x + v) c \; \; \; .
\ee
Because some of the BRS transformations (\ref{brs}) are bilinear in the
classical fields and because in the full quantum theory the classical
fields will become some nontrivial oper\-ators, for a proper definition
of such products of operators in higher orders it is furthermore
necess\-ary to couple all nonlinear BRS transformations to external
sources thereby introducing an additional part
$\Gamma_{ext.f.}$ to the action depending on the external sources:
\be \label{extfp}
\Gamma_{ext.f.} = X (sy) + Y (sx)
\ee
In summary, the complete classical action $\Gamma_{cl}$ is now given by:
\be \label{ssbaccl}
\Gamma_{cl} (x,y,B,c,\overline{c};X,Y) = S(x,y) +
\frac{1}{2} B^2 + \xi m B y - \xi m \overline{c} (x + v) c +
X c (x + v) - Y c y
\ee
(\ref{ssbaccl}) is the fundamental starting point for all considerations
to follow in this section.\\
 In table~1 we list for later convenience  all ``fields''
of the model with their respective dimensions and $\phi \pi$ quantum
numbers $Q_{\phi \pi}$ as well as their behaviour under
charge conjugation\footnote{One easily proves that $\Gamma_{cl}$
(\ref{ssbaccl}) is invariant under $C$. Taking into account the
behaviour under charge conjugation will simplify some 
calculations later on.} \nolinebreak $C$. All these
properties are inherited from the corresponding theory in 
four dimensions.
\par
\vspace*{- 2ex}
\begin{center}
\begin{tabular}{l|c|c|c|c|c|c|c} 
field & $x$ & $y$ & $B$ & $c$ & $\overline{c}$ & $X$ & $Y$ \\ \hline
dim & 1 & 1 & 2 & 0 & 2 & 3 & 3 \\ \hline
$Q_{\phi \pi}$ & 0 & 0 & 0 & 1 & - 1 & - 1 & - 1 \\ \hline
$C$ & + & - & - & - & - & - & +
\end{tabular}
\par
Table~1: Quantum numbers of the fields
\end{center}
From the construction it is clear that $\Gamma_{cl}$ is invariant under
the BRS transformations (\ref{brs}) ($sX = 0 = sY$ by definition). At
the ``functional'' level this invariance under BRS transformations is
expressed in terms of the Slavnov-Taylor identity (STI):
\ba \label{ssbsti}
& {\cal S} (\Gamma_{cl} ) = 0 & \\[.5ex]
& \mbox{with } \; {\cal S} (\Gamma ) \equiv
\displaystyle\frac{\partial \Gamma}{\partial Y} 
\frac{\partial \Gamma}{\partial x} +
\frac{\partial \Gamma}{\partial X} \frac{\partial \Gamma}{\partial y} +
B \frac{\partial \Gamma}{\partial \overline{c}} & \nonumber
\ea
\par
\vspace*{.3cm}
Let us now turn to the second part of this section, namely the recursive
determination of \nolinebreak $\Gamma^{(n)}$. 
We again begin our investigations by
first writing down the generating functional $Z(\rho, \tau, l, \eta,
\overline{\eta}; X, Y)$ of general Green's functions:
\be \label{ssbz}
Z(\rho, \tau, l, \eta, \overline{\eta}; X, Y) = {\cal N}
\int_{- \infty}^{+ \infty} dx \; dy \; dB \; dc \; d\overline{c} \;
\mbox{exp} \left\{ \frac{1}{\hbar} (- \Gamma_{cl} +
\rho x + \tau y + l B + \overline{c} \eta + \overline{\eta} c) \right\}
\ee
$\rho, \tau, l, \eta$ and $\overline{\eta}$ are the five external sources
corresponding to the ``propagating fields'' $x, y, B, \overline{c}$ and $c$,
respectively. Among these sources $\rho, \tau$ and $l$ are ordinary
numbers whereas $\eta$ and $\overline{\eta}$ have to be
Grassmannian (anticommuting) numbers 
with $\phi \pi$-charge +~1 and \linebreak-~1,
respectively. The derivation of the system of differential equations
that $Z$ has to fulfill, copies the by now familar strategy we
already pursued several times. Having five dynamical variables $x, y, B,
c$ and $\overline{c}$ in the game we obtain a system of 
five coupled partial differential equations for $Z$:
\ba \label{ssbdiffz}
\hbar^3 \; \frac{\lambda}{4} \left(
\frac{\partial^3 Z}{\partial \rho^3} +
\frac{\partial^3 Z}{\partial \rho \partial \tau^2} \right) +
\hbar^2 \left[ \frac{\lambda v}{4} \left(
3 \frac{\partial^2 Z}{\partial \rho^2} +
\frac{\partial^2 Z}{\partial \tau^2} \right) + \xi m
\frac{\partial^2 Z}{\partial \eta \partial \overline{\eta}} \right]
& & \nonumber \\
- \hbar \left[ 2 m^2 \frac{\partial Z}{\partial \rho} -
X \frac{\partial Z}{\partial \overline{\eta}} \right] -
\rho Z & = & 0 \nonumber \\[.5ex]
\hbar^3 \; \frac{\lambda}{4} \left(
\frac{\partial^3 Z}{\partial \tau^3} +
\frac{\partial^3 Z}{\partial \rho^2 \partial \tau} \right) +
\hbar^2 \; \frac{\lambda v}{2} \; 
\frac{\partial^2 Z}{\partial \rho \partial \tau} +
\hbar \left( \xi m \frac{\partial Z}{\partial l} -
Y \frac{\partial Z}{\partial \overline{\eta}} \right) -
\tau Z & = & 0 \nonumber \\[.5ex]
\hbar \left( \frac{\partial Z}{\partial l} +
\xi m \frac{\partial Z}{\partial \tau} \right) -
l Z & = & 0 \\[.5ex]
\hbar^2 \; \xi m \; \frac{\partial^2 Z}{\partial \eta \partial \rho} +
\hbar \left( \xi m v \frac{\partial Z}{\partial \eta} -
Y \frac{\partial Z}{\partial \tau} +
X \frac{\partial Z}{\partial \rho} \right) +
\left( v X - \overline{\eta} \right) Z & = & 0 \nonumber \\[.5ex]
\hbar^2 \; \xi m \; 
\frac{\partial^2 Z}{\partial \overline{\eta} \partial \rho} +
\hbar \; \xi m v \;
\frac{\partial Z}{\partial \overline{\eta}} +
\eta Z & = & 0 \nonumber
\ea
The first two equations in (\ref{ssbdiffz}) are direct generalizations of
(\ref{nssbdiffz}) to the case when there are also the variables $B, c$
and $\overline{c}$, whereas the remaining three equations are new
due to the additional presence of the auxiliary field $B$ and the
$\phi \pi$ ghosts $c$ and $\overline{c}$.\\
The system (\ref{ssbdiffz}) now has to be translated first to 
a system of differential
equations for $W(\rho, \tau, l, \eta, \overline{\eta}; X, Y)$ and then
to such a system for $\Gamma (x, y, B, c, \overline{c}; X, Y)$ which
subsequently has to be solved order by order in $\hbar$. We skip the
details of the calculation here and refer the reader once again to the
appendicies, this time to appendix~F, where the computational steps
in between can be found. Instead we just quote some main results.\\[.3cm]
The third equation in (\ref{ssbdiffz}) when expressed for the functional
$\Gamma$ just becomes the gauge fixing condition
\be \label{gfcond}
\displaystyle\frac{\partial \Gamma}{\partial B} = B +
\xi m y \; \; \; ,
\ee
which due to its linearity in the ``propagating fields'' can be integrated
to all orders without any problems. Hence, at least one of the five
equations above is almost trivial reducing that way the complexity of the
system (\ref{ssbdiffz}) a little bit.\\
When rewritten for the function $\Gamma$ the first 
two equations in (\ref{ssbdiffz})
determine the dependence of $\Gamma$ on $x$ and $y$, respectively. Up to
and including 1-loop order we find for these equations (neglecting
momentarily some contributions depending on the ghosts $c$ and $\overline{c}$
which are indicated by the dots):
\ba \label{ssbdiffgh}
\frac{\partial \Gamma}{\partial x} & = & 
\frac{\lambda}{4} \; (x^3 + x y^2 + 3 v x^2 + v y^2) - 2 m^2 x \nonumber \\
& & + \; \hbar \left\{
\frac{\lambda}{4} \; h^{- 1} (x,y) \left[
(x + v) \left( 3 \left( 
\frac{\partial^2 \Gamma}{\partial y^2} - (\xi m)^2 \right) +
\frac{\partial^2 \Gamma}{\partial x^2} \right) -
2 y \frac{\partial^2 \Gamma}{\partial x \partial y} \right] \right.
\nonumber \\
& & \hspace*{1cm} \left. - \xi m \left( 
\frac{\partial^2 \Gamma}{\partial \overline{c} \partial c} \right)^{- 1}
\right\} + \dots + {\cal O} (\hbar^2) \\
\frac{\partial \Gamma}{\partial y} & = & 
\frac{\lambda}{4} (y^3 + x^2 y + 2 v x y) + \xi m B \nonumber \\
& & + \; \hbar \; \frac{\lambda}{4} \; h^{- 1} (x,y) \left\{
- 2 (x + v) \frac{\partial^2 \Gamma}{\partial x \partial y} +
y \left( \left(
\frac{\partial^2 \Gamma}{\partial y^2} - (\xi m)^2 \right) +
3 \frac{\partial \Gamma}{\partial x^2} \right) \right\} \nonumber \\
& & + \dots + {\cal O} (\hbar^2) \nonumber
\ea
The function $h (x,y)$ generalizes the function $\hat{h} (x,y)$ 
(\ref{nssbhh}) of the
previous section and is defined to be:
\be \label{ssbh}
h (x,y) = \left. \frac{\partial^2 \Gamma}{\partial x^2} \left(
\frac{\partial^2 \Gamma}{\partial y^2} - (\xi m)^2 \right) - \left(
\frac{\partial^2 \Gamma}{\partial x \partial y} \right)^2 \;
\right|_{c = 0}
\ee
The modifications brought about by the introduction of the auxiliary
field $B$ and the ghosts $c$ and $\overline{c}$, i.e. by going over from
gauge invariance to BRS invariance, are easily read off by comparing
(\ref{ssbdiffgh}) and (\ref{ssbh}) to (\ref{nssbdiffgh}) and
(\ref{nssbhh}), respectively. In the present approximation these
modifications mainly consist in the following two points:\\
Firstly, in every place of occurence $\frac{\partial^2 \Gamma}{\partial
y^2}$ has to be replaced by $\frac{\partial^2 \Gamma}{\partial y^2} -
(\xi m)^2$, thus allowing for an auxiliary mass $- \frac{1}{2}
(\xi m)^2$ for the Goldstone boson $y$ without breaking invariance
under BRS transformations.\\
Secondly, the first equation in (\ref{ssbdiffgh}) contains the
additional term $- \; \hbar \; \xi m \left(
\frac{\partial^2 \Gamma}{\partial \overline{c} \partial c} \right)^{- 1}$. 
This term
encodes contributions stemming from internal $\langle \overline{c} c 
\rangle$-propagators to scattering processes including at least one
external but amputated Higgs field $x$. These contributions can and will
be present starting from 1 loop on, see also later.\\
Finally, moving on to the recursive solution of (\ref{ssbdiffgh})
(or better to the recursive solution of the complete 
system (\ref{ssbdiffg})) we find -- after the
consistency check $\Gamma^{(0)} = \Gamma_{cl}$ with $\Gamma_{cl}$ 
according to (\ref{ssbaccl}) -- the following expression for
$\Gamma^{(1)}$:
\ba \label{ssbg1}
\Gamma^{(1)} & = & \frac{1}{2} \mbox{ ln } h^{(0)} (x,y) \; - \; 
\mbox{ln } (x + v) \; + \; C^{(1)} \\
& & + \; \overline{c} \; \frac{\xi m}{x + v} \; f^{(0)} (x,y) \; c \; - \;
X \; \frac{f^{(0)} (x,y)}{x + v} \; c \; - \; 
Y \; \frac{\lambda y}{2 \; h^{(0)} (x,y)} \; c \nonumber \\[.7ex]
\mbox{with } \; h^{(0)} (x,y) & = &
\frac{1}{16}(3 \lambda x^2 + \lambda y^2 + 6 \lambda v x - 8 m^2)
(\lambda x^2 + 3 \lambda y^2 + 2 \lambda v x - 4 (\xi m)^2 )
\nonumber \\
& & \hspace*{.5cm} - \; \frac{\lambda^2}{4} (x + v)^2 y^2 \nonumber \\[.7ex]
\mbox{and } \; f^{(0)} (x,y) & = &
\frac{\lambda x^2 + 3 \lambda y^2 + 2 \lambda v x - 4 (\xi m)^2}{4 \;
h^{(0)} (x,y)} \nonumber
\ea
The constant of integration $C^{(1)}$ is fixed by the requirement
$\left. \Gamma \right|_{x = y = B = c = \overline{c} = 0} = 0$ to all orders,
i.e.\footnote{Taking into account this constant of integration the arguments
of the two logarithms in the first line of (\ref{ssbg1}) become dimensionless,
as they should.}
\be \label{ssbint1}
C^{(1)} = - \frac{1}{2} \mbox{ ln } \left( 2 \xi^2 m^4 
\right) \; + \; \mbox{ln } v \; \; \; .
\ee
In contrast to (\ref{nssbg1}) $\Gamma^{(1)}$ in (\ref{ssbg1}) is 
well-defined in the vicinity of $x = y = 0$. 
The same holds, of course, true for all possible 
1~PI $n$-point functions.\\
As a conclusion, we may state that the problems mentioned in the introduction
to this chapter have been cured.\\
Appendix~F also presents the result for the 2-loop approximation 
$\Gamma^{(2)}$ of $\Gamma$. It is clear that due to the recursive
nature of the determination again an implementation of the whole
procedure on the computer would be not only possible but also perhaps
desirable. However, we will abstain from this possibility because in
the present formulation the resulting expressions rapidly blow up
tremendously becoming almost unreadable, a technical difficulty
which hopefully will be overcome when the more appropriate noncommutative
language is used. That indeed the noncommutative version generates
substantial simplification will be the topic of the next chapter.

\newsubsection{Feynman propagators and ``counting'' of diagrams}

\begin{figure}[ht]
\begin{center}
\includegraphics[width=150mm,angle=0]{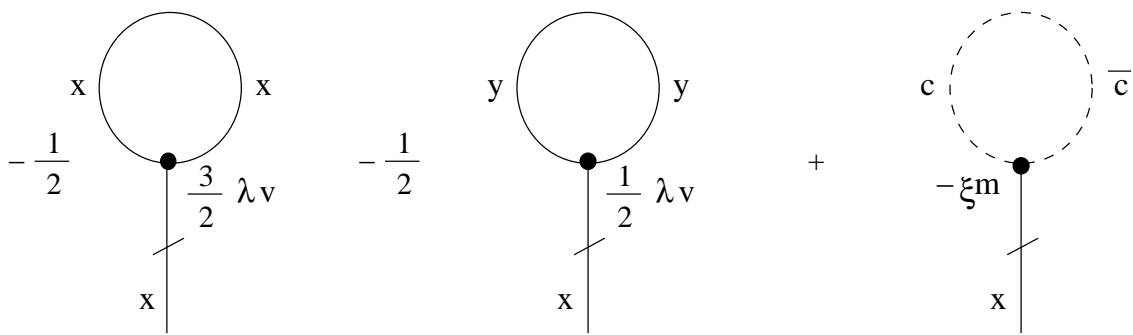}
\end{center}
\caption{\small 1 PI contributions to the Higgs tadpole in 1 loop.
The sloping bar on the external $\langle x x \rangle$-propagators indicates 
that these lines are amputated. Furthermore, the numerical 
factors in front of the diagrams stem from Wick's theorem. \label{subline2}}
\end{figure}

On the basis of the outcome of the previous section we are now in a
position to investigate the question to which extent the counting
argument of Feynman diagrams can be transferred to the model under
consideration including SSB -- at least for the case of 1~PI diagrams.
In order to allow for a detailed comparison between the analytical
results derived above and a direct ``counting'' of Feynman diagrams
we first of all need the free Feynman propagators of the present theory
in zero dimensions. For the calculation of these propagators we start
off, as usual, from the bilinear part
\be \label{ssbacclbi}
\Gamma^{(0)}_{bil} = - m^2 x^2 + \frac{1}{2} B^2 +
\xi m B y - \xi m v \overline{c} c
\ee
of the classical action $\Gamma_{cl}$ (\ref{ssbaccl}). Let us briefly
illustrate another simple method\footnote{Of course, we could equivalently
proceed in the same manner as in chapter~2 by first calculating
\be \label{ssbzfrei}
Z_{free} (\rho, \tau, l, \eta, \overline{\eta} ) = {\cal N} \int
d x \; d y \; d B \; d c \; d \overline{c} \; \mbox{exp} \left\{
\frac{1}{\hbar} \left( - \Gamma^{(0)}_{bil} + \rho x + \tau y +
l B + \overline{c} \eta + \overline{\eta} c \right) \right\} \; \; ,
\ee
with the result,
\be \label{ssbzfreir}
Z_{free} = \left( 1 + \frac{\eta \overline{\eta}}{\xi m v \hbar} \right)
\mbox{ exp} \left\{ \frac{1}{2 \hbar} \left(
- \frac{\rho^2}{2 m^2} - \frac{\tau^2}{\xi^2 m^2} + 
\frac{2 \tau l}{\xi m} \right) \right\} \; \; ,
\ee
and by subsequently using the general definition 
of the Feynman propagator in question which 
uses $Z_{free}$ and looks like (\ref{feynpropd}).}
for finding the propagators by
examining, as an example, the Feynman propagator of the Higgs' variable
$x$: Obviously, we have $\frac{\partial \Gamma^{(0)}_{bil}}{\partial x}
= - 2 m^2 x$ which is coupled to the external source $\rho$. I.e. on the
bilinear part $\rho$ is just a function of $x$ only: $\rho =
\rho (x) = - 2 m^2 x \Leftrightarrow x = - \frac{\rho}{2 m^2}$.
According to the definition of the $\langle x x \rangle$-propagator,
$\Delta^F_{xx} = \hbar \; 
\frac{\partial x}{\partial \rho}$, we hence find the
following answer which could have been expected:
\be \label{propxx}
\Delta^F_{xx} = - \displaystyle\frac{\hbar}{2 m^2}
\ee
In a completely analogous manner also the ghost propagator results:
\be \label{propcc}
\Delta^F_{\overline{c} c} = \displaystyle\frac{\hbar}{\xi m v}
\ee
The derivation of the remaining Feynman propagators in zero dimensions
is slightly more involved but nevertheless follows along the lines indicated
above. We find:
\be \label{proprem}
\Delta^F_{yy} = - \displaystyle\frac{\hbar}{(\xi m)^2} \; \; , \; \; 
\Delta^F_{By} = \frac{\hbar}{\xi m} \; \; , \; \;
\Delta^F_{BB} = 0
\ee
As in the previous basic example of $\varphi^4$-theory, in a next step
we could normalize these propagators to simple numbers by choosing
appropriate values for the parameters $\hbar, m^{(2)}, \xi$ and $v$.
We forgo this possibility here for the following two reasons: On the one
hand, keeping the above parameters explicitly we will have a  better
chance for a more profound check of the many factors in the analytical
expressions. On the other hand, going ahead that way we will also be
able to clarify more accurately the origin of the various terms
contributing to $\Gamma^{(n)}$. In order not to be too academic, we
now turn to some particular examples.\\[1.5ex]
{\bf Example 1: } Higgs tadpole\\[.5ex]
In figure~\ref{subline2} the 1 PI Feynman diagrams contributing in the 1-loop
approximation to the Higgs 1-point function are depicted graphically.\\
\par
\noindent
Figure~\ref{subline2} also contains the information about 
the relevant coupling constants
related to the vertices $x x x, x y y$ and $x \overline{c} c$, as they follow
immediately from the interaction terms in 
$\Gamma_{cl}$ \nolinebreak (\ref{ssbaccl}).
A direct calculation of $\Gamma_x^{(1)}$ is now very easy due to the
fact that in zero dimensions no integrations over internal loops have
to be performed. In other words, the value of each Feynman
diagram in zero dimensions is given by the product of a factor
originating from Wick's theorem, the vertex factors and the propagators
involved in the diagram. For instance, the first contribution in figure~2
is given by ($\hbar \equiv 1$):
\begin{displaymath}
- \frac{1}{2} \cdot \frac{3 \lambda v}{2} \cdot \Delta^F_{xx} =
\frac{3 \lambda v}{8 m^2}
\end{displaymath}
In doing so we altogether obtain\footnote{Please note that for each closed
ghost loop there is an extra factor $- 1$.}
\be \label{ssbgamx1}
\Gamma_x^{(1)} = \frac{3 \lambda v}{8 m^2} +
\frac{\lambda v}{4 (\xi m)^2} + \frac{1}{v} \; \; \; ,
\ee
where the last term comes from internal ghost contributions. This result
has to be compared with the derivative of $\Gamma^{(1)}$
(\ref{ssbg1}) with respect to $x$ subsequently evaluated for all fields
equal to zero. We find complete agreement (up to an overall 
sign\footnote{See footnote~\ref{fn1}; this time, however, the Legendre
transformation is defined in (\ref{ssbleg}).}). 
Furthermore, we observe that
the contribution originating from the internal $\langle \overline{c}
c \rangle$-propagator is encoded in the second term of $\Gamma^{(1)}$,
namely in -~ln$(x + v)$, whereas the other contributions stem from the
first term $\frac{1}{2}$~ln $h^{(0)} (x,y)$.\\[1.5ex]
{\bf Example 2: } Self-energy of the Higgs\\[.5ex]
In 1 loop the following 1~PI diagrams contribute to the self-energy of
the Higgs' variable $x$:\\
\par
\begin{figure}[ht]
\begin{center}
\includegraphics[width=150mm,angle=0]{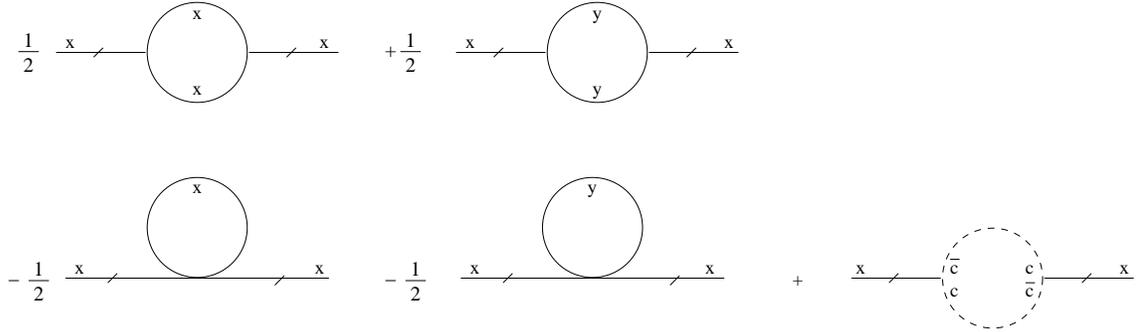}
\end{center}
\caption{\small 1~PI contributions to $\Gamma_{xx}$ at order
$\hbar$. \label{subline3}}
\end{figure}
\noindent
Taking into account the vertex factors $\frac{3}{2} \lambda$ and
$\frac{1}{2} \lambda$ for the vertices $x x x x$ and $x x y y$,
respect\-ively, this time the direct calculation of the above diagrams
yields:
\be \label{ssbgamxx1}
\Gamma_{xx}^{(1)} = \frac{9}{32} \frac{\lambda^2 v^2}{m^4} \; + \;
\frac{1}{8} \frac{\lambda^2 v^2}{(\xi m)^4} \; + \;
\frac{3}{8} \frac{\lambda}{m^2} \; + \;
\frac{1}{4} \frac{\lambda}{(\xi m)^2} \; - \;
\frac{1}{v^2}
\ee
(\ref{ssbgamxx1}) again coincides (up to the same overall sign as in
example~1) with the analytical result, i.e. with the second derivative
of $\Gamma^{(1)}$ (\ref{ssbg1}) with respect to $x$ finally evaluated
for all fields set equal to zero. It also holds true that the contribution
from the internal ghost propagators still stems from the term
-~ln$(x + v)$ in $\Gamma^{(1)}$.\\[1.5ex]
{\bf Example 3:} Higgs tadpole in 2-loop order\\[.5ex]
Let us finally look at an example involving the second order
of the loop expansion. To simplify matters we reconsider the Higgs
1-point function: For 1~PI $n$-point functions different from $\Gamma_x^{(2)}$
an even larger number of Feynman diagrams has to be taken into account in
the 2-loop approximation. As usual, the analytical answer is obtained by
differentiating $\Gamma^{(2)}$ (\ref{ssbg2}) once with respect to $x$
and by subsequently evaluating the resulting expression for $x = 0 = y$.
\par
\vspace*{.5cm}
\begin{figure}[ht]
\begin{center}
\includegraphics[width=150mm,angle=0]{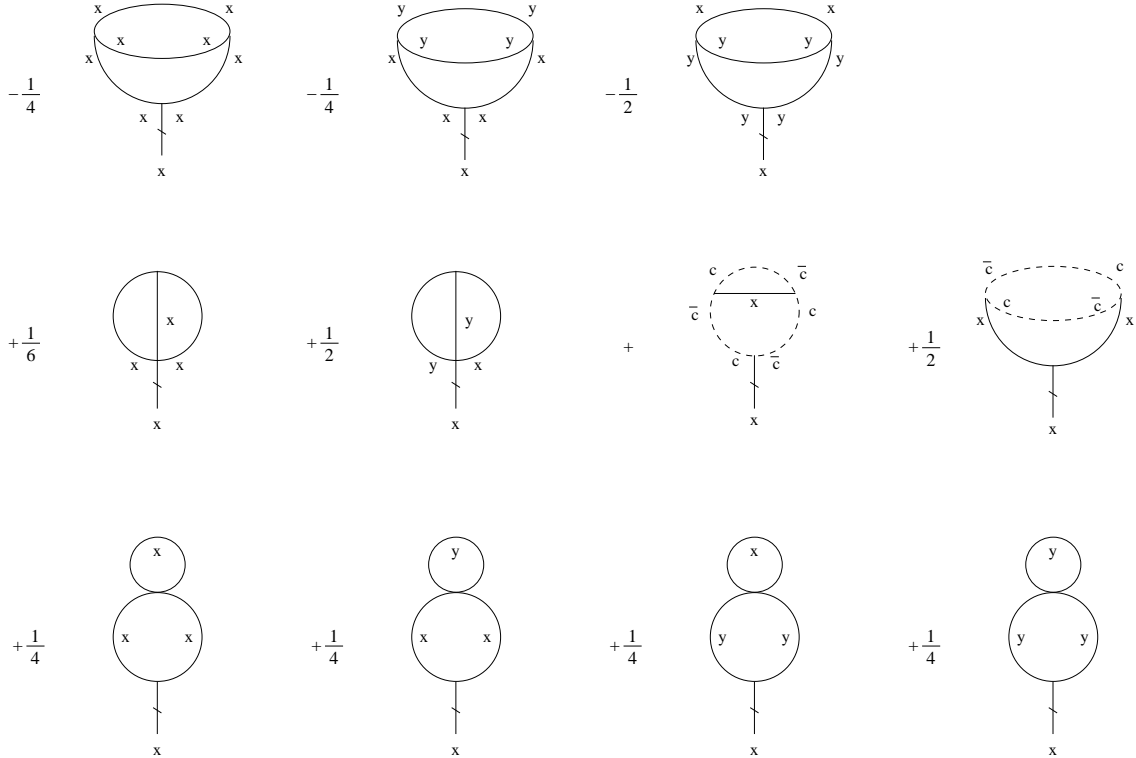}
\end{center}
\caption{\small 1 PI contributions to $\Gamma_x^{(2)}$. \label{subline4}}
\end{figure}
\par
\noindent
This leads to
\ba \label{lasteq}
  \Gamma_x^{(2)} & = & \frac{1}{512 \; m^8 v^3 (\xi m)^6} \left[
  256 \; m^6 (\xi m)^6 - 96 \; \lambda \; m^4 v^2 (\xi m)^6 \right. \\
& & \hspace*{2cm} \left. + \; 12 \; \lambda^2 m^2 v^4 \left( 8 \; m^6 + 
    4 \; m^4 (\xi m)^2 +
    2 \; m^2 (\xi m)^4 + 5 \; (\xi m)^6 \right) \right. \nonumber \\
& & \hspace*{2cm} \left. + \; \lambda^3 v^6 \left( 16 \; m^6 + 
    12 \; m^4 (\xi m)^2 +
    27 \; (\xi m)^6 \right) \right] \; \; \; . \nonumber
\ea
(\ref{lasteq}) again coincides (up to an overall sign) with the direct
calculation of the Feynman diagrams contributing to $\Gamma_x^{(2)}$.
These diagrams are depicted in figure~\ref{subline4} above.\\[.3cm]
In summary, we may draw the following conclusion: Even in models without
SSB the counting of Feynman diagrams, achieved by considering the 
corresponding zero-dimen\-sional theory, is not a naive counting just
yielding the number of diagrams but rather a weighted one: Combinatorial
factors from Wick's theorem as well as possible symmetry factors enter
this counting. As a consequence, the counting argument should be regarded
as a consistency check that in a four-dimensional calculation all
diagrams {\it and} all combinatorial factors {\it and} all symmetry factors
have been taken into account correctly. When taking this point of view, 
the previous examples clearly show that the counting argument
can very well be transferred to models exhibiting spontaneous symmetry
breaking.

\newsubsection{Renormalization}

Suppose that we would like to depict graphically the modifications to
the (zero-dimen\-sional) classical Higgs potential $V(x,y) =
S(x,y) = \left. \Gamma_{cl} \right|_{B = c = \overline{c} = 0}$ stemming
from higher order loop contributions because of the assumption that
the effective Higgs potential $V_{eff}^{(\leq n)} (x,y) = \left.
\Gamma^{(\leq n)} \right|_{B = c = \overline{c} = 0}$ 
(in zero dimensions this is the full physical theory!) should have 
a physical meaning. This immediately raises the question:
What (physical) values do we have to choose for the various parameters
appearing in $\Gamma^{(n)}$? The necessity of finding an answer to this
question indicates nothing else but the necessity of (re)normalizing
the theory, i.e. the necessity of expressing the so far unphysical
parameters of the theory in terms of observable and, hence, physical
quantities (order by order in the loop expansion). When compared to the
first simple example of zero-dimensional $\varphi^4$-theory, in the
present context the necessity of {\it re}normalization (with emphasis put
on the syllable ``re'') becomes even more transparent:
Whereas in the classical approximation the vacuum expectation value
$\langle x \rangle = \Gamma_x^{(0)}$ of the Higgs' variable $x$ is by
construction $0$, example~1 of section~3.3 shows that this statement does
not hold true in higher orders, see (\ref{ssbgamx1}). In other words,
without {\it re}normalization the theory will drift away from the ring
of local minima which, in turn, is disastrous for the stability (and 
definition) of a perturbative treatment of a physical theory. So as to
keep fixed the theory in one of the minima also in higher orders 
(among other things) it is unavoidable to introduce 
appropriate invariant counterterms
to the classical action the coefficients of which have to be 
{\it re}normalized again and again order by order in the loop expansion.\\
In order not to forget a possible counterterm (i.e. in order to control
{\it all} of them explicitly) we choose the common approach to such a task
relying on the defining property of the theory, namely the fundamental
notion of symmetry which in the current investigation is BRS symmetry
expressed in terms of the STI (\ref{ssbsti}). Hence, first of all we have
to look for the most general classical solution $\Gamma_{cl}^{gen}$ of
(\ref{ssbsti}). An inspection of table~1, the table of quantum numbers 
of all the fields involved, yields
the following most general ansatz for $\Gamma_{cl}^{gen}$ (already
exploiting the gauge fixing condition (\ref{gfcond}) and making use of
the facts that $\Gamma_{cl}^{gen}$ has to be $\phi \pi$-neutral,
invariant under charge conjugation $C$ and bounded in dimension by $4$):
\ba \label{ssbggencla}
\Gamma_{cl}^{gen} & = & a_1 x + a_2 x^2 + a_3 y^2 + a_4 x^3 +
a_5 x y^2 + a_6 x^4 + a_7 x^2 y^2 + a_8 y^4 \nonumber \\
& & + b_1 c \overline{c} + b_2 c \overline{c} x +
b_3 c \overline{c} x^2 + b_4 c \overline{c} y^2 +
b_5 c X + b_6 c X x + b_7 c Y y \nonumber \\
& & + \frac{1}{2} B^2 + \xi m B y
\ea
So far $a_1, \dots, a_8$ and $b_1, \dots, b_7$ are unconstrained 
free parameters.
Inserting this ansatz into the STI (\ref{ssbsti}) and comparing coefficients
of independent ``field'' monomials will lead to some restrictions among
the formerly free parameters. After a straightforward but a little bit
cumbersome calculation we obtain the result (upon renaming parameters):
\ba \label{ssbggencl}
\Gamma_{cl}^{gen} & = & \frac{1}{2} \hat{\mu} \left(
2 \hat{v} x + z_x x^2 + z_y y^2 \right) +
\frac{\hat{\lambda}}{16} \left(
2 \hat{v} x + z_x x^2 + z_y y^2 \right)^2 \\
& & + \frac{1}{2} B^2 + \xi m B y - \xi m \overline{c} (z_x x + \hat{v} ) c +
X(z_x x + \hat{v} ) c - Y z_y y c  \nonumber
\ea
From (\ref{ssbggencl}) we easily can read off that there are five free
parameters within the theory, namely $z_x, z_y, \hat{\mu}, \hat{v}$ and
$\hat{\lambda}$. These parameters are not prescribed by BRS symmetry and,
hence, have to be fixed elsewhere. Due to the method of determination
these parameters, furthermore, are in one-to-one correspondence with all
possible invariant counterterms in higher orders.\\
Because of the additional freedom related to the additional free
parameter $\hat{\mu}$ in (\ref{ssbggencl}) when compared to
$\Gamma_{cl}$ (\ref{ssbaccl}), this parameter $\hat{\mu}$ multiplying
a term {\it linear} in $x$, and because of the remarks just given, already
at this point we can foresee that by an appropriate choice of $\hat{\mu}$
in higher orders it will be possible to hold the theory strictly in one
of the minima of the effective potential.\\
As usual, the free parameters will be fixed by normalization conditions.
Dealing with {\it formal} power series in $\hbar$ the only requirement
that these normalization conditions have to fulfill is 
the requirement of uniquely
determining the free parameters in lowest order. Apart from this
requirement the actual choice of the normalization conditions is
to a certain extent a matter of taste at least in the present 
context\footnote{See also the concluding remark at the end of this section.}. 
We propose the following ones ($m_{phys}^2 > 0$):
\be \label{ssbnormc}
\Gamma_x = 0 \; , \; \Gamma_{xx} = \; 2 m_{phys}^2 \; , \;
\Gamma_{xxxx} = \frac{3}{2} g_{phys} \; , \;
\Gamma_{yyyy} = \frac{3}{2} g_{phys} \; , \;
\Gamma_{c X} = \sqrt{\frac{4 m_{phys}^2}{g_{phys}}}
\ee
($\Gamma_x$ denotes $\frac{\partial \Gamma}{\partial x}$ evaluated for all
fields set equal to zero, etc.)\\
Applying these normalization conditions to the most general classical
action (\ref{ssbggencl}) we exactly fall back to $\Gamma_{cl}$
(\ref{ssbaccl}) (with $\lambda \rightarrow g_{phys}$ and
$m^2 \rightarrow - m_{phys}^2$).\\
Now we have everything at our disposal for a detailed study of the
renormalization of the model under consideration. For instance, up to
1~loop at the beginning there are two contributions to the vacuum
expectation value of the Higgs' variable $x$: The first contribution
was already calculated above and is given by (\ref{ssbgamx1})
(times $- 1$) with
$\lambda \rightarrow g_{phys}$ and $m^2 \rightarrow - m_{phys}^2$; the
second contribution represents an invariant counterterm to the 
classical action, see (\ref{ssbggencl}), namely:
\begin{displaymath}
\left( \hat{\mu} \hat{v} \right)^{(1)} = 
\hat{\mu}^{(0)} \hat{v}^{(1)} +
\hat{\mu}^{(1)} \hat{v}^{(0)} =
\hat{\mu}^{(1)} \sqrt{\frac{4 m_{phys}^2}{g_{phys}}}
\end{displaymath}
The last equality holds true because of $\hat{\mu}^{(0)} = 0$ due to
the first normalization condition in (\ref{ssbnormc}) evaluated in lowest
order\footnote{Also the last normalization condition in (\ref{ssbnormc})
is used in order to exclude the case $\hat{v}^{(0)} = 0$.}. 
The sum of these two contributions again has to fulfill the 
normalization condition just mentioned, this time up to 1-loop order. This
determines \nolinebreak $\hat{\mu}^{(1)}$:
\be \label{ssbmu1}
\hat{\mu}^{(1)} = - \; \frac{g_{phys}}{8 m_{phys}^2} \left(
1 + \frac{2}{\xi^2} \right)
\ee
In a similar manner also the higher orders of the other parameters can be
found. Up to and including second order of the loop expansion 
we find:\\[-1.2cm]
\par
{\footnotesize
\ba \label{ssbpar2}
z_x & = & 1 \; + \; \hbar \; \frac{g_{phys}}{16 m_{phys}^4 \xi^8}
(16 - 16 \xi^2 + 29 \xi^8) \nonumber \\
& & + \; \hbar^2 \; \frac{g_{phys}^2}{512 m_{phys}^8 \xi^{16}}
(- 256 + 512 \xi^6 + 3744 \xi^8 - 3648 \xi^{10} + 11435 \xi^{16})
\; + \; {\cal O} (\hbar^3) \nonumber \\[.5ex]
z_y & = & 1 \; - \; \hbar \; \frac{g_{phys}}{4 \xi^2 m_{phys}^4} \; - \;
\hbar^2 \; \frac{g_{phys}^2}{64 m_{phys}^8 \xi^{10}}
(16 - 16 \xi^2 + 4 \xi^6 + 35 \xi^8) \; + \; {\cal O} (\hbar^3) 
\nonumber \\[.5ex]
\hat{\mu} & = & - \; \hbar \; \frac{g_{phys}}{8 m_{phys}^2 \xi^2}
(2 + \xi^2) \; - \; 
\hbar^2 \; \frac{g_{phys}^2}{64 m_{phys}^6 \xi^8}
(8 - 4 \xi^2 - 2 \xi^4 + \xi^6 + 15 \xi^8)
\; + \; {\cal O} (\hbar^3) \\[.5ex]
\hat{v} & = & \sqrt{\frac{4 m_{phys}^2}{g_{phys}}} \left(
1 \; + \; \hbar \;
\frac{g_{phys}}{8 m_{phys}^4} \; + \; 
\hbar^2 \; \frac{g_{phys}^2}{32 m_{phys}^8 \xi^8}
(16 - 16 \xi^2 + 33 \xi^8) \right)
\; + \; {\cal O} (\hbar^3) \nonumber \\[.5ex]
\hat{\lambda} & = & g_{phys} \; + \; \hbar \;
\frac{g_{phys}^2}{16 m_{phys}^4 \xi^4}
(4 + \xi^4) \; + \; 
\hbar^2 \; \frac{g_{phys}^3}{256 m_{phys}^8 \xi^8}
(48 - 32 \xi^2 + 8 \xi^4 + 75 \xi^8)
\; + \; {\cal O} (\hbar^3) \nonumber 
\ea}
\par
\vspace*{-.6cm}
\noindent
The derivation of the results starting from 2~loops on has to be handled
with some caution because, for instance, in the 2-loop approximation
also the 1-loop counterterms will produce nontrivial loop contributions
that have to be taken into account properly. Perhaps the safest way to
do this consists in repeating the analysis of section~3.2 but by using
$\Gamma_{cl}^{gen}$ instead of $\Gamma_{cl}$ as the starting point.
Proceeding that way the parameters $z_x, z_y, \hat{\mu}, \hat{v}$ and
$\hat{\lambda}$ will enter the systems of differential equations for
$Z, W$ and $\Gamma$. Keeping track of the power series of the above
parameters the subsequent recursive solution of the system for $\Gamma$,
carefully and consistently performed order by order in $\hbar$, then
leads to the correct results.\\[.3cm]
Finally, we are able to quote the renormalized effective action of our
zero-dimensional theory. In order not to waste too much space 
we content ourselves with writing down 
the effective Higgs potential (for $\xi = 1$) up to 1 loop:\\[-1cm]
\par
{\footnotesize
\ba \label{ssbhiggseff1}
V_{eff}^{(\leq 1)} & = &
    \frac{g_{phys}}{16} \; (x^2 + y^2) \left( 
    8 \; \sqrt{\frac{m_{phys}^2}{g_{phys}}} \; x + x^2 + y^2 \right) 
    + \; m_{phys}^2 \; x^2 \\
& & - \; \frac{\hbar}{256 \; m_{phys}^4} \left\{
    192 \; g_{phys}^2 \left( 
    \frac{m_{phys}^2}{g_{phys}} \right)^\frac{3}{2} x \; 
    - \; 96 \; g_{phys} \; m_{phys}^2 \; x^2 \right. \nonumber \\
& & \hspace*{3cm} \left. 
    - \; 288 \; g_{phys}^2 \; \sqrt{\frac{m_{phys}^2}{g_{phys}}} \; x^3 \;
    - \; 63 \; g_{phys}^2 \; x^4 \; + \; 48 \; g_{phys} \; m_{phys}^2 \; y^2 
    \right. \nonumber \\
& & \left. - \; 24 \; g_{phys}^2 \; 
    \sqrt{\frac{m_{phys}^2}{g_{phys}}} \; x y^2 \;
    - \; 60 \; g_{phys}^2 \; x^2 y^2 \; + \; 3 \; g_{phys}^2 \; y^4
    \; + \; 128 \; m_{phys}^4 \mbox{ ln} \left(32 \; m_{phys}^4 \right) 
    \right. \nonumber \\
& & \left. - \; 256 \; m_{phys}^4 \mbox{ ln} \left( 
   2 \; \sqrt{\frac{m_{phys}^2}{g_{phys}}} \right)
   + \; 256 \; m_{phys}^4 \mbox{ ln} 
   \left[ 2 \; \sqrt{\frac{m_{phys}^2}{g_{phys}}} + x \right] 
   \right. \nonumber \\
& & \left. - \; 128 \; m_{phys}^4 \mbox{ ln} \left[ 32 \; m_{phys}^4 
   + 3 \; g_{phys}^2 \; (x^2 + y^2) 
   \left( 8 \; \sqrt{\frac{m_{phys}^2}{g_{phys}}} \; x + x^2 + 
   y^2 \right) \right. \right. \nonumber \\
& & \hspace*{3cm} \left. \left. + \; 4 \; g_{phys} \; m_{phys}^2 
   \left( 20 \; \sqrt{\frac{m_{phys}^2}{g_{phys}}} \; x 
   + 17 \; x^2 + 3 \; y^2 \right) \right] \; \right\} \nonumber
\ea
}
\par
\vspace*{-.3cm}
\noindent
\begin{figure}[ht]
\begin{center}
\includegraphics[width=150mm,angle=0]{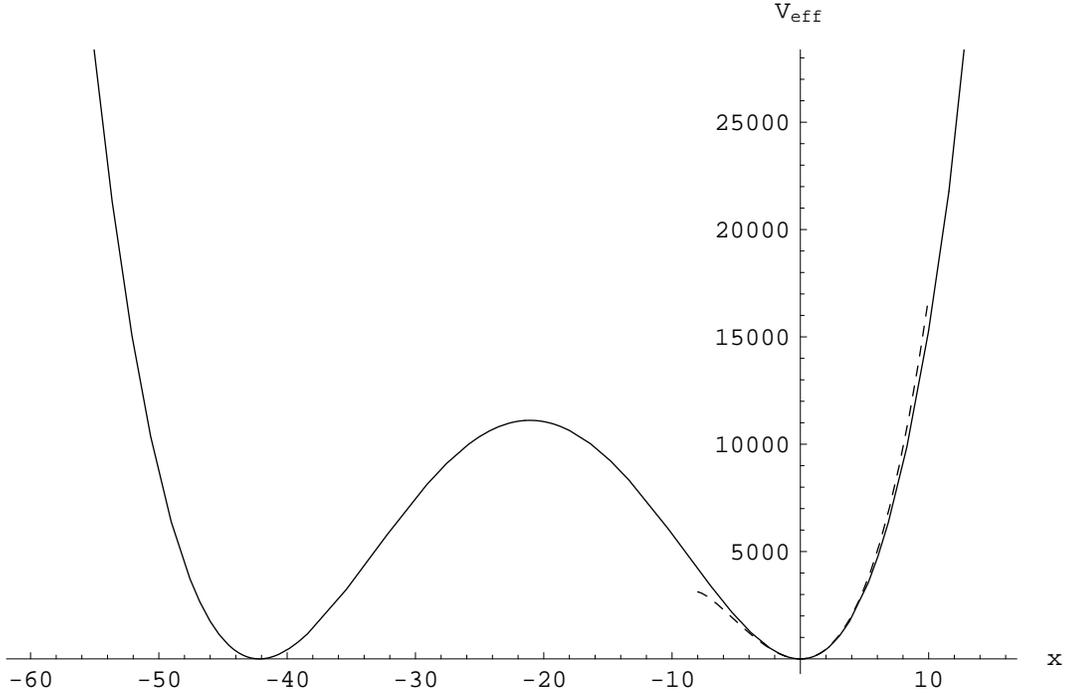}
\end{center}
\caption{\small The effective Higgs potential $V_{eff}$ up to 1-loop
                order. For the sake of clearness we refrain from picturing
the full three-dimensional plot. Instead, we merely
display a two-dimensional section by choosing
$y = 0$. The solid line represents the classical Higgs potential whereas
the dashed line shows $V_{eff} (x, y = 0)$ up to and including 1-loop order,
but only in the region where the formerly derived analytical expression
is well-defined (namely in the vicinity of $x = 0 = y$).
For the physical parameters, as an example, the values $g_{phys} = 0.9$ and
$m_{phys} = 10$ have been chosen. Furthermore, in order to be able to
visualize the tiny modifications of the loop contributions and of
renormalization the unrealistic value $\hbar = 1200$ is assumed.
\label{graph2}}
\end{figure}
\par
\vspace*{-.8cm}
\noindent
In figure~\ref{graph2} below a plot of (\ref{ssbhiggseff1}) (for $y = 0$) is
presented showing also the modifications to the classical Higgs potential 
$V$ due to the loop contributions. $V_{eff}^{(\leq 1)}$ is, however, only
plotted in a neighbourhood of the
minimum $x = 0 = y$ which is the spontaneously chosen vacuum of the theory:
Please note that $\Gamma^{(1)}$ (\ref{ssbg1}) is not defined for
$x = - v$ as well as for the zeros of $h^{(0)} (x,y)$. As a consequence, the
1-loop approximation derived above is only valid in the vicinity of the
origin; in the outside region (containing the singularities)
the above treatment simply breaks 
down\footnote{This observation is also the reason for the fact that we do not
rewrite the higher orders $\Gamma^{(n)}$ of $\Gamma$ in terms of the
unshifted fields $z$ and $\overline{z}$ (\ref{shift}), a strategy which
sometimes would lead to more compact expressions: Such a rewriting would
pretend the validity of perturbatively defined quantities in a region
where they are not valid.}. This breakdown, however, does not pose
a real problem for our purposes which aim at a consistent treatment of
a {\it perturbatively} defined theory. An analogous consideration, of course,
also holds true for $V_{eff}^{(\leq 1)}$ (\ref{ssbhiggseff1}).\\[.3cm]
Let us conclude the present chapter with a remark concerning the control
of gauge par\-ameter dependence: By a slight extension of usual BRS
invariance, namely by allowing (all) the gauge parameter(s) of the theory
also to transform under BRS into (a) Grassmannian variable(s), an
automatic and easily managable control of gauge parameter dependence
of single Green's functions is achieved. In the context of ordinary
gauge theories this method was first introduced in \cite{PS} and
afterwards further elaborated in \cite{KS} and \cite{HKS} for the case
of nonabelian gauge theories without SSB. In addition, the method also
offers an elegant possibility to decide whether or not the normalization
conditions chosen are in agreement with the gauge parameter dependence
of the theory. This last point was discussed
in detail (among other things) in \cite{HK} for the
Abelian Higgs model. Of course, such an analysis could be repeated here,
too. It perhaps would enlighten the trick of a BRS transforming gauge
parameter in the very simple context of a zero-dimensional theory in the
same manner as it is the case for the topic of renormalization.

%% file: chap3.tex
\newpage
\newsection{Noncommutative formulation of a zero-dimensional
            Higgs model and its quantization}

The investigations of the previous chapter, on the one hand, enabled us to
answer the question whether it is possible to apply the counting argument
of Feynman diagrams to theories including spontaneous symmetry breaking,
too, in the affirmative. On the other hand, we were also forced to notice
that the explicit determination of higher orders of $\Gamma$ rapidly becomes
technically rather involved even for the theory in zero dimensions.
We may interpret this drawback as a hint that the existing formulation
using the separate ``fields'' $x$ and $y$ is possibly not a suitable one,
and we, as a consequence, should watch out for an alternative formulation
which is better adapted.\\
During the last decade gauge theories based on (elements of) Noncommutative
Geometry have become more and more important and popular for the description
of elementary interactions. Among many other reasons, the importance of those
models is based on the fact that (in four dimensions) ordinary gauge fields
and Higgs fields are unified into a single object which is a generalized Dirac
operator $D$ or a generalized gauge potential \nolinebreak $\cA$ 
depending on the
particular noncommutative approach chosen. In any case, the marriage of gauge
fields and Higgs fields within these noncommutative models leads to an
automatic occurence of the mechanism of SSB right from the beginning which,
hence, is a built-in feature of these models, i.e it does not have to be
put in by hand as in the usual formulation. In this sense, noncommutative
gauge theories are the appropriate stage both for the study of models
including SSB in general and also for the problem indicated above in the
first lines of this introduction in particular. In other words, what we are
going to do is to repeat the procedure of the previous chapter but this time
by sticking strictly to the noncommutative language, i.e. by formulating
everything in terms of the basic noncommutative object, namely the matrix
$\cA$ comprising the components $x$ and $y$, instead of using the 
component language itself. As it will turn out the noncommutative formulation 
indeed results in a substantial simplification.\\
There is a further reason for our present interest in such a purely
noncommutative description: Up to now the derivation of gauge theories 
within Noncommutative Geometry (NCG) almost always stops at the classical level
by just deducing (in an admittedly elegant way) a classical action from
some first principles. The subsequent quantization is performed according to
common prescriptions thereby entirely forgetting about the initial 
noncommutative structure, so to speak. Of course, such a treatment is not 
satisfactory at all: The ``true'' quantization of noncommutative gauge 
theories has to comprehend the primary noncommutative
structure somehow\footnote{In the meanwhile, there are various attempts
dealing with what roughly could be called noncommutative quantum field
theory. These attempts, however, are not directly related to the
noncommutative model building mentioned above. They rather start off from
an assumed noncommutative behaviour of the underlying spacetime by considering
nonvanishing commutators of the time and position operators, see e.g.
\cite{DFR}, \cite{CHMS} and \cite{GKW}.}. Tackling this problem in full 
generality (i.e. in four dimensions
and for a physical theory like the noncommutative version of the standard
model) is, however, a really hard task. For this reason, it is perhaps
better and more promising to restrict attention to a very simple but
nontrivial example of a noncommutative theory\footnote{Investigations in 
this regard are also undertaken, for example, in \cite{CR} and \cite{MH}
in the context of Connes' formulation of NCG.}. But this is exactly what
we are going to do when we study the quantization of the present
model entirely within the noncommutative language. Our hope, hence, is that
the investigations to follow will possibly serve as a first step towards
a deeper understanding of what a truely noncommutative quantization of
gauge theories within NCG could be.\\
The chapter is organized as follows: After a brief review of model
building within NCG in section~4.1 (where we put
special emphasis on the so-called Marseille-Mainz model which manages
with a minimum of mathematical effort but, nevertheless, includes all the
nice noncommutative implications), in section~4.2 we explicitly define
the zero-dimensional noncommutative Higgs model which will be studied
subsequently. In this context we will avoid to introduce an auxiliary
field $B$ and Faddeev-Popov ghosts $c$ and $\overline{c}$ (necessary for
a proper treatment) in order not to complicate things in a first approach.
Section~4.3 presents the mathematical apparatus needed in the following.
In particular, this section contains the definition of what we mean by
differentiation of a scalar or matrix-valued quantity with respect to
a matrix as well as some useful and frequently required properties of
such a notion. Thus, having at our disposal all mathematical prerequisites,
in section~4.4 we finally are in the position to copy the procedure of the
previous chapter and to derive one after the other the differential
equations for $Z, W$ and $\Gamma$ but this time strictly within the
noncommutative language of matrices. At the end of section~4.4 we also start
the recursive solution of the differential equation for $\Gamma$ and calculate
$\Gamma^{(1)}$ which, quite remarkably, can be expressed as a function of the
generalized field strength $\cal F$ only. This immediately raises the question
whether this property will hold true to all orders of the loop expansion.
The affirmative answer to this question is proven in section~4.5. Please
note that the statement ``To all orders $\Gamma$ is a function of 
$\cal F$ only.'' only makes sense and is only 
accessible within a noncommutative
formulation: In the component language there is no object like a field
strength for the Higgs fields of the theory. Furthermore, the proof of
section~4.5 will be organized in such a way that higher orders of $\Gamma$
are calculated at the same time, too.

\newsubsection{Noncommutative gauge theories}

Already shortly after Alain Connes' pioneering and constitutive work on a 
generalization of ordinary differential geometry resulting in a new
mathematical branch which by now is justly called Connes' Noncommutative
Geometry \cite{AC}, \cite{GVF}, a whole variety of 
attempts emerged dealing with the
application of this promising new tool to fundamental interactions in 
particle physics. Loosely speaking, the common aim of all these attempts
consists of an alternative derivation of the classical action of, for instance,
the standard model of electroweak interactions within the generalized
noncommutative framework thereby trying to overcome
some of the unpleasant features of the ordinary formulation like the
unsettled origin of the Higgs' mechanism and the Yukawa couplings, the
ad-hoc adjustment of quantum numbers, the large number of free parameters
and so on. In general these various attempts may be classified according to
the extent to which they actually incorporate the original mathematical
apparatus of Connes' NCG. On the one hand, there is, of course, the model 
building initiated by Connes and Lott \cite{CL}, \cite{KSch}, \cite{CFF} 
which is directly based on Connes' NCG. On the other hand, there are also
various attempts inspired by elements of Connes' NCG but relying on alternative
approaches to what a noncommutative generalization of ordinary differential
geometry might be. As some prominent examples of this second class let us
mention the models of \cite{DKM}, \cite{MO} and \cite{BGW}.\\[.3cm]
For our purposes it completely suffices to briefly outline the basic
building blocks of the so-called Marseille-Mainz or $su(2|1)$ model which
falls into the second class of models indicated above. Furthermore we
will concentrate on the gauge sector of the model for electroweak interactions;
details concerning the construction as well as the fermionic sector of the
model can be found in \cite{CEV}, \cite{CES}, \cite{HPS1}, \cite{CHPS},
\cite{HPS2}, \cite{CHS} and \cite{H}.\\ 
The construction of the $su(2|1)$ model starts off from the fundamental
observation, very well established by experiment, that electroweak 
interactions strictly distinguish between left- and right-chiral fermionic
fields. Hence, the vector space $C$ of spinor fields carries a $\Z_2$-grading,
$C = C^{(L)} \oplus C^{(R)}$, the grading automorphism being given by 
$\gamma_5$. Clearly, this $\Z_2$-grading is consistent with the 
$\Z_2$-grading of the Clifford algebra $\Gamma$ acting on $C$. Due to the
observation mentioned above for $\psi \in C$ the chiral components $\psi_L$
and $\psi_R$ live in different representations $X^{(L)}$ and
$X^{(R)}$, respectively, of weak isospin $SU(2)_L$ and weak hypercharge 
$U(1)_Y$. As a consequence, the generators $T_k$, $k = 1, 2, 3, 8$, of
$SU(2)_L \times U(1)_Y$ have a block structure of square matrices along
the main diagonal,
\be \label{blockgen}
  T_k = \left( \begin{array}{cc}
  [T_k]_L & 0 \\ 0 & [T_k]_R \end{array} \right) \; \; \; ,
\ee
and act on a space $V = V^{(L)} \oplus V^{(R)}$ with $V^{(L,R)} =
C^{(L,R)} \otimes X^{(L,R)}$ which inherits the $\Z_2$-grading from $C$.
So far everything is standard. But what happens if one replaces the
rectangular zeros in the off-diagonal of (\ref{blockgen}) by something
nontrivial? This leads in a natural way to the notion of the super Lie
algebra $su(2|1)$ which is defined to be the set of all $3 \times 3$
antihermitean matrices
\be \label{defsuper}
  M = \left( \begin{array}{c|c}
  A_{2 \times 2} & C_{2 \times 1} \\ \hline
  D_{1 \times 2} & B_{1 \times 1} \end{array} \right)_{3 \times 3} = \;
  M^{(0)} + M^{(1)} \; \; , \; \; M^\dagger = - M \; \; ,
\ee
with vanishing supertrace: Str$M$ = Tr$A$ $-$ $B$ $= 0$. In (\ref{defsuper})
$M$ is decomposed into an {\it even} part $M^{(0)}$ (containing $A$
and $B$) and an {\it odd} part $M^{(1)}$ (containing $C$ and $D$). Thus,
at this stage another $\Z_2$-grading, the $\Z_2$-grading of block
matrices, enters the construction. In order to complete the definition
of $su(2|1)$ we still need a (generalized) Lie bracket which is given
by a combination of ordinary commutator $[\cdot, \cdot ]$ and anticommutator
$\{ \cdot, \cdot \}$:
\be \label{bracksup}
  [M,N]_g = [M^{(0)}, N^{(0)}] + [M^{(0)}, N^{(1)}] +
            [M^{(1)}, N^{(0)}] + i \; \{ M^{(1)}, N^{(1)} \}
\ee
Now, in close analogy to conventional Yang-Mills theory, the gauge potential
(connection) \nolinebreak $\cal A$ of the $su(2|1)$ model will 
be a {\it super Lie algebra valued generalized
one-form}. Having, however, two gradings within the game, namely the 
$\Z_2$-grading of block matrices and the $\Z$-grading of exterior forms
$\omega \in \Lambda^\star (M)$ (which coincides with the $\Z$-grading of
the Clifford algebra $\Gamma$ due to the vector space isomorphism between
$\Gamma$ and $\Lambda^\star (M)$) we first of all have to merge these two
gradings in order to make sense out of what we mean by a generalized
{\it p-form}. This is achieved by introducing the total grade
$\hat{\partial}$ as the sum of the matrix grade and the exterior form grade:
$\hat{\partial} (M \otimes \omega) = \partial M + \partial \omega$.
Let $\Omega_+, \Omega_-, \Omega'_+ = (\Omega_-)^\dagger$ and
$\Omega'_- = (\Omega_+)^\dagger$ denote the four generators of the odd part
of $su(2|1)$. Then the super gauge potential $\cal A$ can be written 
as\footnote{$I_1, I_2, I_3$ and $Y$ are proportional to $T_1, T_2, T_3$ and
$T_8$, respectively, and, thus, span the even part of $su(2|1)$ as well.}
\be \label{superpot}
  {\cal A} = i \left( \vec{I} \cdot \vec{W} + \frac{Y}{2} W^{(8)} \right)
  + \frac{i}{\mu} \left( \Phi^{(0)} \Omega'_- + \Phi^{(+)} \Omega'_+
  + h.c. \right) \; \; \; .
\ee
In (\ref{superpot}) $\vec{W} = (W^{(1)}, W^{(2)}, W^{(3)})$ are the
$SU(2)$ gauge fields, $W^{(8)}$ is the gauge field associated with $U(1)$
and $\mu$ is a mass scale. Because $\cal A$ has total grade $1$, $\Phi^{(0)}$
and $\Phi^{(+)}$ (and their conjugates) have to be scalar fields (zero-forms).
As the forthgoing construction of the model shows, $\Phi^{(0)}$ and
$\Phi^{(+)}$ are the physical Higgs fields, i.e. the already shifted 
Higgs fields located at one of the minima of the Higgs potential.
Please note that in the present context the Higgs fields become part of
the connection itself.\\
Next, we turn to the field strength (curvature) $\cal F$. In usual
Yang-Mills theories the connection between the field strength and the gauge
potential is (locally) given by Cartan's structure equation, thus
including Cartan's exterior derivative $d_C$ which raises the exterior
form grade by $1$ when acting on $\omega \in \Lambda^k (M)$. Having,
however, merged the exterior form grade and the matrix grade, see above, it
is quite reasonable and natural to enlarge also $d_C$ by a new operation
$d_M$ acting on matrices and raising 
the matrix grade by $1$ (mod~$2$):\linebreak
$d_M$ : $M^{(0),(1)} \rightarrow M^{(1),(0)}$. The {\it matrix derivative}
$d_M$ is combined with the exterior derivative 
\nolinebreak $d_C$ to yield the
generalized exterior derivative $d$, and this is done in such a way 
that the gradings involved are respected:
\ba \label{superd}
  d (M \otimes \omega) & = & d_M M \otimes \omega + (-1)^{\partial M}
                             M \otimes d_C \omega \nonumber \\
  \mbox{with } \; d_M M & = & [\eta, M]_g \; = \;
  [\eta, M^{(0)}] + i \; \{ \eta, M^{(1)} \}
\ea
In (\ref{superd}) $\eta$ has to be an odd element of $su(2|1)$ which without
loss of generality may be chosen to be $\eta = i (\Omega_+ + \Omega'_-)$.
The field strength of the Marseille-Mainz model is then given by generalizing
the ordinary structure equation, viz.
\be \label{supercur}
{\cal F} = d {\cal A} + \frac{1}{2} [\cA, \cA]_g
\ee
Finally, the bosonic Lagrangian of the model is obtained by taking the trace
of the scalar product $\langle \cF^\dagger \cF \rangle$,
\be \label{superbos}
  \cL_{bos} \propto \mbox{ Tr } \langle \cF^\dagger \cF \rangle \; \; \; ,
\ee
where $\langle \cdot \cdot \rangle$ denotes the scalar product in the space
of exterior forms.\\
Let us conclude the present section by summarizing some of the advantages
of the alternative approach sketched above when compared to the usual 
formulation of the electroweak standard model:
\begin{itemize}
\item $\cL_{bos}$ (\ref{superbos}) exactly coincides with the bosonic part
      of the standard model {\it including} the Higgs sector.
However, $\cL_{bos}$ is already given
in the spontaneously broken phase. In particular, the Higgs mechanism is
a built-in feature of the Marseille-Mainz model, and SSB occurs automatically.
The correct shape of the Higgs potential is predicted.
\item Furthermore, the mechanism of SSB gets a geometrical interpretation
      due to the existence of an {\it absolute} element that is invariant
under all rigid $su(2|1)$ transformations, see \cite{CHPS}.
\item The Higgs fields have to form a {\it doublet} with respect to $SU(2)$ 
      because of the minimal extension of $su(2) \times u(1)$ to the super 
      Lie algebra $su(2|1)$, see \cite{HPS2}.
\item The fermionic sector of the $su(2|1)$ model is constructed by using
      the generalized gauged Dirac operator including the super gauge
potential $\cA$, see \cite{CES}. Yukawa couplings occur automatically. 
Due to the existence of a special type of representations of super 
Lie algebras, see e.g. \cite{MS}, \cite{MM}, namely the so-called
reducible but indecomposible representations, there is a natural framework
for generation mixing (both for quarks and neutrinos). This aspect has been
worked out in detail in \cite{HS1}, \cite{HS2}, \cite{HPaS} and \cite{HTh}.
\item The quantum numbers of the fields involved follow directly from and
      are fixed by a natural choice of $su(2|1)$ representations.
\end{itemize}

\newsubsection{Zero-dimensional noncommutative model of SSB and
               motivation for the matrix calculus}

For the sake of simplicity, in the following we will restrict our attention
to the most elementary but, nevertheless, nontrivial noncommutative
gauge model one possibly could think of. That is to say, the model in
question will be obtained from the general noncommutative matrix formalism
of the previous section by shrinking all matrices to the case of
$2 \times 2$ matrices, by setting all the ordinary gauge fields, which
live in the even part of the superconnection $\cal A$, equal to zero
and, further on, by narrowing the support of the scalar fields in the
odd part of $\cal A$, i.e. the support of the Higgs fields, to just
one point. Accordingly, from now on the superpotential $\cal A$ that,
hence, is to be an antihermitean $2 \times 2$ matrix with vanishing
diagonal, is in general
given by
\be \label{ncga}
{\cal A} = \frac{i}{v} \left( \begin{array}{c|c}
0 & x + i y \\ \hline x - i y & 0 \end{array} \right) \; \; \; ,
\ee
where $v$ is a constant (defined in (\ref{absmin})), and $x,y$ 
turn out to be by now the shifted,
i.e. physical, Higgs variables located at one of the minima of the
Mexican hat, see section~4.1 and also later. Analogously, the absolute
element $\eta$, which is responsible for the automatic occurence of SSB in
the noncommutative formulation, reads:
\be \label{ncgeta}
\eta = i \left( \begin{array}{c|c}
0 & 1 \\ \hline 1 & 0 \end{array} \right)
\ee
Already at this point we would like to 
mention that due to $\cal A$ being an
antihermitean $2 \times 2$ matrix with vanishing diagonal 
${\cal A}^2$ is proportional to
$\eins_2$. Of course, the same holds true for $\eta$: $\eta^2 = - \eins_2$.
This particular property of the $2 \times 2$ case will simplify some
considerations substantially later on.\\
Furthermore, in the spirit of what was presented in the previous section, 
in the current investigation we are not going to introduce neither a mass
term for the Goldstone variable \nolinebreak $y$ 
via a gauge fixing term including
an additional auxiliary ``field'' $B$ nor, as an immediate consequence
thereof, Faddeev-Popov ghosts $c$ and $\overline{c}$. Proceeding that way,
however, we have to be aware of the fact that higher orders of the quantized
theory will not be defined properly as was discussed at the end of section~3.1
at length. The only reason for this further 
restriction is due to the intention
not to burden the considerations and calculations to follow by complications
of a more technical nature which eventually could prohibit a deeper insight
into the principles of the matrix approach to be presented below. The
proper handling of the model will be postponed until later.\\
Following up the line of reasoning suggested in section~4.1, the supercurvature
$\cal F$ of our present noncommutative model is defined by generalizing
Cartan's structure equation of ordinary differential geometry thereby
also enlarging the exterior derivative $d_C$ by a matrix derivative $d_M$:
\ba \label{ncgcur}
{\cal F} & = & d_M {\cal A} \; + \; \frac{1}{2} \; [{\cal A},{\cal A}]_g
\; = \; i \left( \eta {\cal A} + {\cal A} \eta + {\cal A}^2 \right) \\[.5ex]
& = & - \; i \; \frac{2}{v} \left[
x + \frac{1}{2 v} (x^2 + y^2) \right] \; \cdot \; \eins_2 \nonumber
\ea
Because $\cal A$ lives on just one point, in (\ref{ncgcur}) no ordinary
exterior derivative $d_C$ may appear. The last equality almost immediately
results upon inserting the explicit expressions for $\cal A$ (\ref{ncga})
and $\eta$ (\ref{ncgeta}).\\
Finally, the classical action $S$ in the noncommutative framework 
is given by:
\be \label{ncgaccl}
S \; = \; - \; \frac{\lambda v^4}{32} \mbox{ Tr } {\cal F}^2 \; = \;
\frac{\lambda v^4}{32} \mbox{ Tr} \left\{
2 \eta \cA \eta \cA + 4 \eta \cA^3 + \cA^4 - 2 \cA \right\}
\ee
It is a very easy matter to check that the noncommutative action $S$
(\ref{ncgaccl}), when expressed in terms of $x$ and $y$, exactly coincides
with $S(x,y)$ (\ref{acssb0})\footnote{In fact, the constant of
proportionality in front of the trace was chosen in such a way that
this statement strictly holds true. In turn, the coincidence of
both actions proves the identity of $x$ and $y$ as the already shifted,
i.e physical, variables.}. According to this 
observation and also due to (\ref{ncgaccl}),
$S$ can be regarded as a function of the generalized field strength 
\nolinebreak $\cal F$, 
or as a function of the superpotential $\cA$ or as a function of the
component fields $x$ and $y$, depending on the 
level of evaluation, so to speak.
In any case, (\ref{ncgaccl}) is the starting point for everything
to follow in this chapter.\\[.3cm]
The next step again consists in writing down the generating functional $Z$
of general Green's functions by coupling $x$ and $y$ to external sources
$\rho$ and $\tau$, respectively. However, having merged the ``fields''
$x$ and $y$ into a single object $\cA$ it certainly makes sense to
proceed in the same way with the sources $\rho$ and $\tau$, thus defining
\be \label{ncgq}
J = - \; i \; \frac{v}{2} \left( \begin{array}{c|c}
0 & \rho + i \tau \\ \hline \rho - i \tau & 0 \end{array} \right) \; \; \; .
\ee
In doing so the source term $\rho x + \tau y$ appearing in the exponent
of the integrand in the integral for $Z$ can be expressed 
as Tr$J \cA$, and, hence, we have:
\be \label{ncgzdef}
Z(J) = {\cal N} \int d x \; d y \mbox{ exp} \left\{
\frac{1}{\hbar} \left( - S(\cA; \eta ) + \mbox{Tr} J \cA \right) \right\}
\ee
In principle, we now could go ahead as in section~3.1 by expressing the
integrand in (\ref{ncgzdef}) in terms of $x$ and $y$ and by subsequently
deriving the system of differential equations for $Z$ in the by now 
well-tried manner. But certainly this should not be our goal! 
Instead of working
out separately the derivatives of (\ref{ncgzdef}) with respect to $x$ and
$y$, we rather should take into account the unification of $x$ and $y$ into
the superpotential $\cA$ which is the basic object of the noncommutative
formulation. Bearing this natural requirement in mind we, hence, should look
out for something like a derivative with repect to the 
matrix \nolinebreak $\cA$ itself
(and - at the level of sources - for a derivative with respect to the
matrix $J$). If
such an operation is available the advantages are obvious: Instead of
handling a system of two coupled partial differential equations for $Z$,
only one (matrix-valued) differential equation containing, nevertheless,
the whole information will have to be taken under inspection. But, of
course, before speculating anymore along these lines we first of all 
have to investigate how 
such a derivative of scalar or matrix-valued expressions 
with respect to a matrix (having, in addition, nice and desirable 
properties) can be defined.
This will be the subject of the next section, which has the character
of a more mathematical interlude.

\newsubsection{The matrix calculus}

Let us begin with some more motivation. In deriving the differential
equations for $Z$ we frequently used the following manipulation which 
in zero dimensions is mathematically rigorous: Each factor $x$ and each
factor $y$ appearing, for instance, in the integral representation for
$\frac{\partial Z}{\partial x}$ in front of the exponential map could be
replaced by $\frac{\hbar \partial}{\partial \rho}$ and 
$\frac{\hbar \partial}{\partial \tau}$, respectively. These partial
derivatives were afterwards taken out of the integral without any
problems, see also (\ref{ident}). If we want to copy somehow this trick
in the present matrix formulation we, hence, have to impose the requirement
that $\frac{\partial}{\partial J}$ Tr $J \cA$ should be defined in 
such a way that the result is just the odd $2 \times 2$ matrix $\cA$ itself.
This immediately suggests the definition
\be \label{ncgdiffq}
\frac{\partial}{\partial J} := \frac{i}{v} \left( \begin{array}{c|c}
0 & \partial_{\rho} + i \partial_{\tau} \\ \hline
\partial_{\rho} - i \partial_{\tau} & 0 \end{array} \right) \; \; \; .
\ee
Actually, it is easily proven that the above requirement fixes not only
the arrangement of the partial derivatives $\partial_{\rho}$ and
$\partial_{\tau}$ in $\frac{\partial}{\partial J}$ but also all factors
{\it uniquely}.\\
In an almost analogous manner we also define
\be \label{ncgdiffa}
\frac{\partial}{\partial \cA} := - \; i \; \frac{v}{2} \left(
\begin{array}{c|c} 0 & \partial_x + i \partial_y \\ \hline
\partial_x - i \partial_y & 0 \end{array} \right) \; \; \; .
\ee
Again, $\frac{\partial}{\partial \cA}$ is uniquely determined by the
requirement $\frac{\partial}{\partial \cA}$ Tr $J \cA$ $=$ $J$ or,
equivalently, by the requirement $\frac{\partial}{\partial \cA} \cA =
\eins_2$. It should be noticed that the action of $\frac{\partial}{\partial
\cA}$ (or $\frac{\partial}{\partial J}$) on a matrix-valued quantity includes
matrix multiplication whereas the action of $\frac{\partial}{\partial \cA}$
(or $\frac{\partial}{\partial J}$) on a scalar expression is defined in an
obvious way in order to yield a matrix-valued quantity of odd degree.\\[.3cm]
What is in need next? When performing calculations involving the matrix
derivatives defined above we certainly do not want to (and, even stronger,
should not) recede every time to the component formulation in terms of $x$
and $y$ (or $\rho$ and $\tau$) but instead should stay within
the language of matrices. Thus, what we have to look for are some rules
for the matrix derivatives $\frac{\partial}{\partial \cA}$ and
$\frac{\partial}{\partial J}$ which allow us to write down {\it directly}
the results of the action of these operators on those scalar
and matrix-valued expressions that we will meet frequently in the upcoming
calculations. For this reason, in the following we are going to summarize
the rules. The proofs of the rules displayed below are
moved to appendix~G in order not to burden the main text unnecessarily.
Just as a remark, however, let us mention that all proofs are
(directly or indirectly by means of other rules) based on the decomposition
\be \label{ncgdiffade}
\frac{\partial}{\partial \cA} = - \; \frac{v}{2} \left(
\eta \; \partial_x \; + \; i \; \gamma \eta \; \partial_y \right)
\ee
of $\frac{\partial}{\partial \cA}$ (and, of course, an analogous one for
$\frac{\partial}{\partial J}$) and the fact that $\partial_x$ and
$\partial_y$ are ordinary partial derivatives. In (\ref{ncgdiffade})
$\gamma$ denotes the grading automorphism of $2 \times 2$ matrices,
\be \label{ncggrad}
\gamma = \left( \begin{array}{c|c}
1 & 0 \\ \hline 0 & - 1 \end{array} \right) \; \; \; ,
\ee
meaning that a $2 \times 2$ matrix $M$ is even iff $[\gamma, M] = 0$ and
is odd iff $\{ \gamma, M \} = 0$.\\
In the list of rules given below we restrict ourselves primarily to the
action of the operator \nolinebreak $\frac{\partial}{\partial \cA}$; 
similar rules for
$\frac{\partial}{\partial J}$ are easily obtained. Every rule will be
followed by one or several example(s) for illustrative purposes.
\begin{description}
\item[{\bf Rule 1: }] 
Let $F(\cA; \eta)$ be any even polynomial in
$\cA$ and $\eta$. Then $\frac{\partial}{\partial \cA}$ Tr $F(\cA; \eta)$
has to be a matrix-valued odd quantity which is given by the following
recipe:\\
Permute successively every $\cA$ in $F(\cA; \eta)$ to the
leftmost position (exploiting
the cyclic property of the trace) and drop it. The sought result is then
the sum of these operations without trace.
\end{description}
As an example let us apply rule~1 to Tr $\eta \cA^3$; we find ($\hat{\cA}$ 
means that $\cA$ has to be dropped):
\begin{displaymath}
\frac{\partial}{\partial \cA} \mbox{ Tr } \eta \cA^3 =
\hat{\cA} \cA \cA \eta + \hat{\cA} \cA \eta \cA + \hat{\cA} \eta \cA \cA =
\cA^2 \eta + \cA \eta \cA + \eta \cA^2 = 2 \cA^2 \eta + \cA \eta \cA
\end{displaymath}
For the last equality we also used $\cA^2 \propto \eins_2$. A further
example which will be of importance in the next section deals with the
derivative of the action $S(\cA; \eta)$ (\ref{ncgaccl}) with respect 
to~$\cA$:
\be \label{ncgsa}
\frac{\partial S}{\partial \cA} = \frac{\lambda v^4}{8} \left(
\eta \cA \eta + \cA^2 \eta + \cA \eta \cA + \eta \cA^2 + \cA^3 - \cA \right)
\ee
\par
\vspace*{.8ex}
\begin{description}
\item[{\bf Rule 2: }] 
Let $M(\cA; \eta)$ be any ({\it not} necessarily even) monomial
in $\cA$ and $\eta$. Then $\frac{\partial}{\partial \cA} M(\cA; \eta)$ is 
given by $(-$ $\eta)$ times a sum of monomials which result from $M(\cA; \eta)$
by successively replacing each $\cA$ that has an even number of (odd) matrices
in front of it by~$\eta$.\\
Due to the linearity of $\frac{\partial}{\partial \cA}$ this rule is
extended straightaway to any polynomial $F(\cA; \eta)$.
\end{description}
Rule~2 sounds complicated but in effect it is not. Look, for instance, at
the derivative of \nolinebreak $\cA^3$ with respect to $\cA$:
\begin{displaymath}
\frac{\partial}{\partial \cA} \cA^3 = - \eta \left(
[\cA \rightarrow \eta] \cA \cA + \cA \cA [\cA \rightarrow \eta] \right) =
- \eta (\eta \cA^2 + \cA^2 \eta) = 2 \cA^2
\end{displaymath}
On the r.h.s. of the first equality only two terms appear in the parentheses:
The second \nolinebreak $\cA$ in $\cA^3$ is not substituted by $\eta$ 
because there is {\it one}
$\cA$ in front of the second $\cA$. For the last equality $\cA^2 \propto
\eins_2$ and $\eta^2 = - \eins_2$ have been taken into 
account\footnote{This example also shows that the matrix derivative in the
way it is defined in some cases differs from what one perhaps would
expect naively ($\frac{\partial}{\partial \cA} \cA^3 \neq 3 \cA^2$).
Let us, however, stress once again that the above definition of
$\frac{\partial}{\partial \cA}$ was guided by the intention to fit
exactly the purposes of the present context. But with this natural
requirement $\frac{\partial}{\partial \cA}$ is completely fixed and there
is no chance to cure the ostensible mismatch with regard to the coefficients
in some derivatives. In any case, the mismatch just mentioned is not a 
real mismatch at all because the only thing that actually counts is to
give a well-defined meaning to $\frac{\partial}{\partial \cA}$ which
certainly has been done.}. By means of
rule~2 we also can differentiate (\ref{ncgsa}) once more with respect to
$\cA$ obtaining
\be \label{ncgsaa}
\frac{\partial^2 S}{\partial \cA^2} = \frac{\lambda v^4}{8} \left(
2 \cA \eta + 2 \eta \cA + 2 \cA^2 - \eins_2 \right) \; \; \; ,
\ee
a result which will be of relevance later on, too.
\begin{description}
\item[{\bf Rule 3: }] {\bf product rule: }
Let $F(\cA; \eta)$ and $G(\cA; \eta)$ be any polynomials in $\cA$ and $\eta$.
Then the following product rules depending on the matrix 
degree of $F$ hold true:
\ba \label{ncgprod}
\frac{\partial}{\partial \cA} (F \cdot G) & = & \left(
\frac{\partial}{\partial \cA} F \right) G - \eta F \eta
\frac{\partial}{\partial \cA} G \mbox{ , if } F 
\mbox{ is even} \nonumber \\
\frac{\partial}{\partial \cA} (F \cdot G) & = & \left(
\frac{\partial}{\partial \cA} F \right) G - \eta F
\frac{\partial}{\partial \cA} \eta G \mbox{ , if } F 
\mbox{ is odd}
\ea
\end{description}
\vspace*{.3cm}
\noindent
Let us again illustrate this rule by means of an example: Choose
$F = \cA$ (i.e. $F$ is odd and, hence, we have to use the second equation
in rule~3) and $G = \cA^2$. Because of $\frac{\partial}{\partial \cA} F =
\frac{\partial}{\partial \cA} \cA = \eins_2$ and $\frac{\partial}{\partial
\cA} \eta G = \frac{\partial}{\partial \cA} \eta \cA^2 = \cA \eta$ we get:
\begin{displaymath}
\frac{\partial}{\partial \cA} (\cA \cdot \cA^2) = \eins_2 \cdot \cA^2 -
\eta \cdot \cA \cdot \cA \eta = \cA^2 - \eta \cA^2 \eta = 2 \cA^2
\end{displaymath}
This result, of course, coincides with the direct calculation of
$\frac{\partial}{\partial \cA} \cA^3$, see above. Further applications
of the product rule~3 can be found in the next section.\\[.5ex]
Furthermore, due to the canonical action of $\frac{\partial}{\partial \cA}$
on scalar expressions it is almost evident that acting with
$\frac{\partial}{\partial \cA}$ on a product of two scalar functions
$f(\cA; \eta)$ and $g(\cA; \eta)$ the usual product rule holds true:
\be \label{ncguprod}
\frac{\partial}{\partial \cA} (f \cdot g) = \left(
\frac{\partial}{\partial \cA} f \right) g + f
\frac{\partial}{\partial \cA} g
\ee
\par
\vspace*{.4ex}
\begin{description}
\item[{\bf Rule 4: }] {\bf chain rule: }
Let $F(J; \eta )$ be any polynomial in $J$ and $\eta$ and let $J(\cA; \eta )$
be an odd polynomial in $\cA$ and $\eta$. Then the following chain rule is
valid:
\be \label{ncgchain}
\frac{\partial}{\partial \cA} F(J(\cA; \eta ); \eta ) =
\frac{\partial J}{\partial \cA} \left.
\frac{\partial F}{\partial J} \right|_{J = J(\cA; \eta )} - \;
\left( \frac{\partial}{\partial \cA} \eta J \right) \left. \left(
\frac{\partial}{\partial J} \eta F \right) \right|_{J = J(\cA; \eta)}
\ee
\end{description}
As an example let us take $F(J; \eta ) = \eta J$ and $J(\cA; \eta ) = \cA^3$.
$F$ as a function of $\cA$ (and $\eta$) is thus given by
$F(J(\cA; \eta); \eta) = \eta \cA^3$, and by means of rule~2 we obtain
$\frac{\partial F}{\partial \cA} = \cA \eta \cA$.\\ 
On the other hand, making 
use of $\frac{\partial J}{\partial \cA} = 2 \cA^2$, 
$\frac{\partial F}{\partial J} = 0$, $\frac{\partial}{\partial \cA}
\eta J = \cA \eta \cA$ and $\frac{\partial}{\partial J} \eta F = - \eins_2$
rule~4 tells us that
\begin{displaymath}
\frac{\partial F}{\partial \cA} = 2 \cA^2 \cdot 0 -
\cA \eta \cA \cdot (- \eins_2) \; \; \; ,
\end{displaymath}
which obviously coincides with the direct calculation.\\[.5ex]
Three remarks may be helpful here: First of all, for the chain rule to hold
true in the above form it is crucial that we have $\cA^2 \propto \eins_2$.
This remark will become important when one wants to look for generalizations
of the matrix calculus just developed to matrices of higher dimensions.
Secondly, the appearance of {\it two} contributions on the r.h.s. of
(\ref{ncgchain}) accounts for the noncommutative nature of matrix
multiplication. And, finally, it is no real restriction that $J(\cA; \eta)$
is to be an {\it odd} matrix, at least for our purposes. For details
about this last remark see also appendix~H.\\[.5ex]
We would like to conclude the present section by establishing a connection
between the matrix derivative $\frac{\partial}{\partial \cA}$ Tr 
$F(\cA; \eta)$ of the scalar quantity Tr $F(\cA; \eta)$ (where $F$ is any
even polynomial in $\cA$ and $\eta$) and the respective usual derivatives
of Tr $F(\cA; \eta)$ with respect to $x$ and $y$. This connection, on the
one hand, offers to us the possibility of various consistency checks 
concerning the validity of the matrix calculus and, on the other hand, will
be of relevance by itself later on. It is summarized in our last rule:
\begin{description}
\item[{\bf Rule 5: }]
Let $F(\cA; \eta )$ be any even polynomial in $\cA$ and $\eta$. Then we have
the following identities:
\ba \label{ncgaxy}
\partial_x \mbox{ Tr } F(\cA; \eta ) & = & \frac{1}{v} \mbox{ Tr }
\eta \; \frac{\partial}{\partial \cA} \mbox{ Tr } F(\cA; \eta ) 
\nonumber \\[.5ex]
\partial_y \mbox{ Tr } F(\cA; \eta ) & = & \frac{i}{v} \mbox{ Str }
\eta \; \frac{\partial}{\partial \cA} \mbox{ Tr } F(\cA; \eta )
\ea
(Str denotes, as usual, the supertrace, i.e. Str $\star$ = Tr $\gamma 
\; \star$.)
\end{description}
\par
\vspace*{.5ex}
\noindent
Choosing, for instance, $\frac{\partial S}{\partial \cA}$ (\ref{ncgsa})
for $\frac{\partial}{\partial \cA}$ Tr $F(\cA; \eta )$ in (\ref{ncgaxy}) 
the r.h.s. of the first equation then reads:
\begin{displaymath}
\frac{1}{v} \mbox{ Tr } \eta \; \frac{\lambda v^4}{8} \;
(\eta \cA \eta + 2 \eta \cA^2 + \cA \eta \cA + \cA^3 - \cA ) \; = \;
\frac{\lambda v^3}{8} \mbox{ Tr } \;
(\eta \cA^3 + \eta \cA \eta \cA - 2 \cA^2 - 2 \eta \cA )
\end{displaymath}
Upon inserting $\cA$ (\ref{ncga}) and $\eta$ (\ref{ncgeta}) this expression
is exactly $\partial_x S(x,y)$ with $S(x,y)$ according to (\ref{acssb0}).

\newsubsection{Differential equations for $Z$, $W$ and $\Gamma$}

After the mathematical preparations of the previous section we are now
able to mimick the treatments of the foregoing chapters without any further
delay. Hence, in order to derive the noncommutative version of the system
of differential equations for the generating functional $Z$ of general
Green's functions let us apply the matrix derivative $\frac{\partial}{\partial
\cA}$ to the integrand of $Z(J)$ (\ref{ncgzdef}) and consider:
\bas
0_2 & = &
{\cal N} \; \int d x \; d y \; \frac{\partial}{\partial \cA}
\mbox{ exp} \left\{ \frac{1}{\hbar} \left(
- S(\cA; \eta ) + \mbox{Tr} J \cA \right) \right\} \\[.5ex]
& = & {\cal N} \; \frac{1}{\hbar} \int d x \; d y \left(
- \frac{\partial S}{\partial \cA} + J \right) \mbox{exp}
\left\{ \frac{1}{\hbar} \left( - S(\cA; \eta ) + \mbox{Tr} J \cA \right)
\right\}
\eas
The last equality directly results from the ordinary chain rule which
naively holds true due to the canonical action of $\frac{\partial}{\partial
\cA}$ on scalar expressions. Inserting $\frac{\partial S}{\partial \cA}$
according to (\ref{ncgsa}) and making use of mathematically rigorous
manipulations of the type
\begin{displaymath}
\int d x \; d y \; \cA \mbox{ exp} \left\{
\frac{1}{\hbar} \left( - S(\cA; \eta ) + \mbox{Tr} J \cA \right) \right\} =
\frac{\hbar \partial}{\partial J} \int d x \; d y \mbox{ exp}
\left\{ \frac{1}{\hbar} \left( - S(\cA; \eta ) + \mbox{Tr} J \cA \right)
\right\}
\end{displaymath}
several times, we thus obtain (also exploiting the fact that
$\frac{\partial^2}{\partial J^2}$ is given by $\eins_2$ times an 
ordinary differential operator):
\be \label{ncgdiffz}
\frac{\lambda v^4}{8} \left\{
\hbar^3 \; \frac{\partial^3}{\partial J^3} \; + \; 
\hbar^2 \left[ 2 \eta \frac{\partial^2}{\partial J^2} +
\frac{\partial}{\partial J} \eta \frac{\partial}{\partial J} \right] \; + \;
\hbar \left[ \eta \frac{\partial}{\partial J} \eta -
\frac{\partial}{\partial J} \right] \right\} Z = J Z
\ee
(\ref{ncgdiffz}) is the differential equation for $Z$ within the noncommutative
language which we are looking for and which resembles the ordinary 
Dyson-Schwinger equations in zero dimensions. Already at this point we
observe an essential advantage of the noncommutative formulation when compared
to the original treatment in terms of the component fields $x$ and $y$:
Instead of being forced to handle a complicated system of (two)
coupled partial differential equations (see (\ref{nssbdiffz})) 
in the present context there is only {\it one} differential equation 
which nevertheless contains the whole information\footnote{One proves
easily that the system (\ref{nssbdiffz}) is totally equivalent to
(\ref{ncgdiffz}). For that multiply (\ref{ncgdiffz}) with $\frac{1}{v} \eta$
(from the left) and take the trace; the result is the first equation
in (\ref{nssbdiffz}). In a similar manner also the second equation in
(\ref{nssbdiffz}) is obtained, namely by multiplication of (\ref{ncgdiffz})
with $\frac{i}{v} \eta$ and subsequent evaluation by means of the
supertrace.}.\\
Perhaps, one could object that this simplification is not at all a real one
because now we have to deal with a matrix-valued equation whereas formerly
there were just scalar equations. But such an objection can be invalidated
easily by the following argument: If we do not only regard (\ref{ncgdiffz}) to
be matrix-valued but also really work out the consequences thereof we
do nothing else but return to the treatment in terms of the components
$x$ and $y$ which, however, is a treatment that we wish to avoid 
strictly in the noncommutative
approach. On the other hand, the rules of the previous section (which are
somehow odd when compared to the usual ones concerning differentiation but
which are nonetheless well-defined) allow us to stay completely within the
formulation that uses just one variable, namely the superpotential $\cA$.
Taking this point of view the fact that $\cA$ is a 
matrix does not play a role.
At any rate, regarding $\cA$ as {\it the} variable of the game fits much more
the intention to unify $x$ and $y$ into a {\it single} 
noncommutative object.\\[.3cm]
Next we are going to convert the differential equation (\ref{ncgdiffz}) for $Z$
into such an equation for the generating functional $W$ of connected
Green's functions according to
\be \label{ncgzw}
Z(J) = \mbox{exp} \left( \frac{1}{\hbar} \; W(J) \right) \; \; \; .
\ee
To this end we have to translate the various (multiple) derivatives of $Z$
with respect to 
\nolinebreak $J$ into the respective derivatives of $W$ with respect to
$J$. Most of these derivatives are obtained easily (using at some places
the ordinary product rule), e.g.:
\be \label{ncgzwder}
\frac{\partial Z}{\partial J} = \frac{1}{\hbar}
\frac{\partial W}{\partial J} Z \; \; , \; \;
\frac{\partial}{\partial J} \eta Z = \frac{1}{\hbar}
\frac{\partial W}{\partial J} \eta Z \; \; , \; \;
\frac{\partial^2 Z}{\partial J^2} = \left[
\frac{1}{\hbar} \frac{\partial^2 W}{\partial J^2} +
\frac{1}{\hbar^2} \left( \frac{\partial W}{\partial J} \right)^2 \right] Z
\ee
There is only one derivative, namely $\frac{\partial^3 Z}{\partial J^3}$,
the conversion of which is a little more involved:
\bas
\frac{\partial^3 Z}{\partial J^3} & = &
\frac{\partial}{\partial J} \frac{\partial^2 Z}{\partial J^2} \; = \;
\frac{\partial}{\partial J} \left[
\frac{1}{\hbar} \frac{\partial^2 W}{\partial J^2} +
\frac{1}{\hbar^2} \left( \frac{\partial W}{\partial J} \right)^2 \right] Z \\
& = & \left\{ \frac{1}{\hbar} \frac{\partial^3 W}{\partial J^3} +
\frac{1}{\hbar^2} \frac{\partial W}{\partial J}
\frac{\partial^2 W}{\partial J^2} +
\frac{1}{\hbar^2} \left[
\frac{\partial}{\partial J} \left(
\frac{\partial W}{\partial J} \right)^2 \right] +
\frac{1}{\hbar^3} \left( \frac{\partial W}{\partial J} \right)^3 \right\} Z
\eas
Hence, it remains to determine $\frac{\partial}{\partial J}
\left( \frac{\partial W}{\partial J} \right)^2$. This determination is another
example for the application of the product rule~3 within the matrix calculus:
Because $W$ is a scalar $\frac{\partial W}{\partial J}$ has to be an odd
matrix and we thus must use the second equation in (\ref{ncgprod}) yielding
\be \label{ncgzwder2}
\frac{\partial}{\partial J} \left(
\frac{\partial W}{\partial J} \right)^2 =
\frac{\partial^2 W}{\partial J^2} \frac{\partial W}{\partial J} -
\eta \frac{\partial W}{\partial J} \frac{\partial}{\partial J} \eta
\frac{\partial}{\partial J} W \; \; \; .
\ee
Collecting everything we end up with the desired differential
equation for $W$:
\ba \label{ncgdiffw}
\frac{\lambda v^4}{8} \left\{
\hbar^2 \; \frac{\partial^3 W}{\partial J^3} \; + \; \hbar \left[
2 \left( \frac{\partial W}{\partial J} + \eta \right)
\frac{\partial^2 W}{\partial J^2} -
\eta \left( \frac{\partial W}{\partial J} + \eta \right)
\frac{\partial}{\partial J} \eta \frac{\partial W}{\partial J}
\right] \right. & & \\[.5ex]
+ \left. \eta \frac{\partial W}{\partial J} \eta +
2 \eta \left( \frac{\partial W}{\partial J} \right)^2 +
\frac{\partial W}{\partial J} \eta \frac{\partial W}{\partial J} +
\left( \frac{\partial W}{\partial J} \right)^3 -
\frac{\partial W}{\partial J} \right\} & = & J \nonumber
\ea
Finally, let us perform a Legendre transformation, this time defined by
\ba \label{ncgleg}
& \Gamma (\cA; \eta ) \; + \; W(J(\cA; \eta ); \eta ) \; - \mbox{ Tr }
J(\cA; \eta ) \cA \; = \; 0 & \\[.5ex]
& \mbox{with } \; \cA := \displaystyle\frac{\partial W}{\partial J} 
\; \mbox{ and } \; J = \frac{\partial \Gamma}{\partial \cA}
\; \; \; , & \nonumber
\ea
in order to go over from $W$ to the generating functional $\Gamma$ of 1~PI
Green's functions. For the translation of (\ref{ncgdiffw}) into a differential
equation for $\Gamma$ we obviously need the Legendre transformations of the
various second and third derivatives of $W$ with respect to $J$ appearing
in (\ref{ncgdiffw}). In analogy to previous investigations of this kind the 
starting point for the determination of these Legendre transformations 
is the identity
\be \label{ncgident}
\cA (J(\cA; \eta ); \eta ) = \cA \; \; \; ,
\ee
which we are going to differentiate with respect to $\cA$. Taking into
account that $J(\cA; \eta )$ has to be an odd matrix (see 
second line of (\ref{ncgleg}))
we can directly apply the chain rule~4 of section~4.3 with the result:
\be \label{ncgleg21}
\eins_2 = \frac{\partial J}{\partial \cA} \frac{\partial \cA}{\partial J}
- \left( \frac{\partial}{\partial \cA} \eta J \right) \left(
\frac{\partial}{\partial J} \eta \cA \right) =
\frac{\partial^2 \Gamma}{\partial \cA^2}
\frac{\partial^2 W}{\partial J^2} - \left(
\frac{\partial}{\partial \cA} \eta \frac{\partial \Gamma}{\partial \cA} \right)
\frac{\partial}{\partial J} \eta \frac{\partial W}{\partial J}
\ee
It is clear that we have to look for a further independent equation 
of this type in order to obtain both second derivatives 
$\frac{\partial^2 W}{\partial J^2}$ and
$\frac{\partial}{\partial J} \eta \frac{\partial W}{\partial J}$ as functions
of $\cA$ and $\eta$. This second equation is deduced in a similar manner by
first multiplying (\ref{ncgident}) from the left by $\eta$ and by subsequently
differentiating the resulting equation with respect to $\cA$
($\frac{\partial}{\partial \cA} \eta \cA = 0$ by means of rule~2):
\be \label{ncgleg22}
0_2 = \frac{\partial J}{\partial \cA}
\frac{\partial}{\partial J} \eta \cA + \left(
\frac{\partial}{\partial \cA} \eta J \right)
\frac{\partial \cA}{\partial J} =
\frac{\partial^2 \Gamma}{\partial \cA^2} 
\frac{\partial}{\partial J} \eta \frac{\partial W}{\partial J} +
\left( \frac{\partial}{\partial \cA} \eta
\frac{\partial \Gamma}{\partial \cA} \right) 
\frac{\partial^2 W}{\partial J^2}
\ee
The solution of the two (matrix-valued) equations (\ref{ncgleg21}) and
(\ref{ncgleg22}) has to be handled with some care because the single factors
in the various terms on the right hand sides do not commute in general:
They are matrices! However, some of them do commute due to
$\frac{\partial^2}{\partial \cA^2}$ and $\frac{\partial^2}{\partial J^2}$ 
being proportional to the unit matrix $\eins_2$, a fact which simplifies 
identifying the solution to quite some extent. We finally get:
\ba \label{ncgleg2}
\frac{\partial^2 W}{\partial J^2} & = &
\frac{\partial^2 \Gamma}{\partial \cA^2} \left[
\left( \frac{\partial^2 \Gamma}{\partial \cA^2} \right)^2 +
\left( \frac{\partial}{\partial \cA} \eta 
\frac{\partial \Gamma}{\partial \cA} \right)^2 \right]^{- 1} \nonumber \\
\frac{\partial}{\partial J} \eta \frac{\partial W}{\partial J} & = & -
\left( \frac{\partial}{\partial \cA} \eta
\frac{\partial \Gamma}{\partial \cA} \right) \left[
\left( \frac{\partial^2 \Gamma}{\partial \cA^2} \right)^2 +
\left( \frac{\partial}{\partial \cA} \eta 
\frac{\partial \Gamma}{\partial \cA} \right)^2 \right]^{- 1}
\ea
In an analogous but technically more tedious way also the Legendre
transformation of the third derivative $\frac{\partial^3 W}{\partial J^3}$
that is still missing can be calculated. We postpone this calculation as well
as the corresponding result to appendix~H.\\[.3cm]
Hence, up to and including 1-loop order\footnote{The full differential
equation for $\Gamma$ valid to all orders of the loop expansion will be
given in appendix~H, too.} we can write down the following differential
equation for $\Gamma$:
\ba \label{ncgdiffg1}
\frac{\partial \Gamma}{\partial \cA} & = &
\eta \cA \eta + 2 \eta \cA^2 + \cA \eta \cA + \cA^3 - \cA \\
& = & + \; \hbar \left[
2 (\cA + \eta ) \frac{\partial^2 \Gamma}{\partial \cA^2} +
\eta (\cA + \eta ) \frac{\partial}{\partial \cA} \eta
\frac{\partial \Gamma}{\partial \cA} \right] \left[
\left( \frac{\partial^2 \Gamma}{\partial \cA^2} \right)^2 +
\left( \frac{\partial}{\partial \cA} \eta 
\frac{\partial \Gamma}{\partial \cA} \right)^2 \right]^{- 1}
+ \; {\cal O} (\hbar^2 ) \nonumber
\ea
This equation (or better the complete one, see (\ref{ncgdiffg})) now has to be
solved recursively order by order in $\hbar$. In zeroth order we obtain
\begin{displaymath}
\frac{\partial \Gamma^{(0)}}{\partial \cA} =
\eta \cA \eta + 2 \eta \cA^2 + \cA \eta \cA + \cA^3 - \cA \; \; \; ,
\end{displaymath}
which, as a matter of consistency, coincides with 
$\frac{\partial S}{\partial \cA}$ (\ref{ncgsa}). For the determination 
of the next order we evaluate (\ref{ncgdiffg1}) consistently in order
$\hbar^1$ and therefore have to take into
account $\frac{\partial^2 S}{\partial \cA^2}$ (\ref{ncgsaa}) and
\be \label{ncgsaetaa}
\frac{\partial}{\partial \cA} \eta \frac{\partial S}{\partial \cA} =
\frac{\lambda v^4}{8} \; (\cA \eta \cA - 2 \cA - \eta ) \; \; \; .
\ee
This last equation results from (\ref{ncgsa}) by means of rule~2. Having by now
all needed ingredi\-ents at our disposal a straightforward calculation
using at several places the definition of the field strength $\cal F$
(\ref{ncgcur}) leads to the answer for 
$\frac{\partial \Gamma^{(1)}}{\partial \cA}$ that we are looking for:
\be \label{ncgg1a}
\frac{\partial \Gamma^{(1)}}{\partial \cA} = \frac{1}{2} \left[
\frac{\partial}{\partial \cA} (2 i \cF - 3 \cF^2 ) \right]
\left( 2 i \cF - 3 \cF^2 \right)^{- 1}
\ee
(\ref{ncgg1a}) can be integrated very easily\footnote{In general, there is no
obvious and simple rule for the integration of a matrix-valued function
$F(\cA; \eta )$ with respect to $\cA$ resulting in a scalar quantity, at least
if $F(\cA; \eta )$ is some function which is more complicated than an
ordinary polynomial in $\cA$ and $\eta$. The problem of finding such a
rule for integration is mainly due to the observation that the kernel
of $\frac{\partial}{\partial \cA}$ is not only generated by constant
matrices but also by some non-constant ones like e.g. $\cA^{- 1}$ and
$\eta \cA$. For our purposes it is, however, completely sufficient to know
the rules for the special kind of integration which is studied in 
section~4.5.} due to the fact that $\cF$ is proportional to the unit
matrix $\eins_2$, see (\ref{ncgcur}) (and also rule~7 in the next
section for details about this special kind of integration), yielding
finally
\be \label{ncgg1}
\Gamma^{(1)} = \frac{1}{2} \mbox{ ln Tr }
(2 i \cF - 3 \cF^2 ) \; \; \; .
\ee
This result is quite noteworthy in two respects: On the one hand, 
(\ref{ncgg1}) represents $\Gamma^{(1)}$ in a much more compact form
than (\ref{nssbg1}). Both expressions, of course,
coincide (up to a constant of integration) as can be seen upon inserting
$\cF$ according to the second line in (\ref{ncgcur}) and evaluating the
trace. On the other hand, and even more remarkably, not only $\Gamma^{(0)} =
S$ but also $\Gamma^{(1)}$ is a simple function of the geometrical field
strength $\cF$ only. Both aspects just mentioned are due to and, even
stronger, are only accessible within the present noncommutative 
formulation.
Two questions arise naturally here: Can the higher orders in $\hbar$ of
$\Gamma$ be calculated along the same lines? And is it true, 
to all orders of the loop expansion,
that $\Gamma$ is a function of $\cF$ only? The following section will
answer the second question in the affirmative. Moreover, the demonstration
is approached in such a way that at the same time the higher orders of
$\Gamma$ are determined, too.

\newsubsection{Recursive determination of $\Gamma$}

The aim of this section is to prove the claim that $\Gamma$ is a function 
of the field strength \nolinebreak $\cF$ only, 
to all orders in $\hbar$. Before turning to
the technical details we first discuss the conceptual
cornerstones on which the proof is based.\\
First of all, the proof below will proceed by induction in the number $n$
of loops thereby vastly relying on the specific structure of the differential
equation (\ref{ncgdiffg}) for $\Gamma$ which means the following: Due to the
observation that every (second or third) derivative of $\Gamma$ with respect
to $\cA$ appearing on the r.h.s. of (\ref{ncgdiffg}) is multiplied by at
least one factor of $\hbar$ the l.h.s. of (\ref{ncgdiffg}) in a given order,
say $n + 1$, is completely prescribed by the lower orders of 
\nolinebreak $\Gamma$ up
to and including order $n$. Indeed, this observation is exactly the one
which always allows us to determine the solution of the differential equation
for $\Gamma$ in a {\it recursive} manner. The starting point for the induction
is $n = 0$. Obviously, $S$ given according to (\ref{ncgaccl}) is a function 
of the field strength $\cF$ only. Hence, in the \nopagebreak 
spirit of proofs based
on induction we may assume from now on that the same holds true for
$\Gamma^{(n)}$, too\footnote{Strictly speaking, (\ref{ncgpras}) does not hold
true for $n = 1$ because of the logarithm in front of the trace, see
(\ref{ncgg1}). So it possibly would be better and more precise to state
the assumption of the inductive proof in the form:
\begin{displaymath}
\Gamma^{(n)} = h^{(n)} (\mbox{Tr } g^{(n)} (\cF )) \; \; \; .
\end{displaymath}
We will, however, not do so for the following two reasons: On the one side,
the 1-loop approximation is rather exceptional and taking along also all
the functions $h^{(n)}$ (actually necessary in a proper and general treatment
although only $h^{(1)}$ is different from the identity map) would overburden
the forthcoming formulae. On the other side, we will only use the
implication (\ref{ncgpras2}) of (\ref{ncgpras}) when proving the statement
in question. And (\ref{ncgpras2}) does hold true also for $n = 1$!}:
\be \label{ncgpras}
\Gamma^{(n)} = \mbox{Tr } g^{(n)} (\cF )
\ee
The second building block needed for the proof rests on the fact that
$\cF = \cF (\cA; \eta )$ is proportional to the unit matrix $\eins_2$,
see (\ref{ncgcur}), a fact which, in turn, implies some very simple rules for
the differentiation of $g(\cF )$ and Tr $g( \cF )$ with respect to $\cA$
summarized in:
\begin{description}
\item[{\bf Rule~6: }] 
Let $\cF (\cA; \eta )$ be the field strength according to (\ref{ncgcur})
and let $g(\cF )$ be any function of $\cF$. Then we have:
\ba \label{ncgrule61}
\frac{\partial}{\partial \cA} \; g(\cF (\cA; \eta )) & = &
\frac{\partial \cF}{\partial \cA} \; \frac{\partial g}{\partial \cF} \; = \;
\frac{\partial g}{\partial \cF} \; \frac{\partial \cF}{\partial \cA} \\[.5ex]
\label{ncgrule62}
\frac{\partial}{\partial \cA} \mbox{ Tr } g(\cF (\cA; \eta )) & = &
2 \; \frac{\partial \cF}{\partial \cA} \; 
\frac{\partial g}{\partial \cF} \; = \;
2 \; \frac{\partial g}{\partial \cF} \; \frac{\partial \cF}{\partial \cA}
\ea
Here $\frac{\partial}{\partial \cF}$ has to be treated just like an ordinary
derivative (because of $\cF \propto \eins_2$).
\end{description}
The easy demonstration of rule~6 is postponed to the first part of
appendix~I.\\
Now, the application of rule (\ref{ncgrule62}) to (\ref{ncgpras}) leads to
\be \label{ncgpras2}
\frac{\partial \Gamma^{(n)}}{\partial \cA} \; = \; 2 \;
\frac{\partial g^{(n)}}{\partial \cF} \;
\frac{\partial \cF}{\partial \cA} \; \; \; .
\ee
It is exactly this form of the inductive assumption that will enter the
technical aspects of the proof in a moment.\\
Finally, let us also mention the third basic ingredient required below.
This last ingredient represents a very special kind of integration within
the matrix calculus and, in fact, is nothing else but the inversion of
(\ref{ncgrule62}):
\begin{description}
\item[{\bf Rule~7: }]
Let $\cF$ and $g$ be as in rule~6. Then it holds true that:
\be \label{ncgrule7}
2 \; \int d \cA \; \frac{\partial g}{\partial \cF} \;
\frac{\partial \cF}{\partial \cA} \; = \; \mbox{Tr }
g(\cF (\cA; \eta )) \; \; \; (+ \mbox{ constant})
\ee
\end{description}
In consideration of the three preparatory remarks just given it should be
clear along which lines we have to move on in order to successfully
complete the demonstration in question. Namely, if in order $n + 1$ we are
able to express the r.h.s. of (\ref{ncgdiffg}) as $f^{(n + 1)} (\cF ) 
\frac{\partial \cF}{\partial \cA}$ where $f^{(n + 1)}$ is an appropriate 
function of $\cF$ only, everything is proven. Of course, to this end we
have to make use of (\ref{ncgpras2}). To simplify matters we will go ahead
bit by bit and also omit the superscripts ``$(n)$'' which anyway can be
reintroduced easily at the end of the calculation.\\[.3cm]
As an illuminative example for the general procedure let us begin with the 
treatment of the term $\Gamma_{\cA \cA}^2 + \Gamma_{\cA \eta \cA}^2$ which
appears frequently on the r.h.s. of (\ref{ncgdiffg}). Setting
\be \label{ncgprabr}
f(\cF ) \; := \; 2 \; \frac{\partial g}{\partial \cF}
\ee
and differentiating (\ref{ncgpras2}) once more with respect to $\cA$
(using the product rule~3 as well as (\ref{ncgrule61})) we get
(with the abbreviatory notation $f' := \frac{\partial f}{\partial \cF}$,
$\cF_\cA := \frac{\partial \cF}{\partial \cA}$ and so on):
\be \label{ncgprgaapr}
\Gamma_{\cA \cA} \; = \; f' \; \cF_\cA^2 \; + \; f \; \cF_{\cA \cA}
\ee
Taking into account that according to (\ref{ncgcur}) and rule~2
\be \label{ncgprf}
\cF_\cA = i (\cA + \eta ) \; \; , \; \;
\cF_\cA^2 = - (\cA + \eta )^2 = \eins_2 + i \cF \; \mbox{ and } \;
\cF_{\cA \cA} = i \eins_2
\ee
we may rewrite (\ref{ncgprgaapr}) as
\be \label{ncgprgaa}
\Gamma_{\cA \cA} \; = \; f' (\eins_2 + i \cF ) \; + \; i f \; \; \; ,
\ee
which evidently shows that $\Gamma_{\cA \cA}$ by itself is a function of
$\cF$ only. In an analogous manner we also calculate $\Gamma_{\cA \eta \cA}$
by differentiating $\eta$ times (\ref{ncgpras2}) once more with respect to
$\cA$ ($\cF_{\cA \eta \cA} = 0$ by means of rule~2):
\be \label{ncgprgaetaa}
\Gamma_{\cA \eta \cA} \; = \; f' \; \cF_\cA \; \eta \; \cF_\cA
\ee
Hence, in contrast to $\Gamma_{\cA \cA}$, $\Gamma_{\cA \eta \cA}$ is not a 
function which only depends on $\cF$. However, $\Gamma_{\cA \eta \cA}^2$
is such a function due to $(\cF_\cA \; \eta \; \cF_\cA )^2 = - \cF_\cA^4 =
- (\eins_2 + i \cF )^2$. Collecting all scattered pieces so far we
have proven that $\Gamma_{\cA \cA}^2 + \Gamma_{\cA \eta \cA}^2$ is given by
\be \label{ncgprgqua}
\Gamma_{\cA \cA}^2 + \Gamma_{\cA \eta \cA}^2 \; = \;
2 i f f' (\eins_2 + i \cF ) \; - \; f^2 \; \; \; ,
\ee
and thus only depends on the field strength $\cF$.\\
Next we turn our attention to the term
\be \label{ncgh1def}
h_1 := 2 (\cA + \eta ) \Gamma_{\cA \cA} +
\eta (\cA + \eta ) \Gamma_{\cA \eta \cA} \; \; \; ,
\ee
which multiplies $\hbar^1$ on the r.h.s. of (\ref{ncgdiffg}) (besides the
factor $\frac{\lambda v^4}{8} \;
[\Gamma_{\cA \cA}^2 + \Gamma_{\cA \eta \cA}^2 ]^{- 1}$). Just upon
inserting (\ref{ncgprgaa}) and (\ref{ncgprgaetaa}) and using
(\ref{ncgprf}) at several places we find
\be \label{ncgh1}
h_1 = h_1^\cF \cdot \cF_\cA \; \; \mbox{ with } \; \;
h_1^\cF := 2 f \; - \; i f' (\eins_2 + i \cF ) \; \; \; .
\ee
As a first conclusion, we may state that the whole term multiplying $\hbar^1$
in (\ref{ncgdiffg}) indeed has the sought form of a function of $\cF$ times
$\cF_\cA$.
It remains to be shown that the same holds true for the term proportional
to $\hbar^2$ in (\ref{ncgdiffg}). The respective calculations are a little
bit more involved and will be skipped here. We just quote the main results
in between which can be obtained along the same lines as above:
\ba \label{ncgprhelp}
\Gamma_{\cA \cA \cA} & = & \left[ f'' (\eins_2 + i \cF ) + 2 i f' \right]
\cF_\cA \nonumber \\[.5ex]
\Gamma_{\cA \eta \cA \eta \cA} & = & f'' \cF_\cA \eta \cF_\cA \eta \cF_\cA 
\nonumber \\[.5ex]
\left[ \Gamma^2_{\cA \cA} + \Gamma^2_{\cA \eta \cA} \right]^2 A & = & 
\eta \left\{ f^2 \left[ 2 i f' + f'' (\eins_2 + i \cF ) \right] +
2 i f'^3 (\eins_2 + i \cF )^2 \right\} \cF_\cA \nonumber \\[.5ex]
\left[ \Gamma^2_{\cA \cA} + \Gamma^2_{\cA \eta \cA} \right]^2 B & = &
\left\{ f^2 f'' - 4 f f'^2 + 2 i f'^3 (\eins_2 + i \cF) \right\}
\cF_\cA \eta \cF_\cA \eta \cF_\cA
\ea
An easy and straightforward calculation using (\ref{ncgprhelp})
and (\ref{ncgprf}) finally proves that the term
\be \label{ncgh2def}
h_2 := - \; \eta \; [ \Gamma_{\cA \cA} \cdot A + \Gamma_{\cA \eta \cA} \cdot
B ] \; [\Gamma^2_{\cA \cA} + \Gamma^2_{\cA \eta \cA} ]^2
\ee
multiplying $\hbar^2$ on the r.h.s. of (\ref{ncgdiffg}) (besides a factor
$\frac{\lambda v^4}{8} \; [ \Gamma^2_{\cA \cA} + \Gamma^2_{\cA \eta \cA} 
]^{- 3}$) can actually be written in the desired form
\ba \label{ncgh2}
h_2 & = & h_2^\cF \cdot \cF_\cA \\[.5ex]
\mbox{with } \; h_2^\cF & := &
- \; 2 f^3 f' \; + \; i f^2 (2 f'^2 + f f'') (\eins_2 + i \cF) \; + \;
2 f f'^3 (\eins_2 + i \cF)^2 \; \; \; , \nonumber
\ea
where $h_2^\cF$ obviously depends on the field strength $\cF$ only. Taking
into account that we furthermore have
\be \label{ncgh0}
h_0 := \eta \cA \eta + 2 \eta \cA^2 + \cA \eta \cA + \cA^3 - \cA =
- \; \cF \cdot \cF_\cA
\ee
(this term multiplying $\frac{\lambda v^4}{8} \; \hbar^0$ on the r.h.s of
(\ref{ncgdiffg})) and recollecting the conceptual considerations presented
above the claim under consideration is truly proven by now.\\[.3cm]
But also all necessary preparations for an eased recursive determination
of the higher orders of $\Gamma$ are already made. Namely, collocating all
the spread results just derived the differential equation (\ref{ncgdiffg}) 
for the vertex functional $\Gamma$ now reads:
\ba \label{ncgdiffgf}
\frac{\lambda v^4}{8} \left\{
\hbar^2 \; h_2^\cF \; + \; 
\hbar \; [2 i f f' (\eins_2 + i \cF) - f^2]^2 \; h_1^\cF \right.
& & \nonumber \\[.5ex]
\left. - \; [2 i f f' (\eins_2 + i \cF) - f^2]^3 \; \cF \right\} \cdot
\cF_\cA & = & [2 i f f' (\eins_2 + i \cF) - f^2]^3 \;
\frac{\partial \Gamma}{\partial \cA}
\ea
In (\ref{ncgdiffgf}) $h_1^\cF$ and $h_2^\cF$ are given according to
(\ref{ncgh1}) and (\ref{ncgh2}), respectively. Using $\frac{\partial
\Gamma}{\partial \cA} \equiv \Gamma_\cA = f \cdot \cF_\cA$ (see
(\ref{ncgpras2}) and (\ref{ncgprabr})) and the fact that
$\cF_\cA = i (\cA + \eta)$ is a non-singular $2 \times 2$
matrix, (\ref{ncgdiffgf}) is equivalent to a highly nonlinear and
involved differential equation for the function $f(\cF )$ which due to its
complexity can not be solved exactly in the sense that the solution is
represented as a closed analytical expression comprising all orders of the
loop expansion. As before, it is, however, possible to solve (\ref{ncgdiffgf})
recursively, i.e. in the shape of a power series in $\hbar$. In fact, this
time the finding of the recursive solution is much simpler when compared
to the respective procedure dealing with the equivalent system 
(\ref{nssbdiffg}) of two coupled partial differential equations for $\Gamma$
because of the following reason: (\ref{ncgdiffgf}) is just {\it one}
differential equation in terms of just {\it one commutative} variable,
namely the field strength $\cF$, which hence may be handled just as an
ordinary variable.\\
The second part of appendix~I contains a short {\it mathematica} routine
for the recursive determination of $\Gamma^{(n)}$. Here, we only quote
the results for 2 and 3 loops:
\ba \label{ncgg2a3}
\Gamma^{(2)} & = & \frac{2}{\lambda v^4} \mbox{ Tr }
\left( 2 i \cF - 13 \cF^2 - 24 i \cF^3 + 15 \cF^4 \right)
\left( 2 i \cF - 3 \cF^2 \right)^{- 3} \\[.5ex]
\Gamma^{(3)} & = & \frac{16}{\lambda^2 v^8} \mbox{ Tr }
\left( 8 \cF^2 + 88 i \cF^3 - 408 \cF^4 - 986 i \cF^5 \right. \nonumber \\
& & \hspace*{2cm} \left. + 1329 \cF^6 +
           936 i \cF^7 - 270 \cF^8 \right) 
\left( 2 i \cF - 3 \cF^2 \right)^{- 6} \nonumber
\ea
It is easily checked (by means of {\it mathematica}) that $\Gamma^{(2)}$
(\ref{ncgg2a3}) exactly coincides (up to a constant of integration) with
$\Gamma^{(2)}$ (\ref{nssbg2}) (upon evaluation of the trace and upon
inserting $v$ according to (\ref{absmin}) into both expressions). However,
(\ref{ncgg2a3}) evidently represents the 2-loop result in a more
compact and elegant form. A similar statement holds true for all higher
orders, too.\\
Let us stress once again that the more compact and elegant form of the
results for $\Gamma^{(n)}$, the simplified determination of the recursive
solution and also the fact that $\Gamma^{(n)}$ is a function of $\cF$ only
is due to and only accessible within the noncommutative language
using matrices.

%% file: chap4.tex
\newpage
\newsection{Ghost contributions within the noncommutative
            formulation}

The certainly existing success of the matrix formalism notwithstanding
so far there is a serious drawback of the whole treatment: The ghost
fields $c, \overline{c}$, the auxiliary field $B$ and also the external
fields $X, Y$ -- together with their contributions to the classical action --
have not been taken into account which, on the other hand, is essential
for a proper treatment of the model. Otherwise we run into intractable
problems with the zero-dimensional analogue of the Goldstone propagator
$\langle y y \rangle$, and the 1 PI $n$-point functions of the theory
are ill-defined in higher orders, see chapter~3. The present chapter is
intended to resolve this problem (once more).\\
The deeper reason for the simplifications implied by the matrix formalism
is founded in the fact that all orders of $\Gamma$ can be written as functions
of the ordinary field strength $\cF$ only which in addition is an even matrix
proportional to $\eins_2$. In particular, the (invariant part of the) classical
action is obtained from the trace of $\cF^2$, see (\ref{ncgaccl}). This 
raises the following natural question: Is it possible to introduce
a reasonable generalization $\tilde{\cF}$ of 
\nolinebreak $\cF$ such that the whole
classical action $\Gamma_{cl}$ (\ref{ssbaccl}) including the $\phi \pi$-term
and the gauge fixing term is given by Tr$\tilde{\cF}^2$ (up to a constant
of proportionality). The possible existence of such a generalized field
strength would simplify matters in a way analogous to the investigations
of section~4.5 and, at the same time, would unify {\it all} fields of the
model in a sense which we would like to call ``strictly noncommutative''.
The inspection of this battery of questions is the subject of
section~5.1.\\
It will turn out, however, that such an incorporation of the unphysical parts
of the classical action is rather difficult to find or even impossible. In any
case, the discussion will point out that $\tilde{\cF}$ (if it exists)
necessarily has to be an {\it unphysical} quantity which contradicts the
common notion of a field strength in physics.\\
Of course, it is always feasible to incorporate the unphysical fields
$c, \overline{c}, B, X$ and $Y$ as ordinary scalars as in chapter~3, thus
not unifying all fields in one noncommutative object. Proceeding this way,
the resulting description of the model is, so to speak, a mixture of a 
noncommutative part (comprising of the gauge potential $\cA$) and a 
conventional part. Due to the negative outcome of section~5.1 we
address ourselves to this admittedly less elegant possibility in
section~5.2 which nevertheless entails {\it some} simplification when
compared to the formulation using exclusively the component fields.\\
The considerations to follow -- in particular those of section~5.1 --
already lead over to the open questions in the outlook.

\newsubsection{Problems with the noncommutative integration of ghost terms}

We begin with a quite general and detailed discussion of the difficulties 
concerning a ``strictly noncommutative incorporation'' of ghost field 
contributions to the classical action. In fact, it will turn out that such 
an incorporation seems to be impossible, at least
as long as the meaning of ``strictly noncommutative incorporation'' is defined
by the following two requirements: Firstly, the Faddeev-Popov term (and
possibly also the gauge fixing term) of the classical action should result
from the trace of the square of a generalized field strength $\tilde{\cF}$
which, in turn, is to be derived from a generalized gauge potential 
$\tilde{\cA}$. And secondly, the generalized construction should fulfill
the more aesthetic qualification of naturalness. The arguments presented 
subsequently are taken to some extent from \cite{HLN} reversing, however, 
the perspective of interpretation given there.\\[.3cm]
In order to have the chance of obtaining the $\phi \pi$-term eventually,
the ghost and anti-ghost fields have to be present right from the beginning.
Otherwise no term proportional to $\overline{c} c$ will ever show up.
Consequently, the gauge potential $\cA$ (\ref{ncga}) has to
be enlarged in such a way that it also includes ghost and anti-ghost fields:
\be \label{ncgagen}
\cA \longrightarrow \tilde{\cA} := \cA + \cC + \overline{\cC}
\ee
In (\ref{ncgagen}), $\cC$ represents a ($2 \times 2$) matrix containing only
ghost fields whereas the elements of $\overline{\cC}$ are anti-ghost fields.\\
Furthermore, it is certainly reasonable to demand that $\tilde{\cA}$ should
carry the definite degree~$1$ as a generalized (exterior) form. At this stage,
however, another $\Z$-grading enters the game due to the appearance of
anticommuting variables, namely the $\Z$-grading of Grassmann algebras
whose several levels are distinguished by $\phi \pi$-charge.
Hence, before talking about generalized (exterior) forms we first have to
merge all gradings floating around. This is achieved by introducing the new
total grade $\tilde{\partial}$ which is defined to be the sum of the matrix
grade $\partial_M$, the exterior form grade $\partial_F$ {\it and} the
Grassmannian grade $\partial_G$ (mod $2$):
\be \label{totgrgen}
\tilde{\partial} \; \cdot \; := \; \partial_M \cdot \; + \; 
\partial_F \cdot \; + \;
\partial_G \cdot \; \; \mbox{ (mod } 2)
\ee
Because of $\tilde{\partial} \tilde{\cA} = 1$ we obtain that $\cC$ as well as
$\overline{\cC}$ have to be {\it even} matrices, i.e.
\be \label{ccbargen}
\cC = a_c \left( \begin{array}{cc} c_1 & 0 \\ 0 & c_2 \end{array} \right)
\; \mbox{ and } \; \overline{\cC} = a_{\overline{c}} \left(
\begin{array}{cc} \overline{c}_1 & 0 \\ 0 & \overline{c}_2 \end{array}
\right) \; \; \; .
\ee
Now, there are obviously {\it two} ghost fields $c_1$ and $c_2$ and
{\it two} anti-ghost fields $\overline{c}_1$ and $\overline{c}_2$ appearing
in (\ref{ccbargen}) and (\ref{ncgagen}) whereas in the present model we
expect only {\it one} ghost field \nolinebreak 
$c$ and {\it one} anti-ghost field
$\overline{c}$ to occur. Hence, there must be relations between
$a_c, c_1, c_2$ and $c$ on the one hand and $a_{\overline{c}},
\overline{c}_1, \overline{c}_2$ and $\overline{c}$ on the other hand which
can be established by the following argument:\\
Within the Marseille-Mainz approach to noncommutative gauge theories which
forms the setting for all considerations of the previous
and the present chapter, infinitesimal gauge transformations are
in general given by (see \cite{CES}, \cite{CHPS}):
\be \label{ncggtrgen}
\cA' = \cA + d {\cal E} + [\cA, {\cal E} ]_g =
\cA + d {\cal E} + [\cA, {\cal E} ] \; \; \;
(\mbox{i.e.} \; \delta_{\cal E} \cA = d {\cal E} + [\cA, {\cal E} ] )
\ee
The second equality (in front of the parantheses) is due to
${\cal E}$ being an {\it even} element of the super Lie algebra involved in
the construction tensorized by $\Lambda^\star (M)$. Because $\cA$ only
contains physical fields (namely gauge fields and Higgs fields) and because
on physical fields BRS transformations coincide with local gauge 
transformations up to the replacement ${\cal E} \rightarrow \cC$, we, hence, 
obtain in our zero-dimensional model:
\be \label{ncgbrsgen}
s \cA = d_M \cC + [\cA, \cC] = [\cA + \eta, \cC]
\ee
On the other hand, we may take for granted the BRS transformations of the
constituents of $\cA$, see (\ref{brs}). This allows us to calculate
\ba \label{ncgbrsgenc}
s \cA & = & \frac{i}{v} \left( \begin{array}{c|c}
0 & s x + i s y \\ \hline s x - i s y & 0 \end{array} \right) \; = \;
\frac{i}{v} \left( \begin{array}{c|c}
0 & - c y + i c (x + v) \\ \hline - c y - i c (x + v) & 0 \end{array} 
\right) \nonumber \\
& = & - i (\cA + \eta ) \gamma c \; = \; 
(\cA + \eta ) (- \frac{i}{2} \gamma c ) +
\frac{i}{2} \gamma c (\cA + \eta ) \; = \; 
[\cA + \eta, - \frac{i}{2} \gamma c ] \; \; \; .
\ea
From (\ref{ncgbrsgen}) and (\ref{ncgbrsgenc}) we immediately deduce
\be \label{ccbargenc}
\cC \; = \; - \; \frac{i}{2} \; c \; \gamma \; \; \; .
\ee
A completely analogous argument based on the anti-BRS transformations (which
we did not consider up to now because they only play an auxiliary role 
in the present section) also fixes $\overline{\cC}$:
\be \label{ccbargencb}
\overline{\cC} \; = \; - \; \frac{i}{2} \; \overline{c} \; \gamma
\ee
As an intermediate result we thus get:
\be \label{ncgagenr}
\tilde{\cA} = \cA - \frac{i}{2} (c + \overline{c}) \gamma
\ee
Let us now turn to the field strength $\tilde{\cF}$ which
should result from $\tilde{\cA}$ by further generalizing Cartan's structure
equation, i.e.
\be \label{ncgcurvgen}
\tilde{\cF} = \tilde{d} \tilde{\cA} + \frac{1}{2}
[\tilde{\cA}, \tilde{\cA} ]_{gen} \; \; \; ,
\ee
where $[\cdot, \cdot ]_{gen}$ denotes the supercommutator with respect
to the {\it total} grade (\ref{totgrgen}):
\be \label{ncggrcgen}
[A,B]_{gen} = A \diamond B - (-1)^{(\tilde{\partial} A) (\tilde{\partial} B)}
B \diamond A
\ee
In the above equation the product $\diamond$ has to be a combination of
matrix multiplication, exterior product and multiplication of the
Grassmann algebra, and, of course, it has to be defined in such a way
that the gradings involved are respected. In the zero-dimensional model
under consideration, for instance, only scalars (zero-forms) appear and, 
hence, we can forget about the exterior form grading. In this 
case the canonical definition of $\diamond$ reads as follows:
\be \label{ncgprodgen}
(M \otimes f) \diamond (N \otimes g) = (-1)^{(\partial_M N) (\partial_G f)}
(M \cdot N) \otimes (fg)
\ee
In (\ref{ncgprodgen}) $M$ and $N$ are ($2 \times 2$) matrices whereas $f$ and 
$g$ represent arbitrary elements of the Grassmann algebra.
Using (\ref{ncggrcgen}), (\ref{ncgprodgen}) and (\ref{ncgagen}) we thus
calculate:
\ba \label{ncggenex}
\frac{1}{2} [\tilde{A}, \tilde{A} ]_{gen} & = & 
\tilde{A} \diamond \tilde{A} \; = \; 
(\cA + \cC + \overline{\cC}) \diamond (\cA + \cC + \overline{\cC}) 
\nonumber \\
& = & \cA \diamond \cA + \cA \diamond (\cC + \overline{\cC}) +
(\cC + \overline{\cC}) \diamond \cA + (\cC + \overline{\cC}) \diamond
(\cC + \overline{\cC}) \nonumber \\[.5ex]
& = & \cA \cdot \cA + \cA \cdot (\cC + \overline{\cC}) -
(\cC + \overline{\cC}) \cdot \cA + 0 \mbox{ (due to } c^2 = 0 =
\overline{c}^2) \nonumber \\[.5ex]
& = & \frac{1}{2} [\cA, \cA]_g + [\cA, \cC + \overline{\cC} ]
\ea
Furthermore, in (\ref{ncgcurvgen}) $\tilde{d}$ stands for a possible 
further enlargement of the already
generalized exterior derivative $d$ ``$=$'' $d_M + d_C$, see (\ref{superd}).
In fact, having merged all three gradings within the game, see 
(\ref{totgrgen}), by the same argument as the one presented in section~4.1
in connection with the introduction of $d$ (\ref{superd}) it is certainly
reasonable to further enlarge $d$ by an operation that acts on the 
Grassmannian degree raising the $\phi \pi$-charge by~$1$ mod~$2$. The
only canonical operations at hand, however, fulfilling the requirement 
just stated and having in addition the property of being idempotent are the 
BRS operator $s$ and the anti-BRS operator $\overline{s}$. As a consequence, 
in the zero-dimensional model where $d_C$ does not contribute (due to the
fact that there is just one point in spacetime) the natural candidate 
for $\tilde{d}$ is given by
\be \label{ncgdgen}
\tilde{d} (M \otimes f) = d_M M \otimes f + 
(-1)^{\partial_M M} M \otimes (s + \overline{s}) f \; \; \; .
\ee
Taking into account (\ref{ncggenex}), (\ref{ncgdgen}) and (\ref{ncgbrsgen})
we finally calculate in a straightforward manner:
\bas
\tilde{\cF} & = & d_M \cA + d_M \cC + d_M \overline{\cC} - 
s \cA - \overline{s} \cA + s \cC + \overline{s} \cC +
s \overline{\cC} + \overline{s} \overline{\cC} \nonumber \\[.5ex]
& & + \frac{1}{2} [\cA, \cA]_g +
\underbrace{[\cA, \cC]}_{= \; s \cA \; - \; d_M \cC} +
\underbrace{[\cA, \overline{\cC}]}_{= \; \overline{s} \cA \; - \; 
d_M \overline{\cC}}
\nonumber \\[.5ex]
& = & d_M \cA + \frac{1}{2} [\cA, \cA]_g \\[.5ex]
& & + [(s \cC + \overline{s} \cC + s \overline{\cC} +
\overline{s} \overline{\cC}) = 0 \mbox{ due to } s \; c = 0 = \overline{s} \;
\overline{c} \mbox{ and } s \; \overline{c} = B = - \overline{s} \; c] 
\nonumber
\eas
Thus, we obtain
\be \label{ncghoriz}
\tilde{\cF} = \cF \; \; \; .
\ee
From (\ref{ncghoriz}) it is obvious that there will be no $\phi \pi$-terms
in the classical action which is proportional to the trace of 
$\tilde{\cF}^2$. In other words, as long as we stick to the two requirements
mentioned in the introduction to this section, these two requirements
specifying the meaning of a ``strictly noncommutative incorporation'' of
ghost contributions, it is effectively impossible to realize the aspired 
goal.\\[.3cm]
In \cite{HLN} the above argument is organized in a logically reversed manner:
With the defi\-nitions (\ref{ncgagen}) and (\ref{ncgcurvgen}) for $\tilde{\cA}$
and $\tilde{\cF}$, respectively, the implementation of the so-called
{\it horizontality condition} (\ref{ncghoriz}), see \cite{NTM},
\cite{BTM}, leads to the recovery of the known BRS and anti-BRS transformations
of all fields involved in the model, which we, in turn, took for granted.\\
Furthermore, concerning the construction of a BRS and anti-BRS invariant
classical action (including $\phi \pi$-terms and gauge fixing terms) \cite{BTM}
and \cite{HLN} quote a recipe which reads as follows when translated to our
language ($\alpha$ is a gauge parameter): 
\be \label{ncggencla}
\Gamma_{cl} = - \frac{1}{4} \mbox{ Tr } \int \left\{
\cF^\dagger \cF - s \overline{s} (\cA^\dagger \cA ) + 
\alpha s (\overline{\cC}^\dagger B ) \right\}
\ee
Neglecting the integral (which in zero dimensions has to be dropped anyway)
(\ref{ncggencla}) evi\-dently differs from Tr$\cF^2$ $=$ Tr$\tilde{\cF}^2$ due
to the second and third term on the r.h.s. which in the present setting are
responsible for the $\phi \pi$-term and the gauge fixing term, respect\-ively.
We may interpret this as an underpinning of our previous observation 
that within the present limited context a ``strictly
noncommutative incorporation'' of ghost
terms (and possibly also of gauge fixing terms) cannot be achieved.\\[.3cm]
From a general and conceptual point of view it could have been clear right 
from the beginning that our intention would meet some difficulties, at least
with regard to questions of physical interpretation: Usually, the field
strength $\cF$ is a {\it physical} object as can be seen, for instance, in 
electrodynamics where $\cF$ contains only {\it physical} quantities, namely
the electric field $\vec{E}$ and the magnetic field $\vec{B}$.
Looking for a generalization of the ordinary field strength $\cF$, denoting
the resulting generalized object by $\tilde{\cF}$ and requiring that 
Tr$\tilde{\cF}^2$ should particularly yield a $\phi \pi$-term (among other
terms), unavoidably implies that $\tilde{\cF}$ itself has to contain somehow 
ghost and anti-ghost fields. But because ghost fields are {\it unphysical}
fields $\tilde{\cF}$ then automatically becomes an {\it unphysical quantity}
itself which contradicts the notion of what a {\it field strength} in physics
is to be. Of course, it is admissible to take a pragmatic point of view and 
say: $\tilde{\cF}$ is merely an {\it auxiliary} quantity possibly useful
for the calculations in between but not physically meaningful by itself.
More precisely, within this interpretation $\tilde{\cF}$ is composed of
a physical part $\cF$ (the original field strength) and an unphysical part
$\cF_{ghost}$ containing the ghost fields $c$ and $\overline{c}$:
$\tilde{\cF} = \cF + \cF_{ghost}$. In fact,
we have (and {\it must} have) this more pragmatic interpretation in mind 
when dealing with the problems of the present section.\\[.3cm]
Next, let us briefly comment on the following question: What happens
if we relax either one of the requirements or even both requirements stated
in the introduction in connection with the definition of the notion ``strictly
noncommutative incorporation''. In particular the second requirement
appealing to the naturalness of the construction is somewhat unprecise and
leaves some scope for possible modifications even though we tried to
emphasize the ``canonical'' nature of all generalizations involved in the
above discussion. In other words, we now would like to pose the question:
Is it possible to find a ``noncommutative incorporation'' of
ghost contributions? By ``noncommutative incorporation'' we mean an
incorporation using the noncommutative matrix formalism in some way but
still in such a manner that the ``square'' of the generalized field strength
$\tilde{\cF} = \cF + \cF_{ghost}$ (which does not necessarily has to
originate from a generalized gauge potential $\tilde{\cA}$ in a
first approach) contains the
$\phi \pi$-term (after the application of trace). Because of the vagueness
of the last statement (definition) there is hardly any hint in which
direction a meaningful search should go ahead. Out of a whole bunch of
possibilities we here mention only two strat\-egies:\\
The first one assumes that the product involved in calculating the ``square''
of $\tilde{\cF}$ has to be modified. So far this product is ordinary
matrix multiplication. For instance, we could try to use the product
(\ref{ncgprodgen}) instead of matrix multiplication or some other product
which is ``polluted'' with the gradings involved in the game.\\
The second strategy is based on the embedding of the $2 \times 2$ matrix
formalism into matrices of some higher dimension. This ansatz is supported
to some extent by the above discussion concerning the unphysical nature of
$\tilde{\cF}$: If $\tilde{\cF}$ is composed of a physical and an unphysical
part, $\tilde{\cF} = \cF + \cF_{ghost}$, then there should be a mathematical
structure which allows the proper distinction between those parts. Within the
language of matrices, a natural candidate for such a structure is a 
$\Z_2$-grading. However, because the original $\Z_2$-grading of $2 \times 2$
matrices cannot produce relief (unless perhaps the first strategy is pursued)
another $\Z_2$-grading comes into the game which, in turn, necessarily entails
matrices of higher dimension.\\
None of these possibilities immediately leads to satisfactory results, at least
not in an obvious manner. First attempts failed or resulted in really odd
and strange expressions which exclude any meaningful interpretation. 
Nevertheless the problem is still unsettled and needs 
further investigation.\\[.3cm]
Let us conclude with a heretical question which admittedly could have been 
posed already at the beginning: Does it make sense at all to look for such an
unphysical ``field strength'' 
\nolinebreak $\tilde{\cF}$ which subsequently should produce
the whole classical action (including $\phi \pi$-terms and gauge fixing terms)
when calculating the trace of the ``square'' of $\tilde{\cF}$?\\
The point here is the following one: The gauge fixing term is to a certain
extent a {\it matter of choice}. As a consequence, also the $\phi \pi$-term
(which has to be chosen in such a way that $s (\phi \pi$-term $+$ gauge
fixing term$) = 0$) is not fixed from the first outset. But if these terms
are a matter of choice in the way just explained, why at all can we hope
to find a well-defined prescription that generalizes $\cF$ and afterwards
yields a unique result for $\Gamma_{cl}$?\\
In this question the answer in support of our investigations is latent:
We do not search for {\it all} admissible $\phi \pi$-terms and gauge fixing
terms. And if there were such a prescription leading to a unique
$\Gamma_{cl}$ this would be
even better because then a specific gauge fixing term would be
singled out as a preferred one.\\
In any case, due to its unphysical nature the generalized quantity
$\tilde{\cF}$ is only an auxiliary one whose existence, however, could 
possibly simplify subsequent calculations tremendously in a way similar to the
role that $\cF$ plays at the end of chapter~4.

\newsubsection{Higher orders of $\Gamma$ including ghost contributions}

In spite of the negative outcome of the previous section it is, of course,
nonetheless possible to incorporate the ghost fields $c, \overline{c}$ and
the auxiliary field $B$ by hand, so to speak. In other words, what we are
going to study in the following is a proper formulation of our zero-dimensional
model where the Higgs fields $x$ and $y$ are unified in the $2 \times 2$
matrix $\cA$ (the gauge potential) and where the unphysical fields are
{\it not} embedded into some matrices but remain (ordinary) scalars. Because
the whole discussion is straightforward and once more copies the by now
well-known procedure we do not have to go into all technical details. The
main reason for this is that due to the scalar nature of $c, \overline{c}$
and $B$ the rules for the matrix calculus obtained in section~4.3 hold
true unchanged or at most have to be modified in a marginal way, see also
below.\\
An easy calculation shows that the classical action $\Gamma_{cl}$
(\ref{ssbaccl}) can be rewritten in a form well adapted for the present
context:
\be \label{ssbaccla}
\Gamma_{cl} = S(\cA ) + \frac{1}{2} B^2 - 
\frac{i v}{2} \xi m \; B \mbox{ Tr } \gamma \eta \cA +
\frac{v}{2} \xi m \; \overline{c} \mbox{ Tr} (\eta \cA - \eins_2 ) \; c -
i \mbox{ Tr } \gamma {\cal X} (\cA + \eta) \; c
\ee
In the above equation $S(\cA )$ is given by (\ref{ncgaccl}), and the external
fields $X$ and $Y$ which couple to the nonlinear BRS transformations of
$y$ and $x$, respectively, have been unified into one matrix-valued external 
field ${\cal X}$ according to
\be \label{ncgextfu}
{\cal X} = - \frac{i v}{2} \left( \begin{array}{c|c}
0 & Y + i X \\ \hline Y - i X & 0 \end{array} \right) \; \; \; ,
\ee
in order to simplify matters.\\[.3cm] 
Next, we determine the set of coupled partial
differential equations for the generating functional $Z = Z(J, l, n,
\overline{n}; {\cal X} )$ by differentiating the integrand of
\be \label{ncggenzdef}
Z(J, l, n, \overline{n}; {\cal X} ) = {\cal N} \int_{- \infty}^{+ \infty}
dx \; dy \; dB \; dc \; d\overline{c} \; \mbox{exp} \left\{ \frac{1}{\hbar} 
(- \Gamma_{cl} + \mbox{Tr} J \cA + lB + \overline{c} n + \overline{n} c )
\right\}
\ee
once with respect to $\cA, B, c$ and $\overline{c}$, respectively. (In 
comparison with the analogous investigations in chapter~3, we have renamed the
sources $\eta$ and $\overline{\eta}$ by $n$ and $\overline{n}$, respectively,
because $\eta$ already has a different meaning, see (\ref{ncgeta}).)
Subsequently, this set of differential equations is rewritten for the
generating functional $W = W(J, l, n, \overline{n}; {\cal X})$ of connected
Green's functions resulting in:
\ba \label{ncgdiffwa}
\hbar^2 \; \frac{\lambda v^4}{8} \; \frac{\partial^3 W}{\partial J^3}
& & \\
+ \; \hbar \left\{ \frac{\lambda v^4}{8} \left[
2 \left( \frac{\partial W}{\partial J} + \eta \right) 
\frac{\partial^2 W}{\partial J^2} - \eta \left(
\frac{\partial W}{\partial J} + \eta \right)
\frac{\partial}{\partial J} \eta \frac{\partial W}{\partial J} \right]
- \frac{v}{2} \xi m \eta 
\frac{\partial^2 W}{\partial n \partial \overline{n}} \right\}
& & \nonumber \\
+ \; \frac{\lambda v^4}{8} \left[
\eta \frac{\partial W}{\partial J} \eta + 
2 \eta \left( \frac{\partial W}{\partial J} \right)^2 +
\frac{\partial W}{\partial J} \eta \frac{\partial W}{\partial J} +
\left( \frac{\partial W}{\partial J} \right)^3 -
\frac{\partial W}{\partial J} \right]
& & \nonumber \\
+ \frac{v}{2} \xi m \eta \frac{\partial W}{\partial \overline{n}}
\frac{\partial W}{\partial n} -
\frac{i v}{2} \xi m \gamma \eta \frac{\partial W}{\partial l} -
i \gamma {\cal X} \frac{\partial W}{\partial \overline{n}} 
& = & J \nonumber \\[.5ex]
\frac{\partial W}{\partial l} - \frac{i v}{2} \xi m
\mbox{ Tr } \gamma \eta \frac{\partial W}{\partial J}
& = & l \nonumber \\[.5ex]
\frac{v}{2} \mbox{ Tr} \left\{ - \; \hbar \; \xi m \eta
\frac{\partial^2 W}{\partial J \partial n} -
\xi m \eta \frac{\partial W}{\partial n} \left( 
\frac{\partial W}{\partial J} + \eta \right) -
\frac{2 i}{v} \gamma {\cal X} \left(
\frac{\partial W}{\partial J} + \eta \right) \right\}
& = & \overline{n} \nonumber \\[.5ex]
\frac{v}{2} \xi m \mbox{ Tr} \left\{
\hbar \; \eta \frac{\partial^2 W}{\partial J \partial \overline{n}} +
\eta \left( \frac{\partial W}{\partial J} + \eta \right)
\frac{\partial W}{\partial \overline{n}} \right\} & = & n \nonumber 
\ea
(\ref{ncgdiffwa}) exactly coincides with (\ref{ssbdiffw}) when evaluating
the traces occuring in (\ref{ncgdiffwa}). The first two equations in
(\ref{ssbdiffw}), however, have been merged into a single (matrix-valued)
equation, namely the first one in (\ref{ncgdiffwa}), which in spite of its
character as an equation of matrices can be treated as {\it one} equation
due to the matrix calculus at hand. This reduces the complexity of the
system to some extent. The modifications brought about by the necessary
inclusion of unphysical fields can be read off from a comparison bewteen
(\ref{ncgdiffwa}) and (\ref{ncgdiffw}).\\
As before, the next step consists of performing a Legendre transformation,
this time with respect to $J, l, n$ and $\overline{n}$:
\ba \label{ncglega}
& \Gamma (\cA, B, c, \overline{c}; {\cal X}) \; + \; 
W(J(\cA, B, c, \overline{c}; {\cal X}), l(\cdots ), n(\cdots ),
\overline{n} (\cdots); {\cal X}) & \nonumber \\[.5ex]
& - \mbox{ Tr} J(\cdots ) \cA \; - \; l(\cdots ) B \; - \;
\overline{n} (\cdots ) c \; - \; \overline{c} n(\cdots ) \; = \; 0 & \\[.5ex]
& \mbox{with: } \cA = \displaystyle\frac{\partial W}{\partial J} \; , \;
B = \frac{\partial W}{\partial l} \; , \;
c = \frac{\partial W}{\partial \overline{n}} \; , \;
\overline{c} = - \frac{\partial W}{\partial n} & \nonumber
\ea
Particularly, we are in need of the Legendre transformations of the various 
second and third derivatives of $W$ appearing in (\ref{ncgdiffwa}), whose
determination again is the hardest piece of work to be done. In principle,
we can go ahead as in appendix~F (as regards the component formalism) or
section~4.4 (as regards the matrix formalism). For instance, the 
determination of the Legendre transformations of $\frac{\partial^2 W}{\partial
J^2}$ and $\frac{\partial}{\partial J} \eta \frac{\partial W}{\partial J}$ 
involves differentiation of the identity
\be \label{ncgexaex}
\cA (J(\cA, B, c, \overline{c}; {\cal X}), l(\cdots ), n(\cdots ),
\overline{n} (\cdots ); {\cal X}) = \cA
\ee
with respect to $\cA$. Now, both matrices and scalars are found in
(\ref{ncgexaex}). In order to be able to treat also such cases we therefore
need a slight modification of the chain rule~4 (\ref{ncgchain}) or, in other
words, a combination of the ordinary chain rule and the chain rule~4 of
matrix calculus which, however, is obtained in a straightforward manner and
summarized in our final rule:
\begin{description}
\item[{\bf Rule 8: }] {\bf generalized chain rule: } 
Let $F(J, l)$ be any function of
the matrix $J$ and the scalar $l$, let $J(\cA, B)$ be an odd polynomial
in $\cA, B$ (and $\eta$) and let $l$ be given as $l =$ Tr $poly(\cA, B)$
where $poly(\cA, B)$ is an even polynomial in $\cA, B$ (and $\eta$). Then
the following chain rules are valid:
\ba \label{ncgchaingen}
\frac{\partial}{\partial \cA} F & = & \left.
\frac{\partial J}{\partial \cA}
\frac{\partial F}{\partial J} -
\left( \frac{\partial}{\partial \cA} \eta J \right)
\left( \frac{\partial}{\partial J} \eta F \right) +
\frac{\partial l}{\partial \cA}
\frac{\partial F}{\partial l}
\; \right|_{J = J(\cA, B) \atop l = l(\cA, B)} \nonumber \\[.5ex]
\frac{\partial}{\partial B} F & = & \left. 
\frac{\partial J}{\partial B}
\frac{\partial F}{\partial J} -
\left( \frac{\partial}{\partial B} \eta J \right)
\left( \frac{\partial}{\partial J} \eta F \right) +
\frac{\partial l}{\partial B}
\frac{\partial F}{\partial l}
\; \right|_{J = J(\cA, B) \atop l = l(\cA, B)}
\ea
These chain rules generalize in an obvious way to the case when $F$ depends
on several scalars $l_j$, $j = 1, \dots, p$ ($p$ arbitrary), which, in turn,
depend on $\cA$ and $B_j$, $j = 1, \dots, p$.
\end{description}
When expressed in words the above rule simply reads: Any admissible
differentiation of $F$ is given by the ordinary chain rule, but with 
the restriction that the chain rule of matrix calculus has to be applied
when the matrix-valued argument of $F$ is involved in the evaluation of the
ordinary chain rule.\\
In all other respects the rules of the matrix calculus developed formerly
remain unchanged except for a cautious but straightforward consideration
of possible signs which might appear due to the presence of anticommuting
variables.\\
By means of rule~8 the Legendre transformations mentioned previous to
(\ref{ncgexaex}) are calculated to be:
\ba \label{ncgleggen1}
\frac{\partial^2 W}{\partial J^2} & \rightarrow &
\left( \frac{\partial^2 \Gamma}{\partial \cA^2} +
\frac{v^2}{4} (\xi m)^2 \right) h_\cA^{- 1} + \dots \nonumber \\
\frac{\partial}{\partial J} \eta \frac{\partial W}{\partial J}
& \rightarrow & - \left( \frac{\partial}{\partial \cA} \eta 
\frac{\partial \Gamma}{\partial \cA} - 
\frac{v^2}{4} (\xi m)^2 \eta \right) h_\cA^{- 1} + \dots \\
\mbox{with} & : & h_\cA (\cA) :=
\left( \frac{\partial^2 \Gamma}{\partial \cA^2} +
\frac{v^2}{4} (\xi m)^2 \right)^2 +
\left( \frac{\partial}{\partial \cA} \eta 
\frac{\partial \Gamma}{\partial \cA} - 
\frac{v^2}{4} (\xi m)^2 \eta \right)^2 \nonumber
\ea
The dots on the right hand sides of the above equation indicate contributions
proportional to the ghost field $c$ whose explicit knowledge is not required
for the procedure to work. In an anlogous manner we also obtain:
\ba \label{ncgleggen2}
\frac{\partial^2 W}{\partial J \partial n} & \rightarrow & 
\frac{1}{\Gamma_1} \left\{
\frac{\partial^2 W}{\partial J^2} \; \frac{\partial}{\partial \cA}
\mbox{ Tr } \Gamma_{23} {\cal X} -
\left( \frac{\partial}{\partial J} \eta
\frac{\partial W}{\partial J} \right) \frac{\partial}{\partial \cA}
\eta \mbox{ Tr } \Gamma_{23} {\cal X} \right\} \nonumber \\
& & + \overline{c} \mbox{-dependent term} \nonumber \\[.5ex]
\frac{\partial^2 W}{\partial J \partial \overline{n}} & \rightarrow &
- \frac{1}{\Gamma_1} \left\{
\frac{\partial^2 W}{\partial J^2} \; \frac{\partial \Gamma_1}{\partial \cA} -
\left( \frac{\partial}{\partial J} \eta
\frac{\partial W}{\partial J} \right)
\frac{\partial}{\partial \cA} \eta \Gamma_1 \right\} c \nonumber \\[.5ex]
\frac{\partial^2 W}{\partial n \partial \overline{n}} & \rightarrow &
- \frac{1}{\Gamma_1} + c \mbox{-dependent term}
\ea
In (\ref{ncgleggen2}) $\Gamma_1$ is uniquely determined by the generally
valid decomposition (see also (\ref{gdecom}))
\ba \label{gdecomgen}
\Gamma (\cA, B, c, \overline{c}; {\cal X}) & = &
\Gamma_0 (\cA, B) + \Gamma_1 (\cA ) c \overline{c} +
\Gamma_2 (\cA ) c X + \Gamma_3 (\cA ) c Y \nonumber \\
& = & \Gamma_0 (\cA, B) + \Gamma_1 (\cA ) c \overline{c} + 
c \mbox{ Tr } \Gamma_{23} (\cA ) {\cal X} \\[.5ex]
\mbox{with } \Gamma_{23} & := & \frac{i}{v} \left( \begin{array}{c|c}
0 & \Gamma_3 + i \Gamma_2 \\ \hline \Gamma_3 - i \Gamma_2 & 0
\end{array} \right) \; \; \; . \nonumber
\ea
The still missing Legendre transformation of $\frac{\partial^3 W}{\partial
J^3}$ is calculated in appendix~J.\\[.5cm]
Now, everything is provided in order to write down explicitly the set of 
differential equations for $\Gamma$, see again appendix~J.
Subsequently, this set of differential equations for $\Gamma$ is solved order
by order in $\hbar$. Of course, as a matter of consistency in zeroth order 
we recover the classical action $\Gamma_{cl}$ (\ref{ssbaccla}) we started
from. In the following, we content ourselves with the specification of the
result in first order:
\ba \label{ncgg1gen}
\Gamma_{23}^{(1)} & = & \frac{i \gamma}{8} \;
\frac{\lambda v^4 (2 \eta \cA \eta + 2 \eta \cA^2 - \cA \eta \cA ) +
      4 v^2 (\xi m)^2 \eta}{\mbox{Tr}(\eta \cA - \eins_2 )} \;
{h_{\cA}^{(0)}}^{- 1} \nonumber \\[.5ex]
\Gamma_1^{(1)} & = & \frac{i v}{2} \; \xi m \mbox{ Tr } 
\Gamma_{23}^{(1)} \gamma \eta \\[.5ex]
\Gamma_0^{(1)} & = & \frac{1}{2} \mbox{ ln } \left| \mbox{ Tr } h_\cA^{(0)} \;
\right| \; - \;
\mbox{ ln } \left| \mbox{ Tr } (\eta \cA - \eins_2 ) \; \right|
\; + \; C^{(1)}_\cA \nonumber \\[.5ex]
\mbox{with} & : & h_{\cA}^{(0)} = \frac{\lambda^2 v^8}{64}
(2 i \cF - 3 \cF^2 ) \nonumber \\
& & \hspace*{1.5cm} + \frac{\lambda v^6}{32} 
(\xi m)^2 (4 \cA^2 - \cA \eta \cA \eta -
\eta \cA \eta \cA + 6 \cA \eta + 6 \eta \cA - 4 \eins_2 ) \nonumber
\ea
This result exactly coincides with (\ref{ssbg1}) if the traces in 
(\ref{ncgg1gen}) are evaluated. However, going explicitly through the
calculations one actually realises some simplification when compared to
the respective calculations leading to (\ref{ssbg1}).\\
Finally, we now could study the renormalization of our zero-dimensional
model. Because this problem was investigated in detail within the component
formulation of the model, see section~3.4, we skip this discussion 
here. We only would like to mention that due to the introduction of the 
matrix-valued external field $\cal X$ the Slavnov-Taylor identity 
(being valid without proof, see final remarks in appendix~F) can be written as
\be \label{ncgstgen}
\cS(\Gamma ) \equiv \mbox{ Tr } \frac{\partial \Gamma}{\partial {\cal X}}
\frac{\partial \Gamma}{\partial \cA} + B \frac{\partial \Gamma}{\partial
\overline{c}} = 0 \; \; \; ,
\ee
which is a form more compact than (\ref{ssbsti}).

%% file: chap5.tex
\newpage
\newsection{Conclusion and outlook}

As we mentioned in the introduction the purpose of this thesis was a threefold
one:\\
First of all, we wanted to answer the question whether the counting argument
of Feynman diagrams relying on the consideration of the zero-dimensional
version of a given theory in four dimensions can be applied to cases when
there is also spontaneous symmetry breaking. To this end we investigated
in detail a simple model of SSB comprising of one complex scalar field
$\Phi$ with a self-interaction of the Mexican hat type but without (ordinary)
gauge fields, living on just one point in spacetime. Within this model we were
able to answer the above question in the affirmative. However, it turned out
that even in zero dimensions a proper definition of the theory unavoidably
requires the inclusion of a gauge fixing term, a Faddeev-Popov term and
an external field part to the classical action. Otherwise higher orders in
the loop expansion of the theory are ill-defined. In particular, ghost fields
$c$ and $\overline{c}$ have to be present although there are no (ordinary)
gauge fields at all. As for theories without SSB the counting argument proved
to be a consistency check that in a four-dimensional calculation all diagrams
{\it and} all combinatorial factors stemming from Wick's theorem ({\it and} 
all possible symmetry factors) are taken into account correctly. All this was
the topic of chapter~3.\\
The second aim was a more pedagogical one: We tried to illustrate the
necessity and the origin of some basic notions in quantum field theory within
the simple context of our zero-dimensional model. In a physical theory in
four dimensions the understanding of those notions sometimes might be hidden
by purely technical complications which, however, in zero dimensions are
absent to a large extent. Some examples of those fundamental notions already
have been indicated above: Let us mention the necessity of gauge fixing
implying, in turn, the extension of (local) gauge invariance to BRS
invariance accompanied by the inclusion of ghost fields and
ghost contributions to the theory, the must of introducing external fields
for all nonlinear BRS transformations in order to control properly these
transformations in higher orders and the role of the functional version
of BRS invariance, namely the role of the Slavnov-Taylor 
identity. Perhaps the most 
important notion in this context is ``renormalization'': Under the assumption
that the zero-dimensional theory has a physical meaning, not only the
necessity of renormalization could be demonstrated in detail but also the
recursive nature of the {\it re}normalization procedure was elaborated
explicitly when calculating the bare parameters in terms of the physical
quantities order by order in the loop expansion. The respective considerations
are not perturbed by infinities due to the fact that in zero dimensions
there do not exist divergencies at all. Thus, the present ``field''
theoretical example is well qualified for a study of renormalization
segregated from baffling questions concerning the existence of the objects
under consideration. This whole complex was discussed at the end of 
chapter~2 (with regard to a model without SSB) and at the end of chapter~3.\\
Last but not least, this thesis was destined to contribute a first tentative
step on the challenging way towards what could be called a really 
noncommutative quantization of models obtained via the model building within
Noncommutative Geometry: In fact, the simple model of SSB in question can
equivalently well be formulated within a special class of noncommutative
models, namely models of the Marseille-Mainz type. In doing so, the generalized
gauge potential $\cA$ becomes the basic object which, in general, unifies
both ordinary gauge fields and Higgs fields into one noncommutative quantity.
This generalized gauge potential should also represent the basis of the
formulation when moving on to the quantization of the model. For these
purposes, in chapter~4 a special matrix calculus was developed which allows
to perform calculations completely within the noncommutative setting 
while $\cA$ is treated as one single variable. The benefit of such a strategy 
lies in a substantial simplification when determining the higher orders 
of the generating functional of 1 PI Green's functions, see 
chapters~4 and~5.\\[.3cm]
Let us now come to some problems which are still open. The investigations
of chapters~4 and~5 clearly pointed out that all calculations in question
become the simpler the more the noncommutative structure is exploited.
In our context this means that we should try to formulate everything, i.e.
the whole theory including gauge fixing and ghost contributions,
within the language of matrices in an uniform way. With regard to the
unphysical ghost fields, so far this goal could not be achieved in a 
satisfactory manner as was discussed at length in section~5.1. Thus,
there is still hope for further improvement when future work referring to
this problem should lead to success.\\
Another open question concerns the incorporation of ordinary gauge fields.
Up to now, only Higgs fields are present in our zero-dimensional model.
Unliterally spoken, it is obvious that for a zero-dimensional
model including both Higgs fields {\it and} ordinary gauge fields just one
point in spacetime will not suffice: Necessarily, the underlying ``spacetime''
must consist of at least two points because otherwise the notion of connection
(gauge field) is meaningless and, in particular, vertices involving derivatives
of the gauge field(s) cannot be taken into account as they should. Work in
this regard dealing with the zero-dimensional version of the Abelian Higgs
model is in progress, see \cite{CP} and \cite{HP}.\\
Subsequently, also the inclusion of matter fields in zero dimensions will have
to be discussed. The solution to this problem, however, will be 
straightforward except for possible technical complications.\\
A proper, uniform and satisfying treatment certainly will require a
further generalization of the existing matrix calculus developed above.
In fact, such a generalization has to allow for various aspects: Due to the
presence of ordinary gauge fields the gauge potential $\cA$ no longer will
be a purely odd matrix but rather a mixture of an even and an odd part. An
analogous statement must hold true for the derivative with respect to $\cA$.
Further on, because of the existence of more than one point in spacetime there
possibly will be several gauge potentials $\cA_i$ (one $\cA_i$ for each point),
i.e. several matrix-valued variables within the game entailing a modification
of the existing chain rule of matrix calculus. At last, the matrix calculus
should be generalized to the case of matrices of higher dimensions in order to
be able to treat also theories with gauge groups different from $U(1)$ like,
for example, the standard model of electroweak interactions. One of the main
obstacles to overcome when heading for such generalizations is the property
that the square of the gauge potential in our simple model of
SSB is proportional to the
unit matrix. This property is essential for the rules of the matrix calculus
to hold true in the form given above.\\
Finally, let us briefly comment on the long-term objective of considerations
of this kind: Suppose that all these questions concerning the generalizations
just outlined will have been settled satisfactorily, then we eventually
will be able to disengage from the case of zero dinemsions thus leaving
the realm of counting of Feynman diagrams and to address ourselves to the
study of really physical theories in four dimensions by translating the
matrix calculus - which by then will be a general calculus for derivatives
of matrix-valued quantities well approved in zero dimensions - to a calculus
of {\it functional} derivatives. In other words, by means of such a 
{\it functional} matrix calculus it hopefully might be possible to approach
the problem of what a really noncommutative quantization of physical gauge
theories within Noncommutative Geometry could be. Of course, in order to 
reach this long-term objective quite some work remains to be done in the
future.

%% file: appa.tex
\newpage
\newsection{Kummer's equation and the function ${}_1 F_1$}

Kummer's differential equation is given by
\be \label{kummer}
z \; y'' (z) \; + \; (c - z) \; y' (z) \; - \; a \; y(z) \; = \; 0 \; \; \; .
\ee
This differential equation has a regular singularity at $z = 0$ and
an irregular singularity at $z = \infty$. If $c$ is not an integer
$(c \notin \Z)$, then the solutions of (\ref{kummer}) read
\be \label{solkum}
y (z) = c_1 \; {}_1 F_1 (a; c; z) + c_2 \; {}_1 F_1 (a - c + 1; 2 - c; z)
\; \; \; .
\ee
Here ${}_1 F_1 (a; c; z)$ denotes, as usual, the confluent hypergeometric
function which can be represented as a power series in $z$ (convergent
for all $z$) as follows:
\ba \label{f11}
{}_1 F_1 (a; c; z) & = & \sum_{n = 0}^\infty 
                         \frac{(a)_n}{(c)_n \; n!} \; z^n \\[.5ex]
\mbox{with } (a)_n & = & \frac{\Gamma (a + n)}{\Gamma (a)}
                         \; \; \mbox{ (Pochhammer's symbol)} \nonumber
\ea
We list some important properties of ${}_1 F_1$:
\ba \label{prop1}
{}_1 F_1 (a; c; z) & = & e^z \; {}_1 F_1 (c - a; c; - z) \\
\label{prop2}
a \; {}_1 F_1 (a + 1; c + 1; z) & = & c \; {}_1 F_1 (a; c; z) \; + \;
(a - c) \; {}_1 F_1 (a; c + 1; z) \\
\label{prop3}
\frac{d^n}{d z^n} \; {}_1 F_1 (a; c; z) & = &
\frac{(a)_n}{(c)_n} \; {}_1 F_1 (a + n; c + n; z)
\ea
Further properties of ${}_1 F_1$ can be found, for example, in \cite{AS}.\\
If $|z|$ is large $(|z| \rightarrow \infty)$, ${}_1 F_1$ possesses the
following asymptotic expansion:
\ba \label{f11asy}
{}_1 F_1 (a; c; z) & \stackrel{|z| \rightarrow \infty}{\longrightarrow} &
\frac{\Gamma (c)}{\Gamma (c - a)} \; e^{i \pi a} \; z^{- a}
\sum_{r = 0}^N (a)_r (1 + a - c)_r \frac{(- z)^{- r}}{r!} 
\; + \; {\cal O} (|z|^{- N - 1}) \nonumber \\[.5ex]
& & + \; \frac{\Gamma (c)}{\Gamma (a)} \; e^z \; z^{a - c}
\sum_{r = 0}^M (c - a)_r (1 - a)_r \frac{z^{- r}}{r!} \; 
+ \; {\cal O} (|z|^{- M - 1})
\ea
(This expansion is valid for $- \frac{\pi}{2} <$ arg$z$ $< \frac{3 \pi}{2}$;
a similar expansion with $e^{i \pi a}$ in the first line replaced 
by $e^{- i \pi a}$ holds for $- \frac{3 \pi}{2} <$ arg$z$ 
$< - \frac{\pi}{2}$.)\\
Among the many relations between ${}_1 F_1$ and other well-known analytical
functions there is one relation connecting ${}_1 F_1$ to Weber's
parabolic cylinder function $U$:
\ba \label{f11u}
U (a, \pm x ) & = &
\frac{\sqrt{\pi} \; 2^{- \frac{1}{4} - \frac{1}{2} a}}{\Gamma 
\left( \frac{3}{4} + \frac{1}{2} a \right) } \;
e^{- \frac{1}{4} x^2} \; {}_1 F_1 
\left( \frac{1}{2} a + \frac{1}{4}; \frac{1}{2}; \frac{1}{2} x^2 \right)
\nonumber \\[.5ex]
& & \mp \; \frac{\sqrt{\pi} \; 2^{\frac{1}{4} - \frac{1}{2} a}}{\Gamma
\left( \frac{1}{4} + \frac{1}{2} a \right) } \; x \; 
e^{- \frac{1}{4} x^2} \; {}_1 F_1
\left( \frac{1}{2} a + \frac{3}{4}; \frac{3}{2}; \frac{1}{2} x^2 \right)
\ea
$U (a, x)$ is one of the standard solutions of the differential
equation
\be \label{parab}
y'' (x) \; - \left( \frac{1}{4} x^2 \; + \; a \right) y (x) \; = \; 0
\; \; \; ;
\ee
for further details see \cite{AS}. Please note that with the help of
(\ref{loesintnfa}) and (\ref{loesintnapp}) we obtain the following
formula for (the asymptotic behaviour\footnote{(\ref{loesintnapp}) only makes
sense for $\lambda$ being small. Hence in this case $\frac{1}{\lambda}$ in
(\ref{loesintnfa}) tends to infinity.} of) $U(\frac{n}{2}, x)$,
\be \label{uasy}
U(\frac{n}{2}, x) = \frac{n!!}{n!} \; x^{- \frac{n + 1}{2}} \;
\mbox{exp}\{ - \frac{x^2}{4} \} \; 
\sum_{p = 0}^\infty \frac{(-1)^p}{(4!)^p}
\frac{(n + 4 p)!}{(n + 4 p)!!} \frac{3^p}{p!} x^{- 2 p} \; \; \; ,
\ee
which in the second part of appendix C is of some relevance.\\
Another important relation that we would like to mention connects ${}_1 F_1$
to the Hermite polynomials $H_n$:
\be \label{fherm}
H_{2m} (x) \; = \; (-1)^m \; \frac{(2m)!}{m!} \; 
{}_1 F_1 (- m; \frac{1}{2}; x^2 )
\ee

%% file: appb.tex
\newsection{Integrals with Gaussian weights}
\label{appendb}

In this appendix we collect some formulae frequently used in the
main text, especially in chapter~2, concerning integrals 
with Gaussian weights.\\
The Gaussian integral in $n$ variables $x_1, \dots, x_n$,
\begin{equation}
\label{gaussd}
  I_G(A,b) = \int_{- \infty}^{+ \infty} d x_1 \cdots d x_n \;
  e^{- \sum_{i,j = 1}^n \frac{1}{2}
  x_i A_{ij} x_j + \sum_{i = 1}^n b_i x_i} \; \; \; ,
\end{equation}
is defined for $A = (A_{ij})$ being a symmetric matrix with eigenvalues
$\lambda_i$ satisfying
\begin{displaymath}
  \mbox{Re }(\lambda_i) \geq 0 \; \; \; , \; \; \; \lambda_i \neq 0
  \; \; \; ,
\end{displaymath}
and its value is given by:
\begin{equation}
\label{gauss}
  I_G(A,b) = (2 \pi)^\frac{n}{2} \;
  (\mbox{det } A)^{- \frac{1}{2}} \;
  e^{\sum_{i,j = 1}^n \frac{1}{2} b_i (A^{-1})_{ij} b_j}
\end{equation}
Especially in the case of just one variable $x_1 \equiv x$ we have:
\begin{equation}
\label{gauss0}
  \int_{- \infty}^{+ \infty} d x \; e^{- \frac{1}{2} a x^2 \; + \;
  b x}
  = \sqrt{\frac{2 \pi}{a}} \; e^{\frac{b^2}{2 a}}
\end{equation}
For the perturbative calculation of $n$-point functions 
in an intermediate step the following integrals are needed:
\begin{eqnarray}
\label{gaussn}
  I_{G, n} (a) := \int_{- \infty}^{+ \infty} d x \; 
  \displaystyle\frac{x^n}{n!} \;
  e^{- \frac{1}{2} a x^2}
  & = & \frac{1}{n!} \; \left( \frac{2}{a} 
        \right)^\frac{n+1}{2} \; 
  \Gamma (\frac{n+1}{2}) \\[.5ex]
  & = & \frac{1}{n!!} \; \sqrt{2 \pi} \; a^{- \frac{n + 1}{2}} \nonumber
\end{eqnarray}
This result holds true for $n$ even; if $n$ is odd, $I_n \equiv 0$ which
is easily seen from the symmetry consideration $x \rightarrow -x$.\\
More precisely, in zero-dimensional $\varphi^4$-theory (chapter~2) 
the $n$-point functions $G_n (\lambda, m^2)$ read
according to (\ref{npoint}) and (\ref{zzznorm})
\be \label{nppdef}
G_n (\lambda, m^2) = \frac{I_n (\lambda, m^2)}{I_0 (\lambda, m^2)}
\ee
with $I_n (\lambda, m^2)$ given by (\ref{intn}):
\be \label{intna}
I_n (\lambda, m^2) = \int_{- \infty}^{+ \infty} \; d y \; y^n \;
\mbox{exp} \left\{ \frac{1}{\hbar} \left(
- \frac{1}{2} m^2 y^2 - \frac{\lambda}{4!} y^4 \right) \right\}
\ee
Hence, in the expansion of $G_n (\lambda, m^2)$
in a power series in the coupling constant $\lambda$ the quantity
\begin{displaymath}
  \left. \frac{d^p I_n (\lambda, m^2)}{d \lambda^p}
  \right|_{\lambda = 0}
\end{displaymath}
will appear frequently. This quantity can be calculated using (\ref{gaussn}):
\ba \label{intnent}
\left. \frac{d^p I_n}{d \lambda^p} \right|_{\lambda = 0} & = &
\frac{(-1)^p}{(4!)^p} \; \hbar^{- p} \; (n + 4p)! \;
\int_{- \infty}^{+ \infty} d y \;
\frac{y^{n + 4p}}{(n + 4p)!} \; e^{- \frac{1}{2} \frac{m^2}{\hbar} y^2} 
\nonumber \\
& = & \frac{(-1)^p}{(4!)^p} \; \hbar^{- p} \; (n + 4p)! \;
      I_{G, n + 4p} (\frac{m^2}{\hbar}) \nonumber \\
& = & \sqrt{2 \pi} \left( \frac{\hbar}{m^2} \right)^\frac{n + 1}{2} \;
      \frac{(-1)^p}{(4!)^p} \; \frac{(n + 4p)!}{(n + 4p)!!} \;
\left( \frac{\hbar}{m^4} \right)^p
\ea
Collecting formulae we end up with the final 
perturbative result for
$G_n (\lambda, m^2)$:
\ba \label{npointp}
G_n (\lambda, m^2) & = & 
\frac{I_n (\lambda, m^2)}{I_0 (\lambda, m^2)} \nonumber \\[.3ex]
& = & \left( \frac{\hbar}{m^2} \right)^{\! \frac{n}{2}}
\frac{\displaystyle\sum_{p = 0}^\infty \displaystyle\frac{(-1)^p}{(4!)^p} 
      \frac{(n + 4p)!}{(n + 4p)!!} \left( \frac{\hbar}{m^4} 
      \right)^p \frac{\lambda^p}{p!}}{1 +
\displaystyle\sum_{q = 1}^\infty 
\displaystyle\frac{(-1)^q}{(4!)^q} \frac{(4q)!}{(4q)!!}
\left( \frac{\hbar}{m^4} \right)^q \frac{\lambda^q}{q!}} \\[.5ex]
\label{loesintnapp}
\mbox{with } \; I_n (\lambda, m^2) & = &
\sqrt{2 \pi} \left( \frac{\hbar}{m^2} \right)^\frac{n + 1}{2}
\sum_{p = 0}^\infty \frac{(-1)^p}{(4!)^p} 
\frac{(n + 4p)!}{(n + 4p)!!} \left( \frac{\hbar}{m^4} \right)^p
\frac{\lambda^p}{p!}
\ea

%% file: appc.tex
\newsection{Determination of $\Gamma$ for $\varphi^4$-theory in zero
            dimensions}

The following {\it mathematica} routine recursively determines the
loop expansion of the generating functional $\Gamma$ up to a prescribed
order {\sf maxord} ({\sf maxord} = 6 in the example below) in the case
of zero-dimensional $\varphi^4$-theory:\\[.5cm]
{\sf 
Clear[$\Gamma$, term1, term2];\\
{\small (* Definition of the depth of recursion *)}\\
maxord = 6;\\
{\small (* Definition of the input: $\Gamma^{(0)} = \Gamma$[0,0],
$\Gamma^{(1)} = \Gamma$[1,0] and derivatives *)}\\
$\Gamma$[0,0] := $\frac{1}{2} m^2 \varphi^2 + \frac{1}{24} \lambda
\varphi^4$;\\
$\Gamma$[0,i\_] := D[$\Gamma$[0,0],\{$\varphi$,i\}];\\
$\Gamma$[1,0] := $\frac{1}{2}$ Log[$(m^2 + \frac{1}{2} \lambda 
\varphi^2)/m^2$];\\
$\Gamma$[1,i\_] := D[$\Gamma$[1,0],\{$\varphi$,i\}];\\
{\small (* Definition of auxiliary quantities *)}\\
term1[0,2] := ($\Gamma$[0,2]$)^2$;\\
term1[n\_,2] := 1/(n $\Gamma$[0,2]) Sum[(3 k - n) $\Gamma$[k,2]
term1[n - k,2],\{k,1,n\}];\\
term1[0,3] := ($\Gamma$[0,2]$)^3$;\\
term1[n\_,3] := 1/(n $\Gamma$[0,2]) Sum[(4 k - n) $\Gamma$[k,2]
term1[n - k,3],\{k,1,n\}];\\
term2[n\_] := Sum[$\Gamma$[n - k,1] term1[k,3],\{k,0,n\}];\\
{\small (* Differential equation to be solved iteratively *)}\\
dgl[n\_] := $- \frac{1}{6} \; \Gamma$[n - 2,3] + $\frac{\lambda}{2} \; \varphi$
term1[n - 1,2] + $\Gamma$[0,1] term1[n,3] - term2[n];\\
{\small (* Recursive calculation of $\Gamma$ *)}\\
For[q = 2,q $\leq$ maxord,q++,\{\\
\hspace*{1cm} $\Gamma$[q,1] = $\Gamma$[q,1]/.Simplify[Solve[dgl[q] == 0,
$\Gamma$[q,1]]];\\
\hspace*{1cm} help[$\varphi$\_] := $\Gamma$[q,1][[1]];\\
\hspace*{1cm} $\Gamma$[q,0] = 
w[$\varphi$]/.Simplify[DSolve[\{D[w[$\varphi$],\{$\varphi$,1\}] == 
help[$\varphi$],\\
\hspace*{7cm} w[0] == 0\},w[$\varphi$],$\varphi$]][[1]];\\
\hspace*{1cm} $\Gamma$[q,2] = D[$\Gamma$[q,0],\{$\varphi$,2\}];\\
\hspace*{1cm} $\Gamma$[q,3] = D[$\Gamma$[q,0],\{$\varphi$,3\}];\\
\hspace*{1cm} Clear[w]\}];
}
\par
\vspace*{.5cm}
\noindent
By means of the above routine we find in addition to (\ref{g0}) --
(\ref{g3}) the expressions for $\Gamma^{(n)}$, $n = 4, 5, 6$, listed
below:\\[-1.3cm]
\par
{\footnotesize
\ba
\Gamma^{(4)} & = & - \left( \lambda^4 \varphi^2 (
176640 m^{16} + 113280 m^{14} \lambda \varphi^2 
+ 175808 m^{12} \lambda^2 \varphi^4 + 133056 m^{10} \lambda^3 \varphi^6
+ 66528 m^8 \lambda^4  \varphi^8 \right. \nonumber \\
& & \left. + 22176 m^6 \lambda^5 \varphi^{10} 
+ 4752 m^4 \lambda^6 \varphi^{12} + 594 m^2 \lambda^7 \varphi^{14}
+ 33 \lambda^8 \varphi{16}) \right)/(288 m^{12} (2 m^2 + 
\lambda \varphi^2)^9) \\[.5ex]
\Gamma^{(5)} & = & \left( \lambda^5 \varphi^2 (
1280000 m^{22} + 145920 m^{20} \lambda \varphi^2 
+ 203673 m^{18} \lambda^2 \varphi^4 + 2164096 m^{16} \lambda^3 \varphi^6
\right. \nonumber \\
& & \left. + 1723392 m^{14} \lambda^4 \varphi^8 
+ 1005312 m^{12} \lambda^5 \varphi^{10} 
+ 430848 m^{10} \lambda^6 \varphi^{12} 
+ 134640 m^8 \lambda^7 \varphi^{14} \right. \nonumber \\
& & \left. + 29920 m^6 \lambda^8 \varphi^{16} 
+ 4488 m^4 \lambda^9 \varphi^{18}
+ 408 m^2 \lambda^{10} \varphi^{20} + 17 \lambda^{11} \varphi^{22}
\right) )/(72 m^{16} (2 m^2 + \lambda \varphi^2)^{12} ) \\[.5ex]
\Gamma^{(6)} & = & - \left( \lambda^6 \varphi^2 (
1765376000 m^{28} - 14837760000 m^{26} \lambda \varphi^2
+ 4835020800 m^{24} \lambda^2 \varphi^4
+ 5159280640 m^{22} \lambda^3 \varphi^6 \right. \nonumber \\
& & \left. + 5701576704 m^{20} \lambda^4 \varphi^8
+ 475 8673920 m^{18} \lambda^5 \varphi^{10}
+305 914 7520 m^{16} \lambda^6 \varphi^{12} \right. \nonumber \\
& & \left. + 1529573760 m^{14} \lambda^7 \varphi^{14}
+ 594834240 m^{12} \lambda^8 \varphi^{16}
+ 178450272 m^{10} \lambda^9 \varphi^{18}
+ 40556880 m^8 \lambda^{10} \varphi^{20} \right. \nonumber \\
& & \left. + 6759480 m^6 \lambda^{11} \varphi^{22}
+ 779940 m^4 \lambda^{12} \varphi^{24}
+ 55710 m^2 \lambda^{13} \varphi^{26} \right. \nonumber \\
& & \left. + 1857 \lambda^{14} \varphi^{28} \right) )/(2880 m^{20}
(2 m^2 + \lambda \varphi^2)^{15})
\ea
}
\par
\vspace*{-.5cm}
\noindent
The preceding considerations allowed for a recursive determination of
$\Gamma^{(n)}$ order by order in the loop expansion. Alternatively,
it is also possible to obtain closed analytical results for each of
the 1 PI $n$-point functions $\Gamma_n$, if $n$ is even. (For $n$ odd,
we have $\Gamma_n \equiv 0$, see above.) This is achieved in the 
following manner:\\
Starting with the expression (\ref{loeszl}) for $Z(j)$ the transition
to $W(j)$ according to (\ref{zw}) immediately yields:
\be \label{loeswl}
W(j) \; = \; \hbar \; \mbox{ln }
\displaystyle\frac{\sum_{n = 0}^\infty 
\displaystyle\frac{(-1)^n}{(4!)^n} \left( \frac{\hbar}{m^4} \right)^n
\frac{(4n)!}{(4n)!!} \; {}_1 F_1 (2 n + \frac{1}{2}; \frac{1}{2};
\frac{j^2}{2 \hbar m^2} ) \; \frac{\lambda^n}{n!}}{\sum_{p = 0}^\infty
\displaystyle\frac{(-1)^p}{(4!)^p} \left( \frac{\hbar}{m^4} \right)^p
\frac{(4p)!}{(4p)!!} \; \frac{\lambda^p}{p!}}
\ee
Hence, $\varphi (j)$ defined in (\ref{legendrey}) is given by
\ba \label{altgam}
\varphi (j) & = & \frac{j}{m^2} \; f(j) \\[.5ex]
\label{deffj}
\mbox{with } f(j) & = & 
\displaystyle\frac{\sum_{n = 0}^\infty
\displaystyle\frac{(-1)^n}{(4!)^n} \left( \frac{\hbar}{m^4} \right)^n
\frac{(4n + 1)!}{(4n)!!} \; {}_1 F_1 (2n + \frac{3}{2}; \frac{3}{2};
\frac{j^2}{2 \hbar m^2} ) \; \frac{\lambda^n}{n!}}{\sum_{p = 0}^\infty
\displaystyle\frac{(-1)^p}{(4!)^p} \left( \frac{\hbar}{m^4} \right)^p
\frac{(4p)!}{(4p)!!} \; {}_1 F_1 (2n + \frac{1}{2}; \frac{1}{2};
\frac{j^2}{2 \hbar m^2} ) \; \frac{\lambda^p}{p!}} \; \; \; ,
\ea
where (\ref{prop3}) has been used. Differentiating (\ref{altgam}) once
with respect to $\varphi$,
\begin{displaymath}
\left. \varphi \; = \; \frac{j(\varphi )}{m^2} \; f(j(\varphi ))
\; \; \right| \; \frac{\partial}{\partial \varphi} \; \; \; ,
\end{displaymath}
leads to
\begin{displaymath}
1 \; = \; \frac{1}{m^2} \frac{\partial j}{\partial \varphi} f \; + \;
\frac{1}{m^2} j \frac{\partial j}{\partial \varphi}
\frac{\partial f}{\partial j} \; \; \; .
\end{displaymath}
Evaluating the last equation at $\varphi = 0$ and using
\begin{displaymath}
\left. \frac{\partial j}{\partial \varphi} \right|_{\varphi = 0} =
\Gamma_2 \; \; , \; \; j = 0 \Leftrightarrow \varphi = 0 \; \; , \; \;
{}_1 F_1 (\star; \star; 0) = 1
\end{displaymath}
determines $\Gamma_2$:
\be \label{twoptpre}
\Gamma_2 = m^2 \displaystyle\frac{\sum_{n = 0}^\infty
\displaystyle\frac{(-1)^n}{(4!)^n} \left( \frac{\hbar}{m^4} \right)^n
\frac{(4n)!}{(4n)!!} \; \frac{\lambda^n}{n!}}{\sum_{p = 0}^\infty
\displaystyle\frac{(-1)^p}{(4!)^p} \left( \frac{\hbar}{m^4} \right)^p
\frac{(4p + 1)!}{(4p)!!} \; \frac{\lambda^p}{p!}}
\ee
Finally, with the help of (\ref{uasy}) $\Gamma_2$ can be rewritten in a
simpler form:
\be \label{twoptf}
\Gamma_2 = \sqrt{\frac{\lambda \hbar}{3}} \; 
\displaystyle\frac{U(0, m^2 \sqrt{\frac{3}{\lambda \hbar}} )}{U(1,
m^2 \sqrt{\frac{3}{\lambda \hbar}} )}
\ee
From here on, the higher 1 PI $n$-point functions $\Gamma_n$ are
recursively obtained by appropriate differentiations of
(\ref{diffg}) with respect to $\varphi$ and subsequent evaluation
at $\varphi = 0$. For example, $\Gamma_4$ results from just one
differentiation of (\ref{diffg}) with respect to $\varphi$ and the
fact that $\Gamma_n \equiv 0$ for $n$ odd:
\be \label{fourptdet}
- \hbar^2 \; \frac{\lambda}{6} \; \Gamma_4 \; + \;
\hbar \; \frac{\lambda}{2} \; (\Gamma_2)^2 \; + \;
m^2 \; (\Gamma_2)^3 \; = \; (\Gamma_2)^4
\ee

%% file: appd.tex
\newsection{Determination of $\Gamma_{ren}$ for 
               zero-dimensional $\varphi^4$-theory}

In the main text (see section~2.3) the first few orders of the loop
expansion of the renormalized action $\Gamma_{ren}$ in the case
of $\varphi^4$-theory in zero dimensions were obtained by first calculating
recursively the unrenormalized action $\Gamma = \sum_{n = 0}^\infty
\hbar^n \Gamma^{(n)}$ according to appendix~C and then by adjusting order
by order in the loop expansion appropriate countertrems to the classical
action $S(\varphi ) \equiv \Gamma^{(0)} (\varphi )$ in order to fulfill the
normalization conditions (\ref{normcm}) and (\ref{normcg}) which
establish the connection between theory and experiment (under the
assumption that zero-dimensional $\varphi^4$-theory is to
be a {\it physical} theory). There is, however, an alternative possibility
for the determination of the loop expansion of $\Gamma_{ren}$ avoiding
the long way round the unrenormalized action $\Gamma$: Namely, we can
directly insert the $\hbar$-expansions of $m^2$ and $\lambda$
(\ref{parpow}) into the differential equation (\ref{diffg}) for
$\Gamma$ and subsequently evaluate this differential equation consistently
order by order in $\hbar$. Of course, both procedures lead to the
same results. The second possibility is automatized in the following
little {\it mathematica} routine yielding $\Gamma_{ren}, m^2$ ({\sf
msq} in the program) and
$\lambda$ ({\sf lam} below) up to an arbitrary prescribed 
order {\sf maxord} of the loop expansion.\\[.5cm]
\newpage
{\sf
Clear[$\Gamma_{ren}$, lam, msq, term1, term2, term3, term4, term5];\\
{\small (* Definition of the depth of recursion *)}\\
maxord = 6;\\
{\small (*Definition of the input quantities*)}\\
$\Gamma_{ren}$[0,0] := $\frac{1}{2}$ m${}_p^2 \; \varphi^2 +
                  \frac{1}{24}$ g${}_p \; \varphi^4$;\\
$\Gamma_{ren}$[0,i\_] := D[$\Gamma_{ren}$[0,0], \{$\varphi$,i\}];\\
$\Gamma_{ren}$[1,0] := $\frac{1}{2}$ Log[(m${}_p^2 + \frac{1}{2}$ g${}_p \;
                                    \varphi^2$)/m${}_p^2$] -
    $\frac{1}{4}$ g${}_p \; \varphi^2$/m${}_p^2$ +
    $\frac{1}{16}$ g${}_p^2 \; \varphi^4$/m${}_p^4$;\\
$\Gamma_{ren}$[1,i\_] := D[$\Gamma_{ren}$[1,0], \{$\varphi$,i\}];\\
msq[0] := m${}_p^2$;\\
msq[1] := - $\frac{1}{2}$ g${}_p$/m${}_p^2$;\\
lam[0] := g${}_p$;\\
lam[1] := $\frac{3}{2}$ g${}_p^2$/m${}_p^4$;\\
{\small (* Definition of auxiliary quantities *)}\\
term1[n\_,2] := Sum[$\Gamma_{ren}$[k,2] $\Gamma_{ren}$[n - k,2], \{k,0,n\}];\\
term1[n\_,3] := Expand[Sum[$\Gamma_{ren}$[n - k,2] term1[k,2], \{k,0,n\}]];\\
term2[n\_] := Sum[lam[n - k] $\Gamma_{ren}$[k,3], \{k,0,n\}];\\
term3[n\_] := Expand[Sum[lam[n - k] term1[k,2], \{k,0,n\}]];\\
term4[n\_] := Expand[Sum[(msq[n - k] $\varphi + \frac{1}{6}$ 
    lam[n - k] $\varphi^3$) term1[k,3], \{k,0,n\}]];\\
term5[n\_] := Expand[Sum[$\Gamma_{ren}$[n - k,1] term1[k,3],\{k,0,n\}]];\\
{\small (* Differential equation to be solved iteratively *)}\\
dgl[n\_] := Expand[- $\frac{1}{6}$ term2[n - 2] + $\frac{1}{2} \;
    \varphi$ term3[n - 1] + term4[n] - term5[n]];\\
{\small (* Recursive calculation of $\Gamma_{ren}$, msq, lam *)}\\
For[q = 2, q $\leq$ maxord, q++, \{\\
\hspace*{1cm} $\Gamma_{ren}$[q,1] = 
    $\Gamma_{ren}$[q,1]/.Simplify[Solve[dgl[q] == 0,
    $\Gamma_{ren}$[q,1]]][[1]];\\
\hspace*{1cm} help1[$\varphi$\_] = $\Gamma_{ren}$[q,1];\\
\hspace*{1cm} $\Gamma_{ren}$[q,0] = 
    w[$\varphi$]/.Simplify[DSolve[\{D[w[$\varphi$], \{$\varphi$,1\}]
    == help1[$\varphi$],\\
\hspace*{7cm} w[0] == 0\}, w[$\varphi$], $\varphi$]][[1]];\\
\hspace*{1cm} $\Gamma_{ren}$[q,2] = D[$\Gamma_{ren}$[q,0], \{$\varphi$,2\}];\\
\hspace*{1cm} help2[$\varphi$\_] = $\Gamma_{ren}$[q,2];\\
\hspace*{1cm} msq[q] = msq[q]/.Simplify[Solve[help2[0] == 0,
    msq[q]]][[1]];\\
\hspace*{1cm} $\Gamma_{ren}$[q,3] = D[$\Gamma_{ren}$[q,0], \{$\varphi$,3\}];\\
\hspace*{1cm} help3[$\varphi$\_] = D[$\Gamma_{ren}$[q,0],\{$\varphi$,4\}];\\
\hspace*{1cm} lam[q] = lam[q]/.Simplify[Solve[help3[0] == 0,
    lam[q]]][[1]];\\
\hspace*{1cm} Clear[w, help1, help2, help3]\}];\\
}

%% file: appe.tex
\newpage
\newsection{Naive treatment of the model with SSB}

In order to allow for an explicit control of the changes originating
when enlarging the naive treatment of the model with SSB (see section~3.1)
to the correct one in section~3.2, in this appendix
we present some more computational details as well as some additional
results that have been omitted in the main text.\\
According to the relation (\ref{nssbzw}) between $Z(\rho, \tau )$ and
$W(\rho, \tau )$ the system of differential equations (\ref{nssbdiffz})
transforms into a set of two coupled differential equations for $W$:
\ba \label{nssbdiffw}
& \hbar^2 \displaystyle\frac{\lambda}{4} \left\{
\frac{\partial^3 W}{\partial \rho^3} +
\frac{\partial^3 W}{\partial \rho \partial \tau^2} \right\} +
\hbar \frac{\lambda}{4} \left\{
\left( \frac{\partial W}{\partial \rho} + v \right)
\left( 3 \frac{\partial^2 W}{\partial \rho^2} +
\frac{\partial^2 W}{\partial \tau^2} \right) +
2 \frac{\partial W}{\partial \tau} 
\frac{\partial^2 W}{\partial \rho \partial \tau} \right\} & \nonumber \\
& + \displaystyle\frac{\lambda}{4} \left\{
\left( \frac{\partial W}{\partial \rho} \right)^3 +
\frac{\partial W}{\partial \rho} 
\left( \frac{\partial W}{\partial \tau} \right)^2 +
3 v \left( \frac{\partial W}{\partial \rho} \right)^2 +
v \left( \frac{\partial W}{\partial \tau} \right)^2 \right\} -
2 m^2 \frac{\partial W}{\partial \rho} = \rho & \nonumber \\[1ex]
& \hbar^2 \displaystyle\frac{\lambda}{4} \left\{
\frac{\partial^3 W}{\partial \tau^3} +
\frac{\partial^3 W}{\partial \rho^2 \partial \tau} \right\} +
\hbar \frac{\lambda}{4} \left\{
2 \left( \frac{\partial W}{\partial \rho} + v \right)
\frac{\partial^2 W}{\partial \rho \partial \tau} +
\frac{\partial W}{\partial \tau} \left(
\frac{\partial^2 W}{\partial \rho^2} +
3 \frac{\partial^2 W}{\partial \tau^2} \right) \right\} & \nonumber \\
& + \displaystyle\frac{\lambda}{4} \left\{
\left( \frac{\partial W}{\partial \tau} \right)^3 +
\frac{\partial W}{\partial \tau} \left(
\frac{\partial W}{\partial \rho} \right)^2 +
2 v \frac{\partial W}{\partial \rho}
\frac{\partial W}{\partial \tau} \right\} = \tau &
\ea
Looking for a perturbative determination of the generating functional
$\Gamma$ of 1~PI Green's functions, in a next step the corresponding
differential equations for $\Gamma$ have to be derived. The connection
between $\Gamma (x,y)$ and $W(\rho, \tau )$ being given by a
Legendre transformation (see (\ref{nssbwg}), (\ref{nssbleg})),
\ba \label{nssblega}
& \Gamma (x,y) \; = \; \rho (x,y) \; x \; + \;
\tau (x,y) \; y \; - \; 
W(\rho (x,y), \tau (x,y)) & \nonumber \\[.5ex]
& \mbox{with } \; x \; := \; \displaystyle\frac{\partial W}{\partial \rho} \;
\mbox{ and } \; y \; := \; \displaystyle\frac{\partial W}{\partial \tau} 
\; \; , &
\ea
we are especially in need of the Legendre 
transformations of the various second
and third derivatives of $W(\rho, \tau )$ with respect to $\rho$ and
$\tau$ appearing in (\ref{nssbdiffw}).
These derivatives are most easily obtained by
appropriately differentiating the identities
\ba \label{xequivx}
x(\rho (x,y), \tau (x,y)) & = & x \\ 
\label{yequivy}
\mbox{and } \; \; y(\rho (x,y), \tau (x,y)) & = & y \; \; \; .
\ea
Because these kind of considerations will frequently bother us also later
on it is perhaps useful to give in the present simple example some more
details on the calculation which hopefully enlighten the conceptual
points without being spoiled by ugly technical complications showing up
in the later correct treatment.\\
For instance, differentiating (\ref{xequivx}) once with respect to
$x$ and $y$, respectively, yields
\ba \label{nssbex1}
1 & = & \frac{\partial \rho}{\partial x} \frac{\partial x}{\partial \rho} +
\frac{\partial \tau}{\partial x} \frac{\partial x}{\partial \tau} \; = \;
\frac{\partial^2 \Gamma}{\partial x^2}
\frac{\partial^2 W}{\partial \rho^2} +
\frac{\partial^2 \Gamma}{\partial x \partial y}
\frac{\partial^2 W}{\partial \rho \partial \tau} \; \; \; , \nonumber \\
0 & = & \frac{\partial \rho}{\partial y} \frac{\partial x}{\partial \rho} +
\frac{\partial \tau}{\partial y} \frac{\partial x}{\partial \tau} \; = \;
\frac{\partial^2 \Gamma}{\partial x \partial y}
\frac{\partial^2 W}{\partial \rho^2} +
\frac{\partial^2 \Gamma}{\partial y^2}
\frac{\partial^2 W}{\partial \rho \partial \tau} \; \; \; .
\ea
For the second equalities we have used the definition of $x$, 
see (\ref{nssblega}), as
well as the reciprocal relations
\be \label{nssblegre}
\rho = \displaystyle\frac{\partial \Gamma}{\partial x} \;
\mbox{ and } \tau = \displaystyle\frac{\partial \Gamma}{\partial y} \; \; \; .
\ee
From (\ref{nssbex1}) some of the desired Legendre transformations
follow immediately:
\ba \label{nssbex2}
\frac{\partial^2 W}{\partial \rho^2} &
\stackrel{Legendre}{\longrightarrow} &
\hat{h}^{- 1} (x,y) \; \frac{\partial^2 \Gamma}{\partial y^2} \; \; \; , 
\nonumber \\[.5ex]
\frac{\partial^2 W}{\partial \rho \partial \tau} &
\stackrel{Legendre}{\longrightarrow} & - \;
\hat{h}^{- 1} (x,y) \; \frac{\partial^2 \Gamma}{\partial x \partial y}
\ea
$\hat{h} (x,y)$ on the right hand sides of (\ref{nssbex2}) denotes the
determinant of Hesse's matrix for $\Gamma (x,y)$, i.e.
\be \label{nssbh}
\hat{h} (x,y) = \frac{\partial^2 \Gamma}{\partial x^2}
\frac{\partial^2 \Gamma}{\partial y^2} -
\left( \frac{\partial^2 \Gamma}{\partial x \partial y} \right)^2 \; \; \; .
\ee
In an analogous manner (by differentiating (\ref{yequivy}) once with
respect to $x$ and $y$, respect\-ively) one also obtains 
the Legendre transformation of
the second derivative of $W$ with respect to $\tau$:
\ba \label{nssbex3}
\frac{\partial^2 W}{\partial \tau^2} &
\stackrel{Legendre}{\longrightarrow} &
\hat{h}^{- 1} (x,y) \; \frac{\partial^2 \Gamma}{\partial x^2}
\ea
For the Legendre transformations of the third derivatives of $W$ one has
to work a little bit more. In fact, differentiating (\ref{nssbex1}) once
more with respect to $x$ and $y$, respectively, we obtain three independent
equations which read when written in matrix form ($W_{\rho \rho} 
\equiv \frac{\partial^2 W}{\partial \rho^2}$ etc.):
\ba \label{nssbex4}
& \hat{M} \; \vec{\hat{v}}_1 \; = \; \vec{\hat{m}}_1 \; \; \; , & \\[.5ex]
& \vec{\hat{m}}_1 = -
(\Gamma_{xxx} W_{\rho \rho} + \Gamma_{xxy} W_{\rho \tau},
\Gamma_{xxy} W_{\rho \rho} + \Gamma_{xyy} W_{\rho \tau},
\Gamma_{xyy} W_{\rho \rho} + \Gamma_{yyy} W_{\rho \tau} )^T
\; \; \; , & \nonumber \\
& \vec{\hat{v}}_1 = 
(W_{\rho \rho \rho}, W_{\rho \rho \tau}, W_{\rho \tau \tau} )^T 
\; \; \; , & \nonumber \\[.5ex]
& \hat{M} = \left( \begin{array}{c|c|c}
\Gamma_{xx}^2 & 2 \Gamma_{xx} \Gamma_{xy} & \Gamma_{xy}^2 \\ \hline
\Gamma_{xx} \Gamma_{xy} & \Gamma_{xx} \Gamma_{yy} + \Gamma_{xy}^2 &
\Gamma_{xy} \Gamma_{yy} \\ \hline
\Gamma_{xy}^2 & 2 \Gamma_{xy} \Gamma_{yy} & \Gamma_{yy}^2
\end{array} \right) & \nonumber
\ea
Please note that $\vec{\hat{m}}_1$ is completely 
determined as a function of $x$
and $y$ by means of (\ref{nssbex2}). Inverting (\ref{nssbex4}) three of the
four Legendre transformations still missing result:
\be \label{nssbex5}
W_J \stackrel{Legendre}{\longrightarrow} 
\displaystyle\frac{\hat{K}_J (x,y)}{\hat{h} (x,y) \; \hat{H} (x,y)} \;
\mbox{ for } \; J \in \{ \rho \rho \rho, \rho \rho \tau, 
\rho \tau \tau \}
\ee
In order to technically simplify the recursive determination of 
$\Gamma^{(n)}$, in (\ref{nssbex5}) we have introduced the following
abbreviations: $\hat{H}$ denotes the determinant of $\hat{M}$,
$\hat{H} = $ det$M$, and $\hat{K}_J$ is $\hat{h}$ times the determinant of the
matrix resulting from $\hat{M}$ by replacing the first or the second or the
third row (depending on $J$) by $\vec{\hat{m}}_1$.\\
By elaborating on all possible but independent second derivatives of
(\ref{yequivy}) with respect to $x$ and $y$ we finally also get the
answer for the Legendre transformation of $W_{\tau \tau \tau}$:
\be \label{nssbex6}
W_{\tau \tau \tau} \stackrel{Legendre}{\longrightarrow}
\displaystyle\frac{\hat{K}_{\tau \tau \tau} 
(x,y)}{\hat{h} (x,y) \hat{H} (x,y)}
\ee
$\hat{K}_{\tau \tau \tau}$ is defined to be $\hat{h}$ times the determinant
of the matrix resulting from $\hat{M}$ by replacing the third row by
$\vec{\hat{m}}_2$ where $\vec{\hat{m}}_2$ is given by 
$\vec{\hat{m}}_1$ (\ref{nssbex4}) with
$W_{\rho \rho} \rightarrow W_{\rho \tau}$ and
$W_{\rho \tau} \rightarrow W_{\tau \tau}$.\\
Collecting everything together we are now in the position to write down
the system (\ref{nssbdiffw}) expressed for $\Gamma$:
\ba \label{nssbdiffg}
\hbar^2 \; \frac{\lambda}{4} \left( 
\hat{K}_{\rho \rho \rho} (x,y) + \hat{K}_{\rho \tau \tau} 
(x,y) \right) & & \nonumber \\
+ \; \hbar \; \frac{\lambda}{4} \; \hat{H} (x,y) \left[
(x + v) \left( 3 \Gamma_{yy} + \Gamma_{xx} \right) -
2 y \Gamma_{xy} \right] & & \nonumber \\
+ \; \hat{h} (x,y) \; \hat{H} (x,y) \left[
\frac{\lambda}{4} (x^3 + x y^2 + 3 v x^2 + v y^2 ) - 2 m^2 x \right] & = &
\hat{h} (x,y) \hat{H} (x,y) \frac{\partial \Gamma}{\partial x}
\nonumber \\[.5ex]
\hbar^2 \; \frac{\lambda}{4} \left(
\hat{K}_{\tau \tau \tau} (x,y) + \hat{K}_{\rho \rho \tau} (x,y) \right) & & \\
+ \; \hbar \; \frac{\lambda}{4} \; \hat{H} (x,y) \left[
- 2 (x + v) \Gamma_{xy} + y \left(
\Gamma_{yy} + 3 \Gamma_{xx} \right) \right] & & \nonumber \\
+ \; \frac{\lambda}{4} \; \hat{h} (x,y) \; \hat{H} (x,y) (
y^3 + x^2 y + 2 v x y ) & = &
\hat{h} (x,y) \hat{H} (x,y) \frac{\partial \Gamma}{\partial y} \nonumber
\ea
Looking for the recursive solution of (\ref{nssbdiffg}) in zeroth order
we fall back to the classical action $S(x,y)$ (\ref{acssb0}), i.e.
$\Gamma^{(0)} = S(x,y)$. The 1-loop approximation $\Gamma^{(1)}$ of
$\Gamma$ was already derived in the main text, see (\ref{nssbg1}). Despite
the fact that $\Gamma^{(1)}$ is ill-defined due to the ``infrared
problem'' related to the Goldstone boson $y$, we also present here for
later convenience the solution of (\ref{nssbdiffg}) in second order
of the loop expansion. For instance, evaluating the first equation
in (\ref{nssbdiffg}) consequently in order $\hbar^2$ and partly using
the results for zero loops and one loop, yields
\be \label{nssbex7}
\Gamma_x^{(2)} = \frac{1}{\hat{h}^{(0)}} \left\{
\frac{\lambda}{4} 
\frac{\hat{K}_{\rho \rho \rho}^{(0)} + 
\hat{K}_{\rho \tau \tau}^{(0)}}{\hat{H}^{(0)}} +
\frac{\lambda}{4} \left[
(x + v) \left( 3 \Gamma_{yy}^{(1)} + \Gamma_{xx}^{(1)} \right) -
2 y \Gamma_{xy}^{(1)} \right] -
\hat{h}^{(1)} \Gamma_x^{(1)} \right\} \; \; .
\ee
According to (\ref{nssbh}) $\hat{h}^{(1)}$ is given by:
\be \label{nssbex8}
\hat{h}^{(1)} (x,y) = \Gamma_{yy}^{(0)} \Gamma_{xx}^{(1)} +
\Gamma_{xx}^{(0)} \Gamma_{yy}^{(1)} -
2 \Gamma_{xy}^{(0)} \Gamma_{xy}^{(1)}
\ee
A further equation analogous to (\ref{nssbex7}) but this time describing
the dependence of $\Gamma^{(2)}$ on \nolinebreak $y$ 
is obtained by examining the
second equation in (\ref{nssbdiffg}) in order $\hbar^2$ in a similar
manner. Finally, we end up with:\\[-1cm]
\par
{\footnotesize
\ba \label{nssbg2}
\Gamma^{(2)} & = & \frac{4}{\lambda
\left[ 8 m^2 (2 v x + x^2 + 3 y^2) + \lambda \left( 4 v^2 (- 3 x^2 + y^2) - 
12 v x (x^2 + y^2) - 3 (x^2 + y^2)^2 \right) \right]^3} 
\cdot \nonumber \\[.5ex]
& & \cdot \left\{ 384 m^6 (2 v x + x^2 - y^2) \right. \nonumber \\
& & \left. + 16 \lambda m^4 \left( x (2 v + x) (4 v^2 - 50 v x - 25 x^2) - 
6 (4 v^2 + 6 v x + 3 x^2) y^2 - 9 y^4 \right) \right. \nonumber \\
& & \left. + 16 \lambda^2 m^2 \left( 54 v x (x^2 + y^2)^2 + 9 (x^2 + y^2)^3 - 
4 v^4 (3 x^2 + y^2) \right. \right. \nonumber \\
& & \hspace*{2cm} \left. \left. + 4 v^3 (15 x^3 + 7 x y^2) + 
v^2 (105 x^4 + 122 x^2 y^2 + 9 y^4) \right) \right. \nonumber \\
& & \left. + \lambda^3 \left( 96 v^5 x (x^2 + y^2) - 
24 v^3 x (17 x^2 - 3 y^2) (x^2 + y^2)
- 12 v^2 (29 x^2 - y^2) 
(x^2 + y^2)^2 \right. \right. \nonumber \\ 
& & \hspace*{2cm} \left. \left. - 120 v x (x^2 + y^2)^3 - 
15 (x^2 + y^2)^4 - 32 v^4 (3 x^4 - 6 x^2 y^2 - y^4) \right) \right\} 
\nonumber \\
& & + \;  C^{(2)}
\ea}
\par
\vspace*{-.5cm}
\noindent
Higher orders of the loop expansion of $\Gamma$ will be determined later on
in the context of the noncommutative approach to the present model
under investigation, see chapter~4.

%% file: appf.tex
\newsection{Correct treatment of the model with SSB}

This appendix is intended to fill some computational gaps left void
in section~3.2 where we considered the recursive determination of $\Gamma$.
As far as possible we will copy the analogous but easier treatment
of appendix~E, thus having the possibility of refering to
that appendix whenever the current calculations copy the
previous ones.\\
Hence, we begin with the system (\ref{ssbdiffz}) of differential equations
translated to the generating functional $W(\rho, \tau, l, \eta,
\overline{\eta}; X, Y)$ of connected Green's functions:\\
\par
{\footnotesize
\ba \label{ssbdiffw}
\hbar^2 \; \displaystyle\frac{\lambda}{4} \left\{
\frac{\partial^3 W}{\partial \rho^3} +
\frac{\partial^3 W}{\partial \rho \partial \tau^2} \right\} & & \nonumber \\
+ \; \hbar \left\{
\displaystyle\frac{\lambda}{4} \left[
\left( \frac{\partial W}{\partial \rho} + v \right) \left(
3 \frac{\partial^2 W}{\partial \rho^2} +
\frac{\partial^2 W}{\partial \tau^2} \right) +
2 \frac{\partial W}{\partial \tau}
\frac{\partial^2 W}{\partial \rho \partial \tau} \right] + \xi m
\frac{\partial^2 W}{\partial \eta \partial \overline{\eta}} \right\}
& & \nonumber \\
+ \; \displaystyle\frac{\lambda}{4} \left[
\left( \frac{\partial W}{\partial \rho} \right)^3 +
\frac{\partial W}{\partial \rho}
\left( \frac{\partial W}{\partial \tau} \right)^2 +
3 v \left( \frac{\partial W}{\partial \rho} \right)^2 +
v \left( \frac{\partial W}{\partial \tau} \right)^2 \right]
- 2 m^2 \frac{\partial W}{\partial \rho} & & \nonumber \\
- \; \xi m \displaystyle\frac{\partial W}{\partial \overline{\eta}} 
\frac{\partial W}{\partial \eta} +
X \frac{\partial W}{\partial \overline{\eta}} & = & \rho \nonumber \\[.5ex]
\hbar^2 \; \displaystyle\frac{\lambda}{4} \left\{
\frac{\partial^3 W}{\partial \tau^3} +
\frac{\partial^3 W}{\partial \rho^2 \partial \tau} \right\}
+ \; \hbar \; \frac{\lambda}{4} \left\{
2 \left( \frac{\partial W}{\partial \rho} + v \right)
\frac{\partial^2 W}{\partial \rho \partial \tau} +
\frac{\partial W}{\partial \tau} \left(
\frac{\partial^2 W}{\partial \rho^2} +
3 \frac{\partial^2 W}{\partial \tau^2} \right) \right\} & & \nonumber \\
+ \; \displaystyle\frac{\lambda}{4} \left[
\left( \frac{\partial W}{\partial \tau} \right)^3 +
\frac{\partial W}{\partial \tau}
\left( \frac{\partial W}{\partial \rho} \right)^2 +
2 v \frac{\partial W}{\partial \rho}
\frac{\partial W}{\partial \tau} \right] \; + \;
\xi m \frac{\partial W}{\partial l} \; - \; 
Y \frac{\partial W}{\partial \overline{\eta}} & = & \tau \nonumber \\[.5ex]
\displaystyle\frac{\partial W}{\partial l} \; + \; \xi m
\frac{\partial W}{\partial \tau} & = & l \\[.5ex]
\hbar \; \xi m 
\displaystyle\frac{\partial^2 W}{\partial \rho \partial \eta}
\; + \; \xi m \frac{\partial W}{\partial \eta} \left(
\frac{\partial W}{\partial \rho} + v \right) \; - \;
Y \frac{\partial W}{\partial \tau} \; + \; 
X \left( \frac{\partial W}{\partial \rho} + v \right) & = & 
\overline{\eta} \nonumber \\[.5ex]
- \; \hbar \; \xi m 
\displaystyle\frac{\partial^2 W}{\partial \rho \partial \overline{\eta}} 
\; - \; \xi m \frac{\partial W}{\partial \overline{\eta}} \left(
\frac{\partial W}{\partial \rho} + v \right) & = & \eta \nonumber
\ea}
\par
\vspace*{-.5cm}
\noindent
(\ref{ssbdiffw}) has to be compared with (\ref{nssbdiffw}) 
in order to study the modifications
brought about by the additional fields $B, c$ and $\overline{c}$.\\
The next step once again consists of performing a Legendre transformation,
this time according to
\ba \label{ssbleg}
& \Gamma (x, y, B, c, \overline{c}; X, Y) = 
\rho (x, y, B, c, \overline{c}; X, Y) x + \tau (\cdots ) y +
l (\cdots ) B + \overline{\eta} (\cdots ) c +
\overline{c} \eta (\cdots ) & \nonumber \\
& \hspace*{1cm} - \; W(\rho (\cdots ), \tau (\cdots ), l (\cdots ),
\eta (\cdots ), \overline{\eta} (\cdots ); X, Y) & \\[.5ex]
& \mbox{with } \; x = \displaystyle\frac{\partial W}{\partial \rho} \; , \;
y = \frac{\partial W}{\partial \tau} \; , \;
B = \frac{\partial W}{\partial l} \; , \;
c = \frac{\partial W}{\partial \overline{\eta}} \mbox{ and } \;
\overline{c} = - \frac{\partial W}{\partial \eta} \; \; . & \nonumber
\ea
In order to obtain explicitly the system (\ref{ssbdiffw}) expressed
for $\Gamma$ there is still lack of the Legendre transformations of the various
second and third derivatives of $W$ with respect to $\rho, \tau, \eta$
and $\overline{\eta}$ appearing in (\ref{ssbdiffw}). In principle, the
determination of those Legendre transformations follows up the line of
argument presented in appendix~E, being, however, more involved due to
the presence of Grassmannian variables. For this reason we are going to
comment briefly on the additional difficulties before stating the results.\\
Because of the quantum numbers of the fields (with respect to $\phi \pi$ and
charge conjugation \nolinebreak $C$) and because of 
the abelian character of the
Grassmannian fields (meaning that the square of each of these fields is
zero) each order of the loop expansion of $\Gamma$ (and hence $\Gamma$
itself) has the decomposition
\be \label{gdecom}
\Gamma (x, y, B, c, \overline{c}; X, Y) =
\Gamma_0 (x, y, B) +
\Gamma_1 (x, y) c \overline{c} +
\Gamma_2 (x, y) c X +
\Gamma_3 (x, y) c Y \; \; .
\ee
An analogous decomposition holds for $x, y$ and $B$ taken as functions of the
sources according to (\ref{ssbleg}) as well as for $W$, e.g.
\be \label{xdecom}
x (\rho, \tau, l, \eta, \overline{\eta}; X, Y) =
x_0 (\rho, \tau, l) +
x_1 (\rho, \tau, l) \eta \overline{\eta} +
x_2 (\rho, \tau, l) \eta X +
x_3 (\rho, \tau, l) \eta Y \; \; .
\ee
Inserting now in (\ref{xdecom}) the sources as functions of the fields and
differentiating the resulting equation for instance with respect to $x$,
on the l.h.s. we evidently obtain 1 which as a number belongs to the
zeroth level $\Lambda^0$ of the Grassmann algebra $\Lambda$, whereas on the
r.h.s. various terms will appear some of which also live in $\Lambda^0$ but
some of which sit in the second level $\Lambda^2$ (namely the ones involving
$x_1, x_2$ and $x_3$ because those terms are multiplied by two
Grassmannian variables). Disentangling this way the resulting equation,
repeating, furthermore, this procedure not only for the derivatives with
respect to $y$ and $B$ but also for the derivatives of the equations
for $y$ and $B$ analogous to (\ref{xdecom}) and also making use of the
gauge fixing condition (\ref{gfcond}) we finally indeed are able to copy the
treatment of appendix~E with the following results ($\frac{\partial^2
W}{\partial \rho^2} \equiv W_{\rho \rho}$ etc.):
\ba \label{ssbex1}
W_{\rho \rho} & \stackrel{Legendre}{\longrightarrow} & h^{- 1} (x,y) \left(
\Gamma_{0_{yy}} - (\xi m)^2 \right) \; + \; 
c\mbox{-dependent term} \nonumber \\[.5ex]
W_{\rho \tau} & \stackrel{Legendre}{\longrightarrow} & -\;  h^{- 1} (x,y) \;  
\Gamma_{0_{xy}} \; + \; c\mbox{-dependent term} \\[.5ex]
W_{\tau \tau} & \stackrel{Legendre}{\longrightarrow} & h^{- 1} (x,y) \;
\Gamma_{0_{xx}} \; + \; c\mbox{-dependent term} \nonumber
\ea
The function $h (x,y)$ is given by (\ref{ssbh}), i.e.
\be \label{ssbha}
h (x,y) = \Gamma_{0_{xx}} \left(
\Gamma_{0_{yy}} - (\xi m)^2 \right) - \Gamma_{0_{xy}}^2 \; \; \; .
\ee
In (\ref{ssbex1}) it suffices to know the contributions
which are independent of $c$ because
all the quantities in (\ref{ssbex1}) only occur in the first two
equations of (\ref{ssbdiffw}) and, furthermore, the $c$-dependent
contributions to $\Gamma$ are completely fixed by the Legendre
transformations of the fourth and fifth
equation in (\ref{ssbdiffw}).\\
In a similar manner we also obtain the Legendre transformations of the
various third derivatives of $W$. Copying the notations of
appendix~E these transformations read
\be \label{ssbex2}
W_J \stackrel{Legendre}{\longrightarrow} 
\displaystyle\frac{K_J (x,y)}{h (x,y) \; H (x,y)} \;
\mbox{ for } \; J \in \{ \rho \rho \rho, \rho \rho \tau,
\rho \tau \tau, \tau \tau \tau \} \; \; \; .
\ee
In the present case $H (x,y)$ is defined to be the determinant of the
$6 \times 6$ matrix M:
\be \label{ssbex3}
M = \left( \begin{array}{c|c|c|c|c|c}
\Gamma_{0_{xx}}^2 & 2 \Gamma_{0_{xx}} \Gamma_{0_{xy}} &
\Gamma_{0_{xy}}^2 & 0 & 0 & 0 \\ \hline
\Gamma_{0_{xx}} \Gamma_{0_{xy}} &
\Gamma_{0_{xx}} \Gamma_{0_{yy}} + \Gamma_{0_{xy}}^2 &
\Gamma_{0_{xy}} \Gamma_{0_{yy}} & \xi m \Gamma_{0_{xx}} &
\xi m \Gamma_{0_{xy}} & 0 \\ \hline
0 & \xi m \Gamma_{0_{xx}} & \xi m \Gamma_{0_{xy}} &
\Gamma_{0_{xx}} & \Gamma_{0_{xy}} & 0 \\ \hline
\Gamma_{0_{xy}}^2 & 2 \Gamma_{0_{xy}} \Gamma_{0_{yy}} &
\Gamma_{0_{yy}}^2 & 2 \xi m \Gamma_{0_{xy}} &
2 \xi m \Gamma_{0_{yy}} & (\xi m)^2 \\ \hline
0 & \xi m \Gamma_{0_{xy}} & \xi m \Gamma_{0_{yy}} &
\Gamma_{0_{xy}} & \Gamma_{0_{yy}} + (\xi m)^2 & \xi m \\ \hline
0 & 0 & (\xi m)^2 & 0 & 2 \xi m & 1 \end{array} \right)
\ee
For $J \in \{ \rho \rho \rho, \rho \rho \tau, \rho \tau \tau \}$ $K_J$ is
given by $h$ times the determinant of the matrix resulting from $M$ by
replacing the first or the second or the third row (depending on $J$)
by $\vec{m}_1$, whereas $K_{\tau \tau \tau}$ is $h$ times the determinant
of the matrix obtained from $M$ by substituting the third row by 
$\vec{m}_2$ with:
\ba \label{ssbex4}
\vec{m}_1 & = & - \left. (
\Gamma_{0_{xxx}} W_{\rho \rho} + \Gamma_{0_{xxy}} W_{\rho \tau},
\Gamma_{0_{xxy}} W_{\rho \rho} + \Gamma_{0_{xyy}} W_{\rho \tau}, 0,
\Gamma_{0_{xyy}} W_{\rho \rho} + \Gamma_{0_{yyy}} W_{\rho \tau}, 0, 0)^T 
\right|_{c = 0} \nonumber \\
\vec{m}_2 & = & \left. \vec{m}_1 
\right|_{W_{\rho \rho} \rightarrow W_{\rho \tau} \mbox{ and }
W_{\rho \tau} \rightarrow W_{\tau \tau}}
\ea
By an analogously careful handling of the different levels of the
Grassmann algebra involved in the respective calculations we finally
get the results for the Legendre transformations still missing, too:
\ba \label{ssbex5}
W_{\rho \overline{\eta}} & \stackrel{Legendre}{\longrightarrow} &
\Gamma_1^{- 1} \; h^{- 1} (x,y) \left(
\Gamma_{0_{xy}} \Gamma_{1_y} - \left( \Gamma_{0_{yy}} - (\xi m)^2 
\right) \Gamma_{1_x} \right) c \nonumber \\[.5ex]
W_{\rho \eta} & \stackrel{Legendre}{\longrightarrow} &
\Gamma_1^{- 1} \; h^{- 1} (x,y) \left\{
\left( \Gamma_{0_{yy}} - (\xi m)^2 \right) \left(
\Gamma_{1_x} \overline{c} + \Gamma_{2_x} X + \Gamma_{3_x} Y \right)
\right. \nonumber \\
& & \hspace*{2cm} \left. - \Gamma_{0_{xy}} \left(
\Gamma_{1_y} \overline{c} + \Gamma_{2_y} X + \Gamma_{3_y} Y \right)
\right\} \nonumber \\[.5ex]
W_{\eta \overline{\eta}} & \stackrel{Legendre}{\longrightarrow} &
- \; \Gamma_1^{- 1} \; + \; c\mbox{-dependent term}
\ea
Cluing all scattered pieces of the puzzle together we now are able to quote
the system (\ref{ssbdiffw}) for $\Gamma$:\\[-1.3cm]
\par
{\footnotesize
\ba \label{ssbdiffg}
\hbar^2 \; \frac{\lambda}{4} \; \Gamma_1 \; 
(K_{\rho \rho \rho} + K_{\rho \tau \tau} ) & & \nonumber \\
+ \; \hbar \; H \left\{
\frac{\lambda}{4} \; \Gamma_1 \left[
(x + v) \left( 3 \left( \Gamma_{0_{yy}} - (\xi m)^2 \right) +
\Gamma_{0_{xx}} \right) - 2 y \Gamma_{0_{xy}} \right] -
\xi m h \right\} & & \nonumber \\
+ \; h \; H \; \Gamma_1 \left[
\frac{\lambda}{4} (x^3 + x y^2 + 3 v x^2 + v y^2) - 2 m^2 x \right] 
& = & h \; H \; \Gamma_1 \; \left. \Gamma_x \right|_{c = 0} \nonumber \\[.5ex]
\hbar^2 \; \frac{\lambda}{4} \; 
(K_{\rho \rho \tau} + K_{\tau \tau \tau} ) & & \nonumber \\
+ \; \hbar \; \frac{\lambda}{4} \; H \left\{
- 2 (x + v) \Gamma_{0_{xy}} + y \left( \left(
\Gamma_{0_{yy}} - (\xi m)^2 \right) + 3 \Gamma_{0_{xx}} \right) 
\right\} & & \nonumber \\
+ \; h \; H \left[ \frac{\lambda}{4}
(y^3 + x^2 y + 2 v x y) + \xi m B \right] & = &
h \; H \; \left. \Gamma_y \right|_{c = 0} \nonumber \\[.5ex]
B \; + \; \xi m y & = & \Gamma_B \\[.5ex]
- \; \hbar \; \xi m \left\{ \left(
\Gamma_{0_{yy}} - (\xi m)^2 \right) \left( 
\Gamma_{1_x} \overline{c} + \Gamma_{2_x} X + \Gamma_{3_x} Y \right) -
\Gamma_{0_{xy}} \left(
\Gamma_{1_y} \overline{c} + \Gamma_{2_y} X + \Gamma_{3_y} Y \right)
\right\} & & \nonumber \\
+ \; h \; \Gamma_1 \left(
\xi m (x + v) \overline{c} + y Y - (x + v) X \right) & = &
h \; \Gamma_1 \; \Gamma_c \nonumber \\[.5ex]
- \; \hbar \; \xi m \left\{
\Gamma_{0_{xy}} \Gamma_{1_y} - \left(
\Gamma_{0_{yy}} - (\xi m)^2 \right) \Gamma_{1_x} \right\} c & & \nonumber \\
- \; \xi m (x + v) \; h \; \Gamma_1 \; c & = &
h \; \Gamma_1 \ \Gamma_{\overline{c}} \nonumber
\ea}
\par
\vspace*{-.6cm}
\noindent
The solution of (\ref{ssbdiffg}) can be obtained recursively by consistently
evaluating (\ref{ssbdiffg}) order by order in $\hbar$. Of course, the
zeroth order $\Gamma^{(0)}$ coincides with $\Gamma_{cl}$ (\ref{ssbaccl}).
The result for $\Gamma^{(1)}$ is already stated in the main text, see
(\ref{ssbg1}). For the second order $\Gamma^{(2)}$ we find after a cumbersome
but straightforward calculation performed by means of 
{\it mathematica}:\\[-1cm]
\par
\input{result2l}

\par
\vspace*{-.5cm}
\noindent
In (\ref{ssbg2}) contributions proportional to $c$ are not specified because
the corresponding expressions are rather lengthy. The knowledge of these
contributions is irrelevant in the present context but as a matter of course
essential if one wants to move on to the determination of higher orders
of $\Gamma$. The constant of integration $C^{(2)}$ in (\ref{ssbg2}) is fixed
as usual by the requirement $\left. \Gamma \right|_{x = y = B = c =
\overline{c} = 0}$
yielding:\\[-.5cm]
\par
{\footnotesize
\be \label{ssbg2c}
  C^{(2)} = \frac{1}{128}
  \left[ \frac{32}{m^2 v^2} - \lambda \left( 
  \frac{6}{m^4} + \frac{24}{(\xi m)^4} + 
  \frac{8}{m^2 (\xi m)^2} \right)
  - \lambda^2 v^2 \frac{4 m^4 + 3 (\xi m)^4}{m^6 (\xi m)^4} \right]
\ee
}
\par
\noindent
We would like to close this appendix by two remarks both of which concern
the STI (\ref{ssbsti}):\\
A direct check shows that $\Gamma^{(n)}$ (derived above for $n = 0, 1, 2$)
satisfies the STI (\ref{ssbsti}) taken in the corresponding order, e.g. in
1-loop order
\be \label{ssbsti1}
\frac{\partial \Gamma^{(0)}}{\partial Y} 
\frac{\partial \Gamma^{(1)}}{\partial x} +
\frac{\partial \Gamma^{(1)}}{\partial Y}
\frac{\partial \Gamma^{(0)}}{\partial x} +
\frac{\partial \Gamma^{(0)}}{\partial X} 
\frac{\partial \Gamma^{(1)}}{\partial y} +
\frac{\partial \Gamma^{(1)}}{\partial X}
\frac{\partial \Gamma^{(0)}}{\partial y} +
B \frac{\partial \Gamma^{(1)}}{\partial \overline{c}} = 0 \; \; .
\ee
This has to be so because in the zero-dimensional theory all quantities
are well-defined and finite and, hence, no (possibly symmetry violating)
regularization procedure is needed when higher orders are calculated.\\
The second remark emphasizes once again the necessity of introducing
the external fields $X$ and $Y$. According to the STI, in all orders
the BRS transformations of $x$ and $y$ are given by $\frac{\partial
\Gamma}{\partial Y}$ and $\frac{\partial \Gamma}{\partial X}$,
respectively. These derivatives, however, are far from being trivial
as can be seen, for instance, from (\ref{ssbg1}). Without 
introducing the external
fields $X$ and $Y$ it is at least very difficult to control the
deformations of the nonlinear classical BRS transformations in higher
orders correctly.

%% file: result2l.tex
{\footnotesize
\ba \label{ssbg2}
  \left. \Gamma^{(2)} \right|_{c = 0} & = &
  \left\{ - 2 \left[ 4096 m^4 (\xi m)^6 - 256 \lambda m^2 (\xi m)^2 
  \left( 12 m^4 (v + x)^2 \right. \right. \right. \nonumber \\
& & \left. \left. \left. + (\xi m)^4 (3 v^2 + 30 v x + 15 x^2 + 4 y^2)
    + 4 m^2 (\xi m)^2 (v^2 + 8 v x + 4 x^2 + 9 y^2) \right) 
    \right. \right. \nonumber \\ 
& & \left. \left. + \lambda^5 \left( 192 v^7 x (x^2 + y^2) - 
    3 (x^2 + y^2)^4 (13 x^2 + 9 y^2) - 
    6 v x (x^2 + y^2)^3 (65 x^2 + 49 y^2) \right. \right. \right. \nonumber \\
& & \left. \left. \left. - 6 v^2 (x^2 + y^2)^2 
    (261 x^4 + 214 x^2 y^2 - 7 y^4) + 64 v^6 (3 x^4 + 12 x^2 y^2 + 
    y^4) \right. \right. \right. \nonumber \\
& & \left. \left. \left. - 48 v^5 (27 x^5 - 10 x^3 y^2 - 5 x y^4) -  
    24 v^4 (135 x^6 + 
    127 x^4 y^2 - 7 x^2 y^4 + y^6) \right. \right. \right. \nonumber \\
& & \left. \left. \left. - 48 v^3 (66 x^7 + 125 x^5 y^2 
    + 56 x^3 y^4 - 3 x y^6) \right) + 
    32 \lambda^2 \left( 24 m^6 (v + x)^2 (2 v x + x^2 - y^2) 
    \right. \right. \right. \nonumber \\
& & \left. \left. \left. + (\xi m)^6 \left( -12 v^4 - 30 v^3 x + 
    15 x^4 + 15 x^2 y^2 + 2 y^4 + 
    30 v x (2 x^2 + y^2) \right. \right. \right. \right. \nonumber \\
& & \left. \left. \left. \left. + 3 v^2 (15 x^2 + y^2) \right) + 
    4 m^4 (\xi m)^2 \left( -4 v^4 + 46 v^3 x + 33 x^4 + 
    57 x^2 y^2 + 54 y^4 \right. \right. \right. \right. \nonumber \\
& & \left. \left. \left. \left. + v^2 (155 x^2 + 21 y^2) + 
    6 v (22 x^3 + 19 x y^2) \right) + 
    2 m^2 (\xi m)^4 \left( 42 v^3 x + 57 x^4 
    \right. \right. \right. \right. \nonumber \\
& & \left. \left. \left. \left. + 131 x^2 y^2 + 36 y^4 + 
    v^2 (249 x^2 + 11 y^2) + v (228 x^3 + 262 x y^2) \right) \right) 
    \right. \right. \nonumber \\
& & \left. \left. + 8 \lambda^3 \left( 3 (\xi m)^4 \left( 24 v^5 x - 
    21 x^6 - 56 x^4 y^2 - 41 x^2 y^4 - 6 y^6 + 
    24 v^4 (2 x^2 + y^2) \right. \right. \right. \right. \nonumber \\
& & \left. \left. \left. \left. - 4 v^3 (33 x^3 - 17 x y^2) - 
    v^2 (243 x^4 + 190 x^2 y^2 + 3 y^4) - 2 v (63 x^5 + 112 x^3 y^2 + 
    41 x y^4) \right) \right. \right. \right. \nonumber \\
& & \left. \left. \left. + 4 m^4 \left( 8 v^5 x - 8 v^4 (10 x^2 + 3 y^2) - 
    12 v^3 (25 x^3 + 7 x y^2) - 3 v^2 (115 x^4 + 
    62 x^2 y^2 + 3 y^4) \right. \right. \right. \right. \nonumber \\
& & \left. \left. \left. \left. - 18 v (9 x^5 + 8 x^3 y^2 + 7 x y^4) - 
    9 (3 x^6 + 4 x^4 y^2 + 
    7 x^2 y^4 + 6 y^6) \right) \right. \right. \right. \nonumber \\
& & \left. \left. \left. + 4 m^2 (\xi m)^2 \left( 24 v^5 x - 
    212 v^3 x (3 x^2 + y^2) - 8 v^4 (15 x^2 + 2 y^2) 
    \right. \right. \right. \right. \nonumber \\
& & \left. \left. \left. \left. + v^2 (-789 x^4 - 922 x^2 y^2 + 3 y^4) - 
    6 v (63 x^5 + 136 x^3 y^2 + 65 x y^4) 
    \right. \right. \right. \right. \nonumber \\
& & \left. \left. \left. \left. - 3 (21 x^6 + 68 x^4 y^2 + 65 x^2 y^4 + 
    18 y^6) \right) \right) - 
    4 \lambda^4 \left( 3 (\xi m)^2 \left( 16 v^6 (3 x^2 + y^2) 
    \right. \right. \right. \right. \nonumber \\
& & \left. \left. \left. \left. + 16 v^5 (3 x^3 + 7 x y^2) -  
    (x^2 + y^2)^2 (21 x^4 + 38 x^2 y^2 + 9 y^4) 
    \right. \right. \right. \right. \nonumber \\
& & \left. \left. \left. \left. + v^4 (-324 x^4 + 136 x^2 y^2 + 12 y^4) - 
    16 v^3 (42 x^5 + 35 x^3 y^2 - 3 x y^4) 
    \right. \right. \right. \right. \nonumber \\
& & \left. \left. \left. \left. - 4 v^2 (126 x^6 + 235 x^4 y^2 + 
    100 x^2 y^4 - y^6) - 
    8 v (21 x^7 + 60 x^5 y^2 + 53 x^3 y^4 + 14 x y^6) \right) 
    \right. \right. \right. \nonumber \\
& & \left. \left. \left. + 4 m^2 \left( 8 v^6 (3 x^2 + y^2) - 
    8 v^5 (9 x^3 + 5 x y^2) - 
    3 (x^2 + y^2)^2 (7 x^4 + 12 x^2 y^2 + 9 y^4) 
    \right. \right. \right. \right. \nonumber \\
& & \left. \left. \left. \left. - 2 v^4 (237 x^4 + 166 x^2 y^2 + 9 y^4) - 
    24 v^3 (31 x^5 + 39 x^3 y^2 + 4 x y^4) 
    \right. \right. \right. \right. \nonumber \\
& & \left. \left. \left. \left. - 6 v^2 (86 x^6 + 169 x^4 y^2 + 
    88 x^2 y^4 - 3 y^6) - 
    12 v (14 x^7 + 39 x^5 y^2 + 40 x^3 y^4 + 15 x y^6)
    \right) \right) \right] \right\} / \nonumber \\
& & \left\{ (v + x)^2 \left[ 32 m^2 (\xi m)^2 + \lambda^2 \left( 4 v^2 
    (3 x^2 - y^2) + 
    12 v x (x^2 + y^2) + 3 (x^2 + y^2)^2 \right) \right. \right. \nonumber \\
& & \left. \left. - 4 \lambda \left( (\xi m)^2 (6 v x + 3 x^2 + y^2) + 2 m^2 
    (2 v x + x^2 + 3 y^2) \right) \right]^3 \right\} \nonumber \\
& & + C^{(2)}
\ea
}

%% file: appg.tex
\newsection{Proofs of the rules for the matrix derivative}

We already indicated in the main text that most of the rules for the matrix
derivative $\frac{\partial}{\partial \cA}$ given in section~4.3 are
easily proven by exploiting the decomposition
\ba \label{ncgdiffadea}
& \displaystyle\frac{\partial}{\partial \cA} = - \; \frac{v}{2} \left(
\eta \; \partial_x \; - \; i \; \eta \gamma \; \partial_y \right)
\; = \; - \; \eta \; \frac{1}{2} \left( {\cal D}_x + {\cal D}_y \right)
& \nonumber \\[.5ex]
& \mbox{with } \; {\cal D}_x := v \; \eins_2 \; \partial_x \;
\mbox{ and } \; {\cal D}_y := - i v \; \gamma \; \partial_y &
\ea
of $\frac{\partial}{\partial \cA}$ with respect to the ordinary partial
derivatives $\partial_x$ and $\partial_y$ contained in 
$\frac{\partial}{\partial \cA}$. A further cornerstone of the proofs to follow
will be the likewise trivial observation that we have
\be \label{ncgaxody}
\partial_x \cA = \frac{1}{v} \; \eta \; \mbox{ and } \;
\partial_y \cA = \frac{i}{v} \; \gamma \eta \; \; \; .
\ee
To simplify matters we begin with rule~2.\\[1ex]
{\bf Proof of rule~2: } 
Let $M(\cA; \eta )$ be any monomial in the variables $\cA$ and $\eta$, i.e
$M$ can be regarded as a ``word'' of arbitrary length composed of the
``letters'' $\cA$ and $\eta$ in an arbitrary way. Because $\partial_x$
is the ordinary partial derivative with respect to $x$ the action of
${\cal D}_x$ on $M$ will result in $v$ times a sum of new ``words'':
These new ``words'' are obtained by replacing successively each $\cA$ in
$M$ by $\partial_x \cA = \frac{1}{v} \eta$ due to the ordinary product 
rule for \nolinebreak $\partial_x$. Hence, all the factors 
$\frac{1}{v}$ cancel with the overall factor $v$, and ${\cal D}_x M$ reads
in a symbolic notation:
\begin{displaymath}
{\cal D}_x M = \sum_{\cA \in M} M(\cA, \cA \rightarrow \eta; \eta)
\end{displaymath}
In the same manner we have to study the action of ${\cal D}_y$ on $M$: This
time we get $- i v \gamma$ times a sum of different new ``words'' resulting
from $M$ by substituting successively each $\cA$ in $M$ by
$\partial_y \cA = \frac{i}{v} \gamma \eta$. Again, all the factors
$\frac{i}{v}$ cancel with the overall factor $- i v$, and using the same
symbolic notation as above in an intermediate step ${\cal D}_y M$ is given
by:
\begin{displaymath}
{\cal D}_y M = \gamma \cdot \sum_{\cA \in M} M(\cA, \cA \rightarrow 
\gamma \eta; \eta )
\end{displaymath}
Now we should take into account that $\gamma$ is the grading automorphism
for $2 \times 2$ matrices. In the present context this implies the following:
Whenever in the calculation of ${\cal D}_y M$
there is an {\it even} number of (necessarily odd) ``letters''
in front of the $\cA$ to be replaced by $\gamma \eta$ 
the $\gamma$ of the replacement $\gamma \eta$ can be permuted
to the leftmost position without producing a sign 
because a product of an even number
of odd matrices obviously is even. However, if the number of ``letters''
in front of the $\cA$ which is substituted by $\gamma \eta$ is odd the
permutation of $\gamma$ to the leftmost position 
results in a minus sign. Denoting
by $\nu (\cA )$ the number of (odd) matrices in front of the $\cA$ to
be replaced and using $\gamma^2 = \eins_2$ we, hence, obtain for
${\cal D}_y M$:
\begin{displaymath}
{\cal D}_y M = \sum_{\stackrel{\cA \in M}{\nu (\cA ) even}}
M(\cA, \cA \rightarrow \eta; \eta ) \; -
\sum_{\stackrel{\cA \in M}{\nu (\cA ) odd}}
M(\cA, \cA \rightarrow \eta; \eta )
\end{displaymath}
Putting everything together according to the first line of 
(\ref{ncgdiffadea}) rule~2 follows immediately:
\bas
& \frac{\partial}{\partial \cA} M(\cA; \eta ) = - \; \eta \;
\sum_{\stackrel{\cA \in M}{\nu (\cA ) even}} M(\cA, \cA \rightarrow \eta;
\eta ) & \\
& & \hspace{3.3cm} \Box
\eas
\par
\vspace*{.3cm}
\noindent
{\bf Proof of rule~3: }
First of all, due to the linearity of the operator $\frac{\partial}{\partial
\cA}$ it suffices to prove rule~3 for arbitrary {\it monomials}
$F(\cA; \eta )$ and $G(\cA; \eta )$ (instead of polynomials 
$F$ and $G$) only.\\
For the first part we furthermore assume $F(\cA; \eta )$ to be an even
matrix.\\
Clearly $F \cdot G$ is the ``word'' which is trivially obtained by
appending the ``word'' $G$ to the ``word'' $F$. According to rule~2,
$\frac{\partial}{\partial \cA} (F \cdot G)$ is then given by $- \eta$
times a sum of new ``words'' resulting from $F \cdot G$ by successively
replacing each $\cA$ in $F \cdot G$ with $\nu (\cA )$ even 
by \nolinebreak $\eta$.
($\nu (\cA )$ denotes the number of (odd) matrices in $F \cdot G$ which
stand in front of $\cA$, see also proof of rule~2, whereas $\nu_F (\cA ),
\nu_G (\cA )$ will be the numbers of matrices in $F, G$ in front of $\cA$,
respectively.) In other words, at first all the $\cA$'s in $F$ with
$\nu_F (\cA )$ even and afterwards all the $\cA$'s in $G$ with
$\nu (\cA )$ even are substituted. Because $F$ is even by assumption for
all $\cA$'s in $G$ we have $(\nu_G (\cA ) \; mod \; 2) =
(\nu (\cA ) \; mod \; 2)$ and, hence, we get in the symbolic notation of the
previous proof:
\bas
\frac{\partial}{\partial \cA} (F \cdot G) & = & - \; \eta \! \! \! \!
\sum_{\stackrel{\cA \in F}{\nu_F (\cA ) even}} \! \! \! \!
F(\cA, \cA \rightarrow \eta; \eta) \cdot G(\cA; \eta ) \; - \; \eta \! \! \! \!
\sum_{\stackrel{\cA \in G}{\nu_G (\cA ) even}}
\! \! \! \! F(\cA; \eta ) \cdot G(\cA, \cA \rightarrow \eta; \eta ) \\[.5ex]
& = & [ \; - \; \eta \! \! \! \!
\sum_{\stackrel{\cA \in F}{\nu_F (\cA ) even}} \! \! \! \! F(\cdots ) 
\; ] \cdot
G(\cdots ) \; - \; \eta F(\cdots ) \eta \cdot [ \; - \; \eta \! \! \! \!
\sum_{\stackrel{\cA \in G}{\nu_G (\cA ) even}} \! \! \! \! G(\cdots ) \; ]
\eas
The last equality holds true due to $\eta^2 = - \eins_2$. But this already
completes the proof of the first claim in rule~3.\\
Let us now assume that $F(\cA; \eta )$ is an odd matrix. The proof just given
indicates by now what is ``wrong'' if $F$ is odd, namely for all
$\cA$'s in $G$ the counting of matrices has to be modified:
\begin{displaymath}
(\nu_G (\cA ) \; mod \; 2) =
((\nu (\cA ) \; + \; 1) \; mod \; 2)
\end{displaymath}
Nevertheless, the present case can be reduced to the first one in a very
simple manner by taking into account the following two trivial observations:
If $F$ is odd then $F \cdot \eta$ is even. Furthermore, rule~2 tells us that
$\frac{\partial}{\partial \cA} (F \cdot \eta) =
(\frac{\partial}{\partial \cA} F) \cdot \eta$ holds true. With this in mind
we calculate:
\bas
\frac{\partial}{\partial \cA} (F \cdot G) & = & - \;
\frac{\partial}{\partial \cA} \; [(F \eta) \cdot (\eta G)] \\[.5ex]
& \stackrel{\mbox{first part}}{=} & 
- \left[ \frac{\partial}{\partial \cA} (F \eta) \right] \cdot \eta G +
\eta (F \eta) \eta \cdot \frac{\partial}{\partial \cA} \eta G \\[.5ex]
& = & \frac{\partial F}{\partial \cA} \cdot G -
\eta F \cdot \frac{\partial}{\partial \cA} \eta G \\
& & \hspace{9.1cm} \Box
\eas
\par
\vspace*{.5ex}
\noindent
Next we are going to demonstrate rule~5. Please note that the 
quite short proof given below by no means makes use of rule~1 which
will be proven subsequently.\\[1ex]
{\bf Proof of rule~5: } 
According to (\ref{ncgdiffadea}) we find ($\eta^2 = - \eins_2$):
\begin{displaymath}
\eta \; \frac{\partial}{\partial \cA} \mbox{ Tr } F \; = \;
\frac{v}{2} \; \eins_2 \; \partial_x \mbox{ Tr } F \; - \;
i \; \frac{v}{2} \; \gamma \; \partial_y \mbox{ Tr } F
\end{displaymath}
Due to Tr$\eins_2 = 2$, Tr$\gamma = 0$, Str$\eins_2 = 0$ and
Str$\gamma = 2$ we, hence, get:
\bas
\frac{1}{v} \mbox{ Tr } \eta \frac{\partial}{\partial \cA}
\mbox{ Tr } F \; = \; 
\frac{1}{v} \; \frac{v}{2} \; \partial_x \mbox{ Tr } F
\mbox{ Tr } \eins_2 -
\frac{1}{v} \; i \; \frac{v}{2} \; \partial_y \mbox{ Tr } F
\mbox{ Tr } \gamma & = & \partial_x \mbox{ Tr } F \\[.5ex]
\frac{i}{v} \mbox{ Str } \eta \frac{\partial}{\partial \cA}
\mbox{ Tr } F \; = \; 
\frac{i}{v} \; \frac{v}{2} \; \partial_x \mbox{ Tr } F
\mbox{ Str } \eins_2 -
\frac{i}{v} \; i \; \frac{v}{2} \; \partial_y \mbox{ Tr } F
\mbox{ Str } \gamma & = & \partial_y \mbox{ Tr } F \\
& & \hspace{1cm} \Box
\eas
\par
\vspace*{.3cm}
\noindent
{\bf Proof of rule~1: }
Unfortunately, the following proof is a little bit lengthy. For this
reason we will try to concentrate on the main conceptual steps
skipping some technical details in between in order to keep the proof in
a readable form. In any case, the computational gaps can be filled very
easily.\\
We start with two observations: The first one concerns the structure of
the result when applying $\frac{\partial}{\partial \cA}$ to Tr $F(\cA; \eta )$.
Namely, by the very definition of $\frac{\partial}{\partial \cA}$ this result
has to be an odd $2 \times 2$ matrix:
\begin{displaymath}
\frac{\partial}{\partial \cA} \mbox{ Tr } F(\cA; \eta ) =
\left( \begin{array}{cc} 0 & a \\ b & 0 \end{array} \right)
\end{displaymath}
Hence, if the direct calculation of the entries $a$ and $b$ (only relying
on the explicit definition of $\frac{\partial}{\partial \cA}$) 
and the calculation
of $a$ and $b$ according to the prescription of rule~1 lead to the same
results, everything is proven. However, instead of computing that way
$a$ and $b$ by themselves we will do so for two independent linear
combinations of $a$ and $b$ suggested by rule~5, viz
\be \label{ncgrule5}
\partial_x \mbox{ Tr } F = \frac{1}{v} \mbox{ Tr } \eta \; 
\frac{\partial}{\partial \cA}
\mbox{ Tr } F \; \; \mbox{ and } \; \;
\partial_y \mbox{ Tr } F = \frac{i}{v} \mbox{ Tr } \gamma \eta \; 
\frac{\partial}{\partial \cA} \mbox{ Tr } F \; \; \; .
\ee
The second observation is based on the nature of $F$ as an even matrix. Because
of $\cA^2 \propto \eins_2$ and $F$ being even there are only two possible
basic types of monomials out of which the polynomial $F$ is built up:
\be \label{ncgbaseven}
\cA^{2k} (\eta \cA )^l \; \mbox{ and } \;
\cA^{2k} \cA (\eta \cA )^l \eta \; \mbox{ with }
k,l \in \N \mbox{ arbitrary}
\ee
Due to the linearity of $\frac{\partial}{\partial \cA}$ we are allowed to
restrict the whole proof to these basic monomials. Let us, for instance,
choose the first one, i.e. $F = \cA^{2k} (\eta \cA)^l$ from now on.
Using the recipe of rule~1 we, hence, obtain
\ba \label{ncgrule1ex1}
\frac{\partial}{\partial \cA} \mbox{ Tr } \cA^{2k} (\eta \cA )^l & = &
\cA^{2k - 1} (\eta \cA )^l + \cA^{2k - 2} (\eta \cA )^l \cA +
\cdots + \cA^0 (\eta \cA )^l \cA^{2k - 1} \nonumber \\
& & + (\eta \cA )^{l - 1} \cA^{2k} \eta +
(\eta \cA )^{l - 2} \cA^{2k} (\eta \cA ) \eta + \cdots +
(\eta \cA )^0 \cA^{2k} (\eta \cA )^{l - 1} \eta \nonumber \\[.5ex]
& = & \cA^{2k - 2} \left(
k \cA (\eta \cA )^l + (k + l) (\eta \cA )^l \cA \right) \; \; ,
\ea
where for the last equality $\cA^2 \propto \eins_2$ has been taken into
account several times.\\
Inserting (\ref{ncgrule1ex1}) into the r.h.s. of the first equation in
(\ref{ncgrule5}) we thus get (utilizing again $\cA^2 
\propto \eins_2$ frequently
as well as $\eta (\eta \cA )^l = - \cA (\eta \cA )^{l -1}$ and the cyclic
property of the trace):
\be \label{ncgrule1ex2}
\frac{1}{v} \mbox{ Tr } \eta \; \frac{\partial}{\partial \cA} \; F =
\frac{1}{v} \left\{ k \mbox{ Tr } \cA^{2k - 2} (\eta \cA )^{l + 1} \; - \;
(k + l) \mbox{ Tr } \cA^{2k} (\eta \cA )^{l - 1} \right\}
\ee
This result has to be compared with the direct calculation of
$\partial_x$ Tr $F$ ($\partial_x \cA = \frac{1}{v} \eta$, see above):
\ba \label{ncgrule1ex3}
\partial_x \mbox{ Tr } F & = & \mbox{Tr } \partial_x \cA^{2k} 
(\eta \cA )^l \nonumber \\[.5ex]
& = & \frac{1}{v} \mbox{ Tr } \left\{
\eta \cA^{2k - 1} (\eta \cA )^l + \cA \eta \cA^{2k - 2} (\eta \cA )^l +
\cdots + \cA^{2k - 1} \eta (\eta \cA )^l \right. \nonumber \\
& & \left. + \cA^{2k} \eta \eta (\eta \cA )^{l - 1} +
\cA^{2k} (\eta \cA ) \eta \eta (\eta \cA )^{l - 2} + \cdots +
\cA^{2k} (\eta \cA )^{l - 1} \eta \eta \right\} \nonumber \\[.5ex]
& = & \frac{1}{v} \mbox{ Tr } \left\{
k \cA^{2k - 2} (\eta \cA )^{l + 1} - k \cA^{2k} (\eta \cA )^{l - 1} -
l \cA^{2k} (\eta \cA )^{l - 1} \right\}
\ea
Needless to say that for the last equality we used $\cA \propto \eins_2$
and $\eta^2 = - \; \eins_2$ several times. Obviously, the two results 
(\ref{ncgrule1ex2}) and (\ref{ncgrule1ex3}) coincide. 
In almost the same manner the second equation in (\ref{ncgrule5}) 
is verified: However, the quite analogous calculation will be skipped
here in order not to protract the proof unnecessarily by presenting
nothing else but trivial computations.\\
Finally, repeating the procedure just outlined also for the other basic
monomial in (\ref{ncgbaseven}) everything is proven.\\
\hspace*{14.3cm} $\Box$\\[2ex]
{\bf Proof of rule~4: }
The following proof proceeds by direct calculation, i.e. in a manner
analogous to the one which already helped us to demonstrate rule~1.
This time, however, even more cases have to be distinguished.\\
First of all, we would like to state that for an odd polynomial
$J(\cA; \eta )$ due to $\cA^2 \propto \eins_2$ there are only two basic
types of monomials out of which $J$ is built up, namely:
\be \label{ncgbasodd}
\cA^{2m} (\cA \eta)^n \cA \; \; \mbox{ and } \; \;
\cA^{2m} (\eta \cA)^n \eta \; \; \mbox{ with }
m, n \in \N \mbox{ arbitrary}
\ee
Because of the linearity of the matrix derivatives $\frac{\partial}{\partial
\cA}$ and $\frac{\partial}{\partial J}$ it again suffices to restrict the
proof to the basic monomials. Taking into account (\ref{ncgbaseven}) and
the fact that $F(J; \eta )$ is allowed to be either even or odd (whereas
$J(\cA; \eta )$ has to be odd) we, hence, have to look at eight different
cases:
\begin{displaymath}
2 \; (F \mbox{ even or odd) } \times \; 2 \mbox{ (each with 2 basic 
monomials) } \times \; 2 \mbox{ (2 basic monomials for } J)
\end{displaymath}
Let us single out just one of these cases as an example:
\be \label{ncgrule4ex}
F(J; \eta ) = J^{2k} J (\eta J)^l \eta \; \; \mbox{ and }
\; \; J(\cA; \eta ) = \cA^{2m} (\eta \cA )^n \eta
\ee
All the other cases can be treated in a completely analogous way.\\
For the direct calculation of $\frac{\partial F}{\partial \cA}$ we first
need $F(J(\cA; \eta ); \eta )$:
\bas
F(J(\cA; \eta ); \eta ) & = & 
(\; \cA^{2m} (\eta \cA)^n \eta \; )^{2k + 1} \;
(\; \eta \cA^{2m} (\eta \cA)^n \eta \; )^l \; \eta \\[.5ex]
& = & (-1)^{k(n + 1) + l + 1} \; \cA^{2[m(2k + l + 1) + kn]} \;
(\eta \cA)^{n(l + 1)}
\eas
The last equality is again due to $\cA^2 \propto \eins_2$, a property that
will be used frequently also in the following without being mentioned every 
time. The above equation now has to be differentiated with respect to $\cA$.
In order to simplify this calculation we first observe that for arbitrary
$p, q \in \N$ we have by means of rule~2:
\ba \label{ncgrule2aex1}
\frac{\partial}{\partial \cA} \cA^{2p} (\eta \cA )^q & = &
- \eta \left\{ \eta \cA^{2p -1} (\eta \cA )^q +
\cA^2 \eta \cA^{2p - 3} (\eta \cA )^q + \cdots +
\cA^{2p - 2} \eta \cA (\eta \cA )^q \right\} \nonumber \\[.5ex]
& = & p \cA^{2p -1} (\eta \cA )^q
\ea
Please note that none of the $\cA$'s appearing in the factors $\eta \cA$
in $\cA^{2p} (\eta \cA )^q$ is replaced by 
\nolinebreak $\eta$ because there is always
an odd number of matrices in front of those $\cA$'s. Hence, the direct
calculation of $\frac{\partial F}{\partial \cA}$ leads to
\be \label{ncgugly1}
\frac{\partial F}{\partial \cA} = (- 1)^{k(n + 1) + l + 1}
[m(2k + l + 1) + kn] \cA^{2[m(2k + l + 1) + kn] - 1}
(\eta \cA )^{n(l + 1)} \; \; \; .
\ee
This result has to be compared to what we get upon 
inserting (\ref{ncgrule4ex})
into the r.h.s. of (\ref{ncgchain}). To this end we need
\ba \label{ncgrule2aex2}
\frac{\partial}{\partial \cA} \; \cA^{2m} (\eta \cA )^n \eta & = &
m \; \cA^{2m - 1} (\eta \cA )^n \eta \; \; \; \\[.5ex]
\frac{\partial}{\partial \cA} \; \eta \cA^{2m} (\eta \cA )^n \eta & = &
- \; (m + 1) \; \cA^{2m} (\eta \cA )^{n - 1} \eta \; + \; 
(n - 1) \; \cA^{2m + 1} (\eta \cA )^{n - 2} \eta \nonumber
\ea
and
\ba \label{ncgrule2aex3}
\frac{\partial}{\partial J} \; J^{2k + 1} (\eta J)^l \eta & = &
(k + 1) \; J^{2k} (\eta J)^l \eta \; - \; 
l \; J^{2k + 1} (\eta J)^{l - 1} \eta 
\; \; \; , \nonumber \\[.5ex]
\frac{\partial}{\partial J} \; \eta J^{2k + 1} (\eta J)^l \eta & = &
k \; J^{2k - 1} (\eta J)^l \eta \; \; \; .
\ea
All this follows from the application of rule~2. 
Finally, cluing everything together according to the prescription of
rule~4 (expressing, of course, the right hand sides of
(\ref{ncgrule2aex3}) as functions of $\cA$ and $\eta$ by means of 
(\ref{ncgrule4ex})) and keeping carefully 
in view the combinatorics when shrinking the at first lengthy expression
due to $\cA^2 \propto \eins_2$, after quite some effort one indeed
obtains the r.h.s. of (\ref{ncgugly1}) again.\\
\hspace*{14.3cm} $\Box$

%% file: apph.tex
\newsection{The full differential equation for $\Gamma$ within the
            matrix formulation}

In order to be able to write down explicitly the differential equation
for $\Gamma$ valid to all orders of the loop expansion we still need the
Legendre transformation of $\frac{\partial^3 W}{\partial J^3}$:
All other required ingredients already have been 
determined in the main text, see section~4.4.\\
The following calculation of the Legendre transformation in question also
serves as a further example for the application of the rules for the
matrix derivatives $\frac{\partial}{\partial \cA}$ and 
$\frac{\partial}{\partial J}$.\\ 
We begin with
\ba \label{ncgleg3aaa}
\left. \frac{\partial}{\partial \cA} \; \frac{\partial}{\partial J}
\eta \frac{\partial W}{\partial J} \right|_{J = J(\cA; \eta )} & 
\stackrel{\mbox{rule~4}}{=} &
\frac{\partial J}{\partial \cA} \; \frac{\partial^2}{\partial J^2}
\eta \frac{\partial W}{\partial J} \; - \;
\left( \frac{\partial}{\partial \cA} \eta J \right) \left(
\frac{\partial}{\partial J} \eta \frac{\partial}{\partial J} \eta
\frac{\partial W}{\partial J} \right) \nonumber \\[.5ex]
& = & \frac{\partial^2 \Gamma}{\partial \cA^2} \;
\frac{\partial^2}{\partial J^2} \eta \frac{\partial W}{\partial J} \; - \;
\left( \frac{\partial}{\partial \cA} \eta
\frac{\partial \Gamma}{\partial \cA} \right) \left(
\frac{\partial}{\partial J} \eta \frac{\partial}{\partial J} \eta
\frac{\partial W}{\partial J} \right) \nonumber \\[.5ex]
& = & \eta \; \Gamma_{\cA \cA} \; W_{JJJ} \; - \; \Gamma_{\cA \eta \cA} 
\; W_{J \eta J \eta J} \; =: \; A
\ea
For the last equality we used the fact that both $\frac{\partial^2}{\partial
\cA^2}$ and $\frac{\partial^2}{\partial J^2}$ are some differential operators
proportional to the unit matrix $\eins_2$. In order to keep the forthcoming
formulae in a readable form we also introduced an almost
obvious abbreviatory notation for the various multiple derivatives of $\Gamma$
and $W$ occuring in these kinds of calculations:
$\frac{\partial^2 \Gamma}{\partial \cA^2} \equiv \Gamma_{\cA \cA}$,
$\frac{\partial}{\partial \cA} \eta \frac{\partial \Gamma}{\partial \cA} 
\equiv \Gamma_{\cA \eta \cA}$ and so on.\\
In an analogous manner we obtain as well
\be \label{ncgleg3bbb}
\left. \frac{\partial}{\partial \cA} \; \eta \frac{\partial}{\partial J}
\eta \frac{\partial W}{\partial J} \right|_{J = J(\cA; \eta )} \; \;
\stackrel{\mbox{rule~4}}{=} \; \;
\Gamma_{\cA \cA} \; W_{J \eta J \eta J} \; + \;
\Gamma_{\cA \eta \cA} \; \eta \; W_{JJJ} \; =: \; B \; \; \; .
\ee
Now, it is an easy and straightforward undertaking to solve (\ref{ncgleg3aaa})
and (\ref{ncgleg3bbb}) in order to get $W_{JJJ}$ (and $W_{J \eta J \eta J}$):
\be \label{ncgleg3}
W_{JJJ} = - \; \eta \left[ \Gamma_{\cA \cA}^2 + \Gamma_{\cA \eta \cA}^2
\right]^{- 1} \left\{ \Gamma_{\cA \cA} \cdot A \; + \;
\Gamma_{\cA \eta \cA} \cdot B \right\}
\ee
(\ref{ncgleg3}) is the desired result with the restriction, however, that
it still remains to determine $A$ and $B$ as functions of $\cA$ and $\eta$.\\
To this end we observe that due to the definition of $B$ (\ref{ncgleg3bbb}),
due to $\eta$ times the second equation in (\ref{ncgleg2}) 
and due to the product rule~3
(\ref{ncgprod}) we have (noticing that $\eta 
\frac{\partial}{\partial \cA} \eta \frac{\partial \Gamma}{\partial \cA}$ 
necessarily has to be even)
\bas
B & = & \frac{\partial}{\partial \cA} \; \eta \frac{\partial}{\partial J}
\eta \frac{\partial W}{\partial J} \; = \;
- \frac{\partial}{\partial \cA} \left\{
\eta \frac{\partial}{\partial \cA} \eta \frac{\partial \Gamma}{\partial \cA}
\left[ \left( \frac{\partial^2 \Gamma}{\partial \cA^2} \right)^2 +
\left( \frac{\partial}{\partial \cA} \eta
\frac{\partial \Gamma}{\partial \cA} \right)^2 \right]^{- 1} \right\} \\
& = & - \; \Gamma_{\cA \eta \cA \eta \cA} \;
[ \Gamma_{\cA \cA}^2 + \Gamma_{\cA \eta \cA}^2 ]^{- 1} \; - \;
\Gamma_{\cA \eta \cA} \; \eta \; \frac{\partial}{\partial \cA}
[ \Gamma_{\cA \cA}^2 + \Gamma_{\cA \eta \cA}^2 ]^{- 1} \; \; \; .
\eas
Hence, for the determination of $B$ only the derivative
$\frac{\partial}{\partial \cA} [\Gamma_{\cA \cA}^2 + \Gamma_{\cA \eta \cA}^2 
]^{- 1}$ is not evaluated yet. This can be done
by means of the chain rule~4 (\ref{ncgchain}) as follows\footnote{This
calculation also illustrates why it completely suffices to state the
chain rule within the matrix calculus for the case that the function
$J(\cA; \eta )$ is to be odd.}:
\bas
\frac{\partial}{\partial \cA} [\Gamma_{\cA \cA}^2 +
\Gamma_{\cA \eta \cA}^2 ]^{- 1} & = &
\frac{\partial}{\partial \cA} \eta \eta^{- 1} [ \cdots ]^{- 1} \; = \;
\frac{\partial}{\partial \cA} \eta
[(\Gamma_{\cA \cA}^2 + \Gamma_{\cA \eta \cA}^2 ) \eta ]^{- 1} \\[.5ex]
& = & \frac{\partial}{\partial \cA} F(\hat{J} (\cA; \eta); \eta ) \;
\mbox{ ( with } \; F := \eta \hat{J}^{- 1} \; \mbox{ and } \;
\hat{J} = (\Gamma_{\cA \cA}^2 + \Gamma_{\cA \eta \cA}^2 ) \eta \; ) \\[.5ex]
& = & \frac{\partial \hat{J}}{\partial \cA}
\frac{\partial F}{\partial \hat{J}} - \left(
\frac{\partial}{\partial \cA} \eta \hat{J} \right) 
\frac{\partial}{\partial \hat{J}} \eta F \\[.5ex]
& = & - \; [2 \Gamma_{\cA \cA} \Gamma_{\cA \cA \cA} +
\eta \Gamma_{\cA \cA \cA} \Gamma_{\cA \eta \cA} -
\eta \Gamma_{\cA \eta \cA} \Gamma_{\cA \eta \cA \eta \cA} ] \;
[\Gamma_{\cA \cA}^2 + \Gamma_{\cA \eta \cA}^2 ]^{- 2}
\eas
For the last equality we used
\begin{displaymath}
\frac{\partial F}{\partial \hat{J}} = \frac{\partial}{\partial \hat{J}}
\eta \hat{J}^{- 1} = (\hat{J} \eta \hat{J} )^{- 1} \; \; , \; \;
\frac{\partial}{\partial \hat{J}} \eta F =
- \frac{\partial}{\partial \hat{J}} \hat{J}^{- 1} = 0
\end{displaymath}
as well as
\begin{displaymath}
\frac{\partial \hat{J}}{\partial \cA} =
\frac{\partial}{\partial \cA} [(\Gamma_{\cA \cA}^2 + \Gamma_{\cA \eta \cA}^2) 
\eta ] = [2 \Gamma_{\cA \cA} \Gamma_{\cA \cA \cA} +
\eta \Gamma_{\cA \cA \cA} \Gamma_{\cA \eta \cA} -
\eta \Gamma_{\cA \eta \cA} \Gamma_{\cA \eta \cA \eta \cA} ] \eta \; \; \; .
\end{displaymath}
Putting everything together we thus obtain for $B$ the result:
\ba \label{ncgleg3bbbr}
B & = & - \; \Gamma_{\cA \eta \cA \eta \cA} \;
[\Gamma_{\cA \cA}^2 + \Gamma_{\cA \eta \cA}^2 ]^{- 1} \\[.5ex]
& & + \; \Gamma_{\cA \eta \cA} \; \eta \;
[2 \Gamma_{\cA \cA} \Gamma_{\cA \cA \cA} +
\eta \Gamma_{\cA \cA \cA} \Gamma_{\cA \eta \cA} -
\eta \Gamma_{\cA \eta \cA} \Gamma_{\cA \eta \cA \eta \cA} ] \;
[\Gamma_{\cA \cA}^2 + \Gamma_{\cA \eta \cA}^2 ]^{- 2} \nonumber
\ea
In a similar way also $A$ is calculated. We finally get:
\ba \label{ncgleg3aaar}
A & = & - \; \eta \; \Gamma_{\cA \cA \cA} \;
[\Gamma_{\cA \cA}^2 + \Gamma_{\cA \eta \cA}^2 ]^{- 1} \\[.5ex]
& & - \; \eta \; \Gamma_{\cA \eta \cA} \;
[2 \Gamma_{\cA \cA} \Gamma_{\cA \cA \cA} +
\eta \Gamma_{\cA \cA \cA} \Gamma_{\cA \eta \cA} -
\eta \Gamma_{\cA \eta \cA} \Gamma_{\cA \eta \cA \eta \cA} ] \; \eta \;
[\Gamma_{\cA \cA}^2 + \Gamma_{\cA \eta \cA}^2 ]^{- 2} \nonumber
\ea
Ultimately, we are now in the position to present the full differential
equation for the generating functional $\Gamma$ of 1~PI Green's functions
explicitly:
\ba \label{ncgdiffg}
\frac{\partial \Gamma}{\partial \cA} & = &
\frac{\lambda v^4}{8} \left\{
- \; \hbar^2 \; \eta \;
[\Gamma_{\cA \cA}^2 + \Gamma_{\cA \eta \cA}^2 ]^{- 1} \;
[\Gamma_{\cA \cA} \cdot A + \Gamma_{\cA \eta \cA} \cdot B] \right. 
\nonumber \\[.5ex]
& & \left. \hspace{1.5cm} + \; \hbar \;
[2 (\cA + \eta ) \Gamma_{\cA \cA} + \eta (\cA + \eta ) \Gamma_{\cA \eta \cA}]
\; [\Gamma_{\cA \cA}^2 + \Gamma_{\cA \eta \cA}^2 ]^{- 1} \right. 
\nonumber \\[.5ex]
& & \left. \hspace{1.5cm} + \eta \cA \eta + 2 \eta \cA^2 + \cA \eta \cA +
\cA^3 - \cA \right\}
\ea
Of course, in (\ref{ncgdiffg}) $A$ and $B$ have to be inserted according
to (\ref{ncgleg3aaar}) and (\ref{ncgleg3bbbr}), respectively.\\
At first sight, (\ref{ncgdiffg}) looks rather involved due to the
complexity of $A$ and $B$ when expressed in terms of $\cA$ and $\eta$ via
the various multiple derivatives of $\Gamma$ with respect to $\cA$, and it
seems that the former advantage of using the matrix formalism consisting
of the reduction to just one differential equation (instead of two coupled
differential equations in the component formulation) gets lost at the 
present point of the treatment. This naive first observation, however,
is misleading as is demonstrated in section~4.5. As a small comfort
we may state already now that the functions $K_J (x,y)$ and $H(x,y)$
(see (\ref{nssbex5}) and below) appearing in the corresponding differential
equations (\ref{nssbdiffg}) for $\Gamma$ within the language using 
the components $x$ and $y$ are in fact not at all less involved than 
$A$ and $B$ are in the present context.

%% file: appi.tex
\newsection{Higher orders of $\Gamma$ within the matrix formulation}

Before really considering the computer-based determination of the higher
orders $\Gamma^{(n)}$ of 
\nolinebreak $\Gamma$ which according to the proof in section~4.5
are functions of the field strength $\cF$ only let us first turn to the still
outstanding demonstration of rule~6.\\[1.5ex]
{\bf Proof of rule~6: }
We would like to begin with the proof of (\ref{ncgrule61}): Due to the 
linearity of $\frac{\partial}{\partial \cA}$ it completely suffices to show
(\ref{ncgrule61}) for $g(\cF ) = \cF^k$ with $k \in \N$ arbitrary. This
will be done by induction with respect to $k$.\\
For $k = 1$ (i.e. $g(\cF ) = \cF$) the direct calculation of $\frac{\partial
g}{\partial \cA}$ just yields (by definition) $\cF_\cA$ (\ref{ncgprf})
which obviously coincides with the r.h.s. of (\ref{ncgrule61}) because of
$\frac{\partial \cF}{\partial \cF} = \eins$. Hence, we may assume from now
on that the claim in question holds true for $k$:
\begin{displaymath}
\frac{\partial}{\partial \cA} \cF^k =
\cF_\cA \; \frac{\partial}{\partial \cF} \cF^k =
k \; \cF_\cA \; \cF^{k - 1}
\end{displaymath}
By means of the product rule~3 (\ref{ncgprod}) it then follows that
\bas
\frac{\partial}{\partial \cA} \cF^{k + 1} & = &
\frac{\partial}{\partial \cA} (\cF^k \cdot \cF ) \; = \;
\left( \frac{\partial}{\partial \cA} \cF^k \right) \cF -
\eta \cF^k \eta \; \frac{\partial}{\partial \cA} \cF \\
& = & k \; \cF_\cA \; \cF^{k - 1} \; \cF + \cF^k \; \cF_\cA \; = \; 
\cF_\cA \; (k + 1) \; \cF^k \\
& = & \cF_\cA \; \frac{\partial}{\partial \cF} \cF^{k + 1} \; \; \; .
\eas
The second equality in (\ref{ncgrule61}) is simply due to $\frac{\partial
g}{\partial \cF} \propto \eins_2$ and, thus, (\ref{ncgrule61}) is proven
by now.\\
In an almost analogous manner also the demonstration of (\ref{ncgrule62})
follows. We skip the details here (in order not to repeat trivial calculations)
and just mention that the factors of $2$ appearing on the right hand sides
of (\ref{ncgrule62}) are an immediate consequence of the fact that $\cF$ and
hence $g(\cF )$ are proportional to the unit matrix $\eins_2$ in {\it two}
dimensions implying that the trace on the l.h.s. produces these aforesaid 
factors of $2$.\\
\hspace*{14.3cm} $\Box$
\par
\vspace*{.5cm}
\noindent
In the following we present a short computer routine written in
{\it mathematica} which iter\-atively solves the differential equation
(\ref{ncgdiffgf}) order by order in $\hbar$ up to a prescribed maximal
order {\sf maxord} ({\sf maxord} $= 6$ in the example below). Please notice
that the essential part of this routine (namely the last few lines) is indeed
quite simple; the remaining part mainly consists of definitions of some
auxiliary quantities in between which are necessary to 
\linebreak handle the contributing
orders of the various terms occuring in (\ref{ncgdiffgf}) correctly:\\[1.5ex]
{\sf
Clear[f, g];\\
{\small (* Definition of the depth of recursion *)}\\
maxord = 6;\\
{\small (* Definition of the input: $f^{(0)} = f$[0,0],
           $f^{(1)} = f$[1,0] and derivatives *)}\\
$f$[0,0] := - $\frac{\lambda v^4}{8} \cF$;\\
$f$[0,i\_] := D[$f$[0,0],\{$\cF$,i\}];\\
$f$[1,0] := $(i \eins - 3 \cF) (2 i \cF - 3 \cF^2)^{- 1}$;\\
$f$[1,i\_] := D[$f$[1,0],\{$\cF$,i\}];\\
{\small (* Definition of auxiliary quantities *)}\\
help1[n\_] := Sum[$f$[k,0] $f$[n - k,0],\{k,0,n\}];\\
help2[n\_] := Sum[$f$[k,0] $f$[n - k,1],\{k,0,n\}];\\
help3[n\_] := Sum[$f$[k,1] $f$[n - k,1],\{k,0,n\}];\\
help4[n\_] := Sum[$f$[k,0] $f$[n - k,2],\{k,0,n\}];\\
h$_1^\cF$[n\_] := 2 $f$[n,0] - $i (\eins + i \cF)$ $f$[n,1];\\
h$_2^\cF$[n\_] := - 2 Sum[help1[k] help2[n - k],\{k,0,n\}] +\\
\hspace*{1cm} $i (\eins + i \cF)$ Sum[help1[k] (2 help3[n - k]
                                      + help4[n - k]),\{k,0,n\}] +\\
\hspace*{1cm} 2$(\eins + i \cF)^2$ Sum[help2[k] help3[n - k],\{k,0,n\}];\\
help[n\_,1] := 2$i (\eins + i \cF)$ help2[n] - help1[n];\\
help[n\_,2] := Sum[help[k,1] help[n - k,1],\{k,0,n\}];\\
help[n\_,3] := Sum[help[k,2] help[n - k,1],\{k,0,n\}];\\
term1[n\_] := Sum[help[k,2] h$_1^\cF$[n - k],\{k,0,n\}];\\
termr[n\_] := Sum[help[k,3] $f$[n - k,0],\{k,0,n\}];\\
{\small (* Differential equation to be solved iteratively *)}\\
dgl[n\_] := $\frac{\lambda v^4}{8}$ (h$_2^\cF$[n - 2] + term1[n - 1] -
            $\cF$ help[n,3]) - termr[n];\\
{\small (* Recursive calculation of $f^{(n)}$ and thus of $\Gamma^{(n)}$ *)}\\
For[q = 2,q $\leq$ maxord,q ++,\{ \\
\hspace*{1cm} $f$[q,0] = $f$[q,0]/.Simplify[Solve[Simplify[dgl[q]] == 0,
                         $f$[q,0]]][[1]];\\
\hspace*{1cm} $f$[q,1] = D[$f$[q,0],\{$\cF$,1\}];\\
\hspace*{1cm} $f$[q,2] = D[$f$[q,0],\{$\cF$,2\}];\\
\hspace*{1cm} $g$[q] = Simplify[Integrate[$\frac{1}{2}$ $f$[q,0], $\cF$]]\};
}
\par
\vspace*{.5cm}
\noindent
Besides $\Gamma^{(0)} = S$ (\ref{ncgaccl}) and $\Gamma^{(1)}$ which has already
been calculated in the main text, see (\ref{ncgg1}), and which is used as 
an input (in order not to bulge the routine unnecessarily) the above 
routine also yields $\Gamma^{(2)}$ and $\Gamma^{(3)}$, see (\ref{ncgg2a3}), 
as well as:\\[-1cm]
\par
{\footnotesize
\bas
\Gamma^{(4)} & = & \frac{128}{3 \lambda^3 v^{12}} \mbox{ Tr }
\left( - 160 i \cF^3 + 2544 \cF^4 + 18096 i \cF^5 - 75240 \cF^6 -
       200196 i \cF^7 + 353718 \cF^8 \right. \\
& & \left. \hspace*{2cm} + 414018 i \cF^9 - 310635 \cF^{10} - 
    136080 i \cF^{11} + 26487 \cF^{12} \right) \left( 2 i \cF - 3 \cF^2 
    \right)^{-9} \\[.5ex]
\Gamma^{(5)} & = & - \frac{512}{\lambda^4 v^{16}} \mbox{ Tr }
\left( 896 \cF^4 + 18688 i \cF^5 - 179456 \cF^6 - 1047488 i \cF^7 +
       4130992 \cF^8 \right. \\
& & \left. \hspace*{2cm} + 11573760 i \cF^9 - 23606796 \cF^{10} 
    - 35290260 i \cF^{11} +
    38392173 \cF^{12} \right. \\
& & \left. + 29630664 i \cF^{13}
    - 15381900 \cF^{14} - 4825008 i \cF^{15} + 694008 \cF^{16}
    \right) \left( 2 i \cF - 3 \cF^2 \right)^{- 12} \\[.5ex]
\Gamma^{(6)} & = & \frac{8192}{5 \lambda^5 v^{20}} \mbox{ Tr }
\left( 21504 i \cF^5 - 555520 \cF^6 - 6720000 i \cF^7 +
       50456320 \cF^8 + 262734560 i \cF^9 \right. \\
& & \left. \hspace*{2cm} - 1004053872 \cF^{10} - 2906698840 i \cF^{11} +
           6487716060 \cF^{12} \right. \\
& & \left. + 11251183950 i \cF^{13}
    - 15159210795 \cF^{14} - 15735128364 i \cF^{15} +
    12358066725 \cF^{16} \right. \\
& & \left. + 7114184640 i \cF^{17}
    - 2835423630 \cF^{18} - 698965200 i \cF^{19} \right. \\
& & \left. + 80162298 \cF^{20} \right)
    \left( 2 i \cF - 3 \cF^2 \right)^{- 15}
\eas
}
\par
\vspace*{-.5cm}
\noindent
Of course, this list can be continued easily by just raising the value
chosen for {\sf maxord}.

%% file: appj.tex
\newsection{The full set of differential equations for $\Gamma$}

We begin with the determination of the Legendre transformation
of $\frac{\partial^3 W}{\partial J^3}$ which is still missing. In principle,
we can go ahead as in appendix~H taking into account, however, the fact that
in the present treatment there are additional variables representing
the unphysical fields. All formulae given below in the first part of this
appendix are valid up to contributions proportional to $c$ whose explicit 
knowledge is not required.\\
By means of the generalized chain rule~8 (\ref{ncgchaingen}) and
the gauge fixing condition (\ref{ncggfcgen}) we at first calculate
\begin{displaymath}
0 = \frac{\partial}{\partial B} \; \left. \frac{\partial}{\partial J} \eta
\frac{\partial W}{\partial J} \right|_{{J = J(\cdots ) \atop
l = l(\cdots )} \atop \cdots} =
\frac{i v}{2} \; \xi m \; \gamma \; (W_{JJJ} + W_{J \eta J \eta J} ) 
\; + \; \frac{\partial}{\partial l} W_{J \eta J} \; \; \; .
\end{displaymath}
The zero on the l.h.s. directly results from the observation that the
Legendre transformation of $\frac{\partial}{\partial J} \eta
\frac{\partial W}{\partial J}$ does not depend on $B$ at all, see
(\ref{ncgleggen1}). Thus, we obtain:
\be \label{appj1}
\frac{\partial}{\partial l} W_{J \eta J} =
- \frac{i v}{2} \; \xi m \; \gamma \; (W_{JJJ} + W_{J \eta J \eta J} )
\ee
Next, copying the abbreviatory notation of appendix~H as far as possible,
we determine in an analogous manner (inserting (\ref{appj1}) in a step
in between):
\be \label{appj2}
\hat{A} := \frac{\partial}{\partial \cA} \; \left. \frac{\partial}{\partial J}
\eta \frac{\partial W}{\partial J} \right|_{\cdots} = \left(
\Gamma_{\cA \cA} + \frac{v^2}{4} (\xi m)^2 \eins \right) \eta W_{JJJ} -
\left( \Gamma_{\cA \eta \cA} - \frac{v^2}{4} (\xi m)^2 \eta \right)
W_{J \eta J \eta J}
\ee
In a similar way we also get:
\be \label{appj3}
\hat{B} :=  \frac{\partial}{\partial \cA} \; \left. 
\eta \frac{\partial}{\partial J} \eta \frac{\partial W}{\partial J} 
\right|_{\cdots} = \left(
\Gamma_{\cA \cA} + \frac{v^2}{4} (\xi m)^2 \eins \right) 
W_{J \eta J \eta J} + \left( \Gamma_{\cA \eta \cA} - \frac{v^2}{4} 
(\xi m)^2 \eta \right) \eta W_{JJJ}
\ee
From (\ref{appj2}) and (\ref{appj3}) it immediately follows that the
Legendre transformation in question is given by:
\be \label{ncgleggen3}
\frac{\partial^3 W}{\partial J^3} \rightarrow
- \frac{1}{h_\cA} \; \eta \left\{ \left(
\Gamma_{\cA \cA} + \frac{v^2}{4} (\xi m)^2 \eins \right) \hat{A} + \left(
\Gamma_{\cA \eta \cA} - \frac{v^2}{4} (\xi m)^2 \eta \right) \hat{B} \right\}
\ee
However, it remains to calculate $\hat{A}$ and $\hat{B}$. This can be achieved
by mimicking the respective calculations in appendix~H (concerning the
determination of $A$ and $B$). By this means we finally end up with:
\ba \label{appj4}
\hat{A} & = & - \; \eta \; \Gamma_{\cA \cA \cA} \; h_\cA^{- 1} -
\eta \left( \Gamma_{\cA \eta \cA} - \frac{v^2}{4} (\xi m)^2 \eta \right)
[ \cdots ] \; \eta \; h_\cA^{- 2} \nonumber \\[.5ex]
\hat{B} & = & - \; \Gamma_{\cA \eta \cA \eta \cA} \; h_\cA^{- 1} +
\left( \Gamma_{\cA \eta \cA} - \frac{v^2}{4} (\xi m)^2 \eta \right) \eta \;
[ \cdots ] \; h_\cA^{- 2} \\[.5ex]
\mbox{with } [ \cdots ] & = & 2 \Gamma_{\cA \cA \cA} \left(
\Gamma_{\cA \cA} + \frac{v^2}{4} (\xi m)^2 \eins \right) + \eta
\Gamma_{\cA \cA \cA} \left(
\Gamma_{\cA \eta \cA} - \frac{v^2}{4} (\xi m)^2 \eta \right) \nonumber \\
& & - \eta \left(
\Gamma_{\cA \eta \cA} - \frac{v^2}{4} (\xi m)^2 \eta \right)
\Gamma_{\cA \eta \cA \eta \cA} \nonumber
\ea
\par
\vspace{.5cm}
\noindent
Now, all necessary preparations are completed, and we simply could write
down the Legendre transformation of (\ref{ncgdiffwa}). Instead of doing so
we proceed slightly differently by explictly making use of the Slavnov-Taylor
identity (\ref{ncgstgen}) and the fact that the external fields $X$ and $Y$
have been merged into one single object $\cal X$ (\ref{ncgextfu}). The result
will be a simpler set of (almost) algebraic equations which uniquely determine
the various parts $\Gamma_0, \Gamma_1$ and $\Gamma_{23}$ of $\Gamma$, see
(\ref{gdecomgen}), order by order in the loop expansion.\\
Before actually starting with this investigation we first would like to 
mention that the Legendre transformation of the second equation in 
(\ref{ncgdiffwa}) is nothing else but the gauge fixing condition
\be \label{ncggfcgen}
\frac{\partial \Gamma}{\partial B} = B \; - \; \frac{i v}{2} \xi m 
\mbox{ Tr } \gamma \eta \cA \; \; \; ,
\ee
which due to its linearity in propagating fields holds true to all orders and,
thus, can be integrated in a trivial way.\\
Let us now turn to the third equation in (\ref{ncgdiffwa}) which after
Legendre transformation fixes the $c$-dependent part of $\Gamma$ because of
$\overline{n} = - \frac{\partial \Gamma}{\partial c}$. Hence, differentiating
this equation once more with respect to $\cal X$ helps on finding 
$\Gamma_{23}$, viz $\frac{\partial \overline{n}}{\partial {\cal X}} =
- \frac{\partial^2 \Gamma}{\partial {\cal X} \partial c} = - \Gamma_{23}$,
see (\ref{gdecomgen}). At first sight, the difficult part seems to be the
calculation of $\frac{\partial}{\partial {\cal X}} \frac{\partial^2
W}{\partial J \partial n}$. However, by inspection of (the first equation in)
(\ref{ncgleggen2}), an argument concerning the different levels of the
Grassmann algebra involved in the game (which shows that the only
$\cal X$-dependence of $\frac{\partial^2 W}{\partial J \partial n}$ enters
through the {\it explicit} $\cal X$-dependence on the r.h.s. of 
(\ref{ncgleggen2})) and some simple algebraic manipulations, the calculation
in question can be performed easily. Working that way we finally get:
\ba \label{ncgdiffga1}
\Gamma_{23} & = & \hbar \; \frac{v}{2} \frac{\xi m}{\Gamma_1} \left\{
\left[ \eta \frac{\partial^2 W}{\partial J^2} +
\frac{\partial}{\partial J} \eta \frac{\partial W}{\partial J} \right]
\frac{\partial \Gamma_{23}}{\partial \cA} 
- \eta \left[ \eta \frac{\partial^2 W}{\partial J^2} +
\frac{\partial}{\partial J} \eta \frac{\partial W}{\partial J} \right]
\frac{\partial}{\partial \cA} \eta \Gamma_{23} \right\} \nonumber \\[.5ex] 
& & - i \gamma (\cA + \eta)
\ea
Please note that (\ref{ncgdiffga1}) (up to a factor $c$) describes the
deformation of the BRS transformation of $\cA$ in higher orders, see
(\ref{ncgstgen}).\\
Next, we determine the defining equation for $\Gamma_1$. This can be 
achieved in a speedy manner
by means of the ST identity (\ref{ncgstgen}) which is valid to all orders
without proof as is explained at the end of appendix~F. In fact,
differentiating (\ref{ncgstgen}) once with respect to $B$ and making use
of the gauge fixing condition (\ref{ncggfcgen}) and the decomposition
(\ref{gdecomgen}) we find:
\be \label{ncgdiffga2}
\Gamma_1 = \frac{i v}{2} \xi m \mbox{ Tr } \Gamma_{23} \gamma \eta
\ee
The above equation is exactly the zero-dimensional remnant of the so-called
ghost equation.\\
Finally, we are in need of an equation that fixes $\Gamma_0$. This equation
will be the Legendre transformation of the first equation in
(\ref{ncgdiffwa}):
\ba \label{ncgdiffga3}
\hbar^2 \; \frac{\lambda v^4}{8} \frac{\partial^3 W}{\partial J^3} \; + \;
\hbar \left\{ \frac{\lambda v^4}{8} \left[
2 (\cA + \eta ) \frac{\partial^2 W}{\partial J^2} -
\eta (\cA + \eta ) \frac{\partial}{\partial J} \eta
\frac{\partial W}{\partial J} \right] + \frac{v}{2} \xi m 
\frac{\eta}{\Gamma_1} \right\} & & \\[.5ex]
+ \frac{\lambda v^4}{8} (\eta \cA \eta + 2 \eta \cA^2 +
\cA \eta \cA + \cA^3 - \cA ) & = & \left.
\frac{\partial \Gamma_0}{\partial \cA} \right|_{B = 0} \nonumber
\ea
Of course, in (\ref{ncgdiffga3}) $\frac{\partial^3 W}{\partial J^3},
\frac{\partial^2 W}{\partial J^2}$ and $\frac{\partial}{\partial J} \eta
\frac{\partial W}{\partial J}$ have to be replaced according to
(\ref{ncgleggen3}) and (\ref{ncgleggen1}), respectively. One easily assures
oneself of the fact that (\ref{ncgdiffga1}), (\ref{ncgdiffga2}) and
(\ref{ncgdiffga3}) completely determine $\Gamma$ order by order in the
loop expansion when exploited in a recursive manner.

%% file: dank.tex
\newpage
\thispagestyle{empty}
{\bf Acknowledgements}\\[1cm]
I am very grateful to Prof.\,\,Dr.\,\,Florian Scheck, both for his constant and
encouraging support during the development of this thesis and for the fruitful
collaboration over the past years which I enjoyed a lot.\\
Thanks are also due to Dipl.\,\,phys.\,\,Christian 
P\"oselt with whom I had many
helpful discussions on topics related to the present work.\\
In general, the stimulating atmosphere of the whole theory group 
in Mainz always made
a pleasant working possible: I owe my thanks to all the members of the 
THEP.\\[1ex]
I also would like to thank Prof.\,\,Dr.\,\,Robert Coquereaux for drawing my 
attention to questions concerning field theories on zero-dimesional space as
well as for some very interesting discussions.\\[1ex]
Last but not least, I am indebted to my wife, Annette Bohlinger, for her
patience and love.

%% file: paper.bbl
\begin{thebibliography}{99}
\addcontentsline{toc}{section}{References} 
\bibitem[AC]{AC}{\begin{tabbing}
                 \hspace*{1cm} \= \kill
                 \> A. Connes,\\ 
                 \> {\it Noncommutative Geometry},\\
                 \> Academic Press Inc. 1994
                 \end{tabbing}}
\bibitem[AS]{AS}{\begin{tabbing}
                 \hspace*{1cm} \= \kill
                 \> M. Abramowitz and I. Stegun (editors),\\
                 \> {\it Handbook of Mathematical Functions},\\
                 \> Dover Publications, New York 1970
                 \end{tabbing}}
\bibitem[BGW]{BGW}{\begin{tabbing}
                   \hspace*{1cm} \= \kill
                   \> B.S. Balakrishna, F. G\"ursey, and K.C. Wali,\\
                   \> {\it Non-commutative Geometry and Higgs Mechanism
                           in the Standard Model},\\
                   \> Phys. Lett. B254 (1991) 430\\[.5ex]
                   \> {\it Towards an Unified Treatment of Yang-Mills and
                           Higgs Fields},\\
                   \> Phys. Rev. D44 (1991) 3313
                   \end{tabbing}}
\bibitem[BTM]{BTM}{\begin{tabbing}
                   \hspace*{1cm} \= \kill
                   \> L. Baulieu and J. Thierry-Mieg,\\
                   \> {\it The Principle of BRS Symmetry: An Alternative
                           Approach to Yang-Mills}\\
                   \> {\it Theories},\\
                   \> Nucl. Phys. B197 (1982) 477
                   \end{tabbing}}
\bibitem[CES]{CES}{\begin{tabbing}
                   \hspace*{1cm} \= \kill
                   \> R. Coquereaux, G. Esposito-Far\a`ese, and F. Scheck,\\
                   \> {\it Noncommutative Geometry and Graded Algebras
                           in Electroweak Interactions},\\
                   \> Int. J. Mod. Phys. A7 (1992) 6555
                   \end{tabbing}}
\bibitem[CEV]{CEV}{\begin{tabbing}
                   \hspace*{1cm} \= \kill
                   \> R. Coquereaux, G. Esposito-Far\a`ese, and G. Vaillant,\\
                   \> {\it Higgs Fields as Yang-Mills Fields and
                           Discrete Symmetries},\\
                   \> Nucl. Phys. B353 (1991) 689
                   \end{tabbing}}
\bibitem[CFF]{CFF}{\begin{tabbing}
                   \hspace*{1cm} \= \kill
                   \> A.H. Chamseddine, G. Felder, and J. Fr\"ohlich,\\
                   \> {\it Unified Gauge Theories in Non-commutative
                           Geometry},\\
                   \> Phys. Lett. B296 (1992) 109
                   \end{tabbing}}
\bibitem[CHMS]{CHMS}{\begin{tabbing}
                     \hspace*{1cm} \= \kill
                     \> S. Cho, R. Hinterding, J. Madore, and H. Steinacker,\\
                     \> {\it Finite Field Theory on 
                             Noncommutative Geometries},\\
                     \> Int. J. Mod. Phys. D9 (2000) 161
                     \end{tabbing}} 
\bibitem[CHPS]{CHPS}{\begin{tabbing}
                     \hspace*{1cm} \= \kill
                     \> R. Coquereaux, R. H\"au\ss ling, N.A. Papadopoulos,
                        and F. Scheck,\\
                     \> {\it Generalized Gauge Transformations and
                             Hidden Symmetry in the Standard}\\
                     \> {\it Model},\\
                     \> Int. J. Mod. Phys. A7 (1992) 2809
                     \end{tabbing}}
\bibitem[CHS]{CHS}{\begin{tabbing}
                   \hspace*{1cm} \= \kill
                   \> R. Coquereaux, R. H\"au\ss ling, and F. Scheck,\\
                   \> {\it Algebraic Connections on Parallel Universes},\\
                   \> Int J. Mod. Phys. A10 (1995) 89
                   \end{tabbing}}
\bibitem[CL]{CL}{\begin{tabbing}
                 \hspace*{1cm} \= \kill
                 \> A. Connes and J. Lott,\\
                 \> {\it Particle Models and Noncommutative Geometry},\\
                 \> Nucl. Phys B (Proc. Suppl.) 18 (1990) 29
                 \end{tabbing}}
\bibitem[CLP]{CLP}{\begin{tabbing}
                   \hspace*{1cm} \= \kill
                   \> P. Cvitanovi\a`c, B. Lautrup, and R. B. Pearson,\\
                   \> {\it Number and Weights of Feynman Diagrams},\\
                   \> Phys. Rev. D 18 (1978) 1939
                   \end{tabbing}}
\bibitem[CP]{CP}{\begin{tabbing}
                 \hspace*{1cm} \= \kill
                 \> C. P\"oselt, PhD thesis, Mainz 2001
                 \end{tabbing}}
\bibitem[CR]{CR}{\begin{tabbing}
                 \hspace*{1cm} \= \kill
                 \> C. Rovelli,\\
                 \> {\it Spectral Noncommutative Geometry and Quantization:
                         A Simple Example},\\
                 \> Phys. Rev. Lett. 83 (1999) 1079
                 \end{tabbing}}
\bibitem[DFR]{DFR}{\begin{tabbing}
                   \hspace*{1cm} \= \kill
                   \> S. Doplicher, K. Fredenhagen, and J.E. Roberts,\\
                   \> {\it The Quantum Structure of Space-Time at the
                           Planck Scale and Quantum}\\
                   \> {\it Fields},\\
                   \> Commun. Math. Phys. 172 (1995) 187
                   \end{tabbing}}
\bibitem[DKM]{DKM}{\begin{tabbing}
                   \hspace*{1cm} \= \kill
                   \> M. Dubois-Violette, R. Kerner, and J. Madore,\\
                   \> {\it Non-commutative Differential Geometry and
                           New Models of Gauge Theory},\\
                   \> J. Math. Phys. 31(2) (1990) 323
                   \end{tabbing}}
\bibitem[GKW]{GKW}{\begin{tabbing}
                   \hspace*{1cm} \= \kill
                   \> H. Grosse, Th. Krajewski, and R. Wulkenhaar,\\
                   \> {\it Renormalization of Noncommutative Yang-Mills
                           Theories: A Simple}\\
                   \> {\it Example},\\
                   \> hep-th/0001182
                   \end{tabbing}}
\bibitem[GVF]{GVF}{\begin{tabbing}
                   \hspace*{1cm} \= \kill
                   \> J.M. Gracia-Bondia, J. Varilly, and H. Fiqueroa,\\
                   \> {\it Elements of Noncommutative Geometry},\\
                   \> Birkh\"auser 2000
                   \end{tabbing}}
\bibitem[H]{H}{\begin{tabbing}
               \hspace*{1cm} \= \kill
               \> R. H\"au\ss ling,\\
               \> {\it The $su(2|1)$ Model of Electroweak Interactions and
                       its Connection to}\\
               \> {\it Noncommutative Geometry},\\
               \> Invited talk given at the Hesselberg Workshop 1998
               \end{tabbing}}
\bibitem[HK]{HK}{\begin{tabbing}
                 \hspace*{1cm} \= \kill
                 \> R. H\"au\ss ling and E. Kraus,\\
                 \> {\it Gauge Parameter Dependence and Gauge Invariance
                         in the}\\
                 \> {\it Abelian Higgs Model},\\
                 \> Zeitschr. f. Physik C 75 (1997) 739\\[.5ex]
                 \> R. H\"au\ss ling and S. Kappel,\\
                 \> {\it Complete Control of Gauge Parameter
                         Dependence in the}\\
                 \> {\it Abelian Higgs Model},\\
                 \> Eur. Phys. J. C 4 (1998) 543
                 \end{tabbing}}
\bibitem[HKS]{HKS}{\begin{tabbing}
                   \hspace*{1cm} \= \kill
                   \> R. H\"au\ss ling, E. Kraus, and K. Sibold,\\
                   \> {\it Gauge Parameter Dependence in the Background
                           Field Gauge and the}\\
                   \> {\it Construction of an Invariant
                           Charge},\\
                   \> Nucl. Phys. B 539 (1999) 691
                   \end{tabbing}}
\bibitem[HLN]{HLN}{\begin{tabbing}
                   \hspace*{1cm} \= \kill
                   \> C.-Y. Lee, D.S. Hwang, and Y. Ne'eman,\\
                   \> {\it BRST Quantization of Gauge Theory in Noncommutative
                           Geometry:}\\
                   \> {\it Matrix Derivative Approach},\\
                   \> J. Math. Phys. 37 (1996) 3725
                   \end{tabbing}}
\bibitem[HP]{HP}{\begin{tabbing}
                 \hspace*{1cm} \= \kill
                 \> R. H\"au\ss ling and C. P\"oselt,
                    in preparation
                 \end{tabbing}}
\bibitem[HPaS]{HPaS}{\begin{tabbing}
                     \hspace*{1cm} \= \kill
                     \> R. H\"au\ss ling, M. Paschke, and F. Scheck,\\
                     \> {\it Leptonic Generation Mixing, Noncommutative
                             Geometry and Solar}\\
                     \> {\it Neutrino Fluxes},\\
                     \> Phys. Lett. B417 (1998) 312
                     \end{tabbing}}
\bibitem[HPS1]{HPS1}{\begin{tabbing}
                     \hspace*{1cm} \= \kill
                     \> R. H\"au\ss ling, N.A. Papadopoulos, and F. Scheck,\\
                     \> {\it $su(2|1)$ Symmetry, Algebraic Superconnections
                             and a Generalized Theory of}\\
                     \> {\it Electroweak 
                             Interactions},\\
                     \> Phys. Lett B260 (1991) 125
                     \end{tabbing}}
\bibitem[HPS2]{HPS2}{\begin{tabbing}
                     \hspace*{1cm} \= \kill
                     \> R. H\"au\ss ling, N.A. Papadopoulos, and F. Scheck,\\
                     \> {\it Supersymmetry in the Standard Model of
                             Electroweak Interactions},\\
                     \> Phys. Lett B303 (1993) 265
                     \end{tabbing}}
\bibitem[HS1]{HS1}{\begin{tabbing}
                   \hspace*{1cm} \= \kill
                   \> R. H\"au\ss ling and F. Scheck,\\
                   \> {\it Quark Mass Matrices and Generation Mixing in the
                           Standard Model with}\\
                   \> {\it Noncommutative Geometry},\\
                   \> Phys. lett B336 (1994) 477\\[.5cm]
                   \end{tabbing}}
\bibitem[HS2]{HS2}{\begin{tabbing}
                   \hspace*{1cm} \= \kill
                   \> R. H\"au\ss ling and F. Scheck,\\
                   \> {\it Triangular Mass Matrices of Quarks and
                           Cabibbo-Kobayashi-Maskawa Mixing},\\
                   \> Phys. Rev. D57 (1998) 6656
                   \end{tabbing}}
\bibitem[HTh]{HTh}{\begin{tabbing}
                   \hspace*{1cm} \= \kill
                   \> R. H\"au\ss ling,\\
                   \> {\it Das Ph\"anomen der Flavormischung 
                           im $su(2|1)$-Modell
                           der elektroschwachen}\\
                   \> {\it Wechselwirkung},\\
                   \> PhD theses, Mainz 1994
                   \end{tabbing}}
\bibitem[IZ]{IZ}{\begin{tabbing}
                 \hspace*{1cm} \= \kill
                 \> C. Itzykson and J.-B. Zuber,\\
                 \> {\it Quantum Field Theory},\\
                 \> McGraw-Hill 1980
                 \end{tabbing}}
\bibitem[KA]{KA}{\begin{tabbing}
                 \hspace*{1cm} \= \kill
                 \> E. Kamke,\\
                 \> {\it Differentialgleichungen},\\
                 \> Teubner, Stuttgart, 1983
                 \end{tabbing}}
\bibitem[KS]{KS}{\begin{tabbing}
                 \hspace*{1cm} \= \kill
                 \> E. Kraus and K. Sibold,\\
                 \> {\it Yang-Mills Theories in a Non-linear Gauge:
                         Quantization and Gauge}\\
                 \> {\it Independence},\\
                 \> Nucl. Phys. 331 (1990) 350\\[.5ex]
                 \> {\it Gauge Parameter Dependence in Gauge Theories},\\
                 \> Nucl. Phys. B 37B (1994) 120
                 \end{tabbing}}
\bibitem[KSch]{KSch}{\begin{tabbing}
                     \hspace*{1cm} \= \kill
                     \> D. Kastler and Th. Sch\"ucker,\\
                     \> {\it The Standard Model A La Connes-Lott},\\
                     \> J. Geom. Phys. 24 (1997) 1,\\
                     \> and references therein
                     \end{tabbing}}
\bibitem[MH]{MH}{\begin{tabbing}
                 \hspace*{1cm} \= \kill
                 \> M. Hale,\\
                 \> {\it Path Integral Quantization of Finite Noncommutative
                         Geometries},\\
                 \> gr-qc/0007005
                 \end{tabbing}}
\bibitem[MM]{MM}{\begin{tabbing}
                 \hspace*{1cm} \= \kill
                 \> M. Marcu,\\
                 \> {\it The Representations of $spl(2,1)$ - an Example
                         of Representations of Basic}\\
                 \> {\it Superalgebras},\\
                 \> J. Math. Phys. 21 (1980) 1277\\[.5ex]
                 \> {\it The Tensor Product of Two Irreducible Representations
                         of the $spl(2,1)$}\\
                 \> {\it Superalgebra},\\
                 \> J. Math. Phys. 21 (1980) 1284
                 \end{tabbing}}
\bibitem[MO]{MO}{\begin{tabbing}
                 \hspace*{1cm} \= \kill
                 \> K. Morita and Y. Okumura,\\
                 \> {\it Weinberg-Salam Theory in Noncommutative Geometry},\\
                 \> Prog. Theor. Phys. 91 (1994) 959\\[.5ex]
                 \> Y. Okumura,\\
                 \> {\it Standard Model in Differential Geometry on Discrete
                         Space $M(4) \times Z(3)$},\\
                 \> Prog. Theor. Phys. 92 (1994) 625\\[.5ex]
                 \> K. Morita and Y. Okumura,\\
                 \> {\it Gauge Theory and Higgs Mechanism Based on Differential
                         Geometry on}\\
                 \> {\it Discrete Space $M(4) \times Z(3)$},\\
                 \> Phys. Rev. D50 (1994) 1016
                 \end{tabbing}}
\bibitem[MS]{MS}{\begin{tabbing}
                 \hspace*{1cm} \= \kill
                 \> M. Scheunert,\\
                 \> {\it The Theory of Lie Superalgebras},\\
                 \> Lecture Notes in Mathematics, Vol. 716, Springer 1979
                 \end{tabbing}}
\bibitem[NTM]{NTM}{\begin{tabbing}
                   \hspace*{1cm} \= \kill
                   \> Y. Ne'eman and J. Thierry-Mieg,\\
                   \> {\it Exterior Gauging of an Internal Supersymmetry and
                           $SU(2|1)$ Quantum}\\
                   \> {\it Asthenodynamics},\\
                   \> Proc. Natl. Acad. Sci. USA 79 (1982) 7068
                   \end{tabbing}}
\bibitem[PR]{PR}{\begin{tabbing}
                 \hspace*{1cm} \= \kill
                 \> P. Ramond,\\
                 \> {\it Field Theory: A Modern Primer},\\
                 \> Frontiers in Physics 51, Benjamin/Cummings 1981
                 \end{tabbing}}
\bibitem[PS]{PS}{\begin{tabbing}
                 \hspace*{1cm} \= \kill
                 \> O. Piguet and K. Sibold,\\
                 \> {\it Gauge Independence in Ordinary Yang-Mills Theories},\\
                 \> Nucl. Phys. B 253 (1985) 517
                 \end{tabbing}}
\bibitem[ZJ]{ZJ}{\begin{tabbing}
                 \hspace*{1cm} \= \kill
                 \> J. Zinn-Justin,\\
                 \> {\it Quantum Field Theory and Critical Phenomena},\\
                 \> Oxford University Press, 1996
                 \end{tabbing}}
\end{thebibliography}
